\documentclass[useAMS,usenatbib,usegraphicx]{mn2e}

\usepackage[paper=a4paper,totalwidth=480pt,totalheight=680pt]{geometry}
\usepackage{amsmath}
\usepackage{subfig}
\usepackage{yhmath}
\usepackage{mathrsfs} 
\usepackage[squaren]{SIunits}

\title[Angular Power Spectra with Finite Counts]{Angular Power Spectra with Finite Counts}
\author[S.~S.~Campbell]{Sheldon~S.~Campbell\\
Center for Cosmology and AstroParticle Physics (CCAPP) and Department of Physics, The Ohio State University\\
191 W. Woodruff Ave., Columbus, OH 43210}
\date{In original form 2014 October 23}

\begin{document}

\label{firstpage}

\newcommand{\bn}{\mathbf{n}}
\newcommand{\der}{\mathrm{d}}
\newcommand{\wignerthreej}[6]{\begin{pmatrix}#1&\!\!#2&\!\!#3\\#4&\!\!#5&\!\!#6\end{pmatrix}}
\newcommand{\eqn}[1]{Eqn.~(\ref{#1})}
\newcommand{\ssec}[1]{Sec.~\ref{#1}}
\newcommand{\mean}[1]{\left\langle #1\right\rangle}
\newcommand{\meang}[1]{\mean{#1}_{\!\mathcal{G}}}
\newcommand{\meangb}[1]{\big\langle #1 \big\rangle_{\!\mathcal{G}}}
\newcommand{\psfmean}[1]{\overline{#1}}
\newcommand{\sumthree}{\!\!\!\!\!\!\!\!\!\!\sum_{\ \ \ \ \ i_3\notin\{i_1,i_2\}}}
\newcommand{\sumfour}{\!\!\!\!\!\!\!\!\!\!\!\!\sum_{\ \ \ \ \ \ \ i_4\notin\{i_1,i_2,i_3\}}}
\newcommand{\var}[1]{\mathscr{V}\!\!\text{\it ar}\,[#1]}
\newcommand{\varb}[1]{\mathscr{V}\!\!\text{\it ar}\left[#1\right]}
\newcommand{\wvar}[1]{\widehat{\mathscr{V}\!\!\text{\it ar}}\,[#1]}
\newcommand{\wvarb}[1]{\widehat{\mathscr{V}\!\!\text{\it ar}}\,\left[#1\right]}
\newcommand{\scw}{\mathscr{W}}

\maketitle

\begin{abstract}
Angular anisotropy techniques for cosmic diffuse radiation maps are powerful probes, even for quite small data sets. A popular observable is the angular power spectrum; we present a detailed study applicable to any unbinned source skymap $S(\bn)$ from which $N$ random, independent events are observed. Its exact variance, which is due to the finite statistics, depends only on $S(\bn)$ and $N$; we also derive an unbiased estimator of the variance from the data. First-order effects agree with previous analytic estimates. Importantly, heretofore unidentified higher-order effects are found to contribute to the variance and may cause the uncertainty to be significantly larger than previous analytic estimates---potentially orders of magnitude larger. Neglect of these higher-order terms, when significant, may result in a spurious detection of the power spectrum. On the other hand, this would indicate the presence of higher-order spatial correlations, such as a large bispectrum, providing new clues about the sources. Numerical simulations are shown to support these conclusions. Applying the formalism to an ensemble of Gaussian-distributed skymaps, the noise-dominated part of the power spectrum uncertainty is significantly increased at high multipoles by the new, higher-order effects. This work is important for harmonic analyses of the distributions of diffuse high-energy $\gamma$-rays, neutrinos, and charged cosmic rays, as well as for populations of sparse point sources such as active galactic nuclei.
\end{abstract}


\section{Introduction}
An important experiment in astronomy is measuring the angular distribution of points on the sky from the arrival directions of incident radiation. When distance information for sources is unavailable (or too unreliable), then techniques for quantifying their two-dimensional angular distribution become essential. These methods are important for analyzing distributions of point sources, as well as incident radiation from diffuse sources (or those that appear to be, due to insufficient angular resolution).

The information contained in angular distributions depends on the application. Temperature anisotropies of the cosmic microwave background (CMB) contains information about primordial fluctuations at the epoch of last-scattering, as well as the distribution of matter and ionized gas at subsequent epochs that affected the propagation of the microwaves \citep[e.g.,][]{WMAP2013,Planck2014o}. The distribution of galaxies \citep[e.g.,][]{Hayes2012,Ho2012} or quasars \citep{Leistedt2013,Ho2013} probes the large scale structure of matter. High-energy messengers, the focus of this article, have special challenges: charged cosmic rays are deflected, $\gamma$-rays can be attenuated, and neutrinos are detected only in small numbers. Directional detection of these messengers allows inferences about their sources and propagation effects.

One popular measure of the angular distribution is its power spectrum $C_\ell$, the mean-square-amplitudes of fluctuations with wavelength $\pi/\ell$ radians, specified with a basis of spherical harmonics $Y_\ell^{\,m}$. These are particularly convenient observables because they characterize the angular scales of anisotropy and they are statistically orthogonal, $\mean{C_\ell C_{\ell'}}-\mean{C_\ell}\mean{C_{\ell'}}\propto\delta_{\ell\ell'}$, for full-sky Gaussian-distributed skymaps. Since the $C_\ell$ are 2-point functions, they represent the lowest order deviations from isotropy. Such harmonic analyses are also applied to the clustering of three-dimensional cosmological data in thin spherical shells to simplify redshift distortion effects \citep*[e.g.,][]{Fisher1994,Percival2004}.

One intent for analyses of galaxy surveys is to determine the statistical properties of the large scale structure in terms of their typical spatial correlations. These correlations contain information about the physics of the early Universe, the expansion history of the Universe, and the gravitational collapse of matter.

Rather than measure the statistical properties of the spatial distribution, the intent of this paper is to focus on the particular spatial distribution of observations at our particular position in the Universe we live in. We treat the power spectrum as an observable property of the source skymap itself, instead of a measurement of its statistical properties. One reason to do this is because it provides potentially important information: as the number of detected events increases, the power spectrum of high-energy gamma-rays and neutrinos will converge precisely to the flux power spectrum of their sources. In general, harmonic decompositions are a convenient way to express information about the spatial structure of astrophysical observations.

A crucial step in extracting the information contained in these measurements is assessing their uncertainty. In general, the statistical variance must be determined from the data itself.

In some cases, one can determine the variance directly from the mean or from other measurements, due to theoretical considerations. One important example is the variance of the primordial CMB temperature power spectrum, which is strongly constrained to be Gaussian-distributed, is measured with incredibly large photon counts, and where the goal is to constrain cosmological parameters. In this case, the limiting statistical uncertainty of the power spectrum is the cosmic variance over the ensemble of Universes with the same cosmology and different initial conditions. Proportional to $C_\ell^2$, this uncertainty is estimated with the theoretical power spectrum that best fits the data. If we didn't have a cosmological theory of the CMB (but still knew it to be Gaussian-distributed), then the cosmic variance would need to be approximated from an estimate of $C_\ell^2$ using the data alone. When determining the statistical uncertainty from the data in this way, one must ensure there is enough data to guarantee that the power spectrum's error estimate is sufficiently precise.

If one wants to know the CMB power spectrum in \emph{our specific sky}, perhaps to compare with the power spectrum of another related distribution such as known X-ray galaxies, its statistical uncertainty is smaller than the cosmic variance, and the limiting uncertainty is instrument noise.

Error analyses that require spatial modelling of the data (such as assuming Gaussianity, for example) may not be applicable or be easy to properly apply to data that is not well understood. There are many examples of this. The sources of ultra-high energy cosmic rays are unknown and their propagation has uncertainties that make modeling the expected signal difficult. As such, quantifying the statistical significance of observed features, such as hotspots \citep{TA2014}, or dipole phase transitions \citep*[e.g.,][]{Giaccari2013,Ivanov2014}, require a spatial-model independent approach \citep[such as in][]{Linsley1975}. Likewise for anisotropy features in high-energy astrophysical ensembles of observed charged cosmic rays \citep*[e.g.,][]{Westerhoff2013,Santander2013,Desiati2014,Tamburro2014}, neutrinos \citep*{IC2013,IC2014a,IC2014b,IC2014c}, and diffuse $\gamma$-rays \citep{FLAT2012}. Spatial-model independent approaches are also useful for verifying the applicability of more efficient model-dependent techniques, and providing a fall-back standard to apply in regimes where those models prove inadequate. A current example of such a modeling challenge is non-Gaussianities in the galaxy distribution induced at small distance scales (including the baryon acoustic oscillation scale) by non-linear gravitational collapse \citep[e.g.,][]{SDSS2006,Harnois-Deraps2013}.

Thus, there is a need for statistical uncertainty estimates of the power spectrum using data alone, independent of information about the radiation's source. In particular, there has been less development of power spectrum measurements for experiments with a relatively small number of counts. Such experiments do not require data compression techniques to be applied---the low counts of high energy data may allow exact methods to be computationally tractable, without the need to bin the data into sky-pixels. \emph{For these experiments, new techniques are now needed, including methods to determine the variance of observables such as the power spectrum.}

A first precise formulation of this kind is presented in this paper for the angular power spectrum of finite counts. The statistical framework is minimal and widely applicable, containing only two postulates:
\begin{enumerate}
\item stationarity---the data consists of point events on a sphere sourced from a stationary skymap distribution $S(\bn)$, and
\item independence---each observed event is statistically independent of the other events (Poisson process), with its spatial probability density function proportional to $S$.
\end{enumerate}

The stationarity postulate is only broken when the data contains significant contributions from transients on time scales of the order of the experiment livetime. The independence postulate is broken for any beams that exhibit quantum boson bunching (such as for the CMB) or fermion antibunching (as expected in the cosmic neutrino background), known as the HBT effect \citep{Brown1957,Zmuidzinas2003,Jeltes2007}. Such effects are left for future studies.

With this framework, we show how to estimate the power spectrum without bias\footnote{It is worth pointing out that the presented estimators in this work are unbiased with respect to a fixed number of observed events, instead of a fixed exposure. Indeed, it is unclear how to determine an unbiased estimator of power spectrum from fixed exposure as long as there is a finite probability of receiving less than two events, in which case the estimate of the power spectrum is undefined.} for any finite set of data with at least two events, and show how its variance depends on the source distribution. In particular, we find new contributions to the statistical variance of the power spectrum, due to the source distribution's power spectrum, bispectrum, and trispectrum. We provide new unbiased estimates of the variance derived from the data, applicable for any distribution of sources. This work is immediately important for the high-energy $\gamma$-ray anisotropy analyses carried out with \emph{Fermi-LAT} data \citep{FLAT2012}.

For simplicity of presentation in this first paper, the analysis is presented for uniform-exposure full-sky observations.

The rest of this paper is organized as follows. In \ssec{sec:cvar}, we briefly discuss some of the current methods of power spectrum estimation, review previous analytic uncertainty estimates of the power spectrum and preview the new results of this paper for comparison. \ssec{sec:clvar} contains the detailed derivation of the new results. Arguments are given for the specific case of a $\gamma$-ray analysis. In \ssec{sec:comp}, we present a detailed discussion of the results, and verify the new effects in some initial numerical simulations. A discussion for the applicability of the results to other experiments appears in \ssec{sec:exp}. We conclude in \ssec{sec:conc}.

The error analysis of $C_\ell$ is most easily expressed in terms of rotation-invariant spherical tensors of higher order than the angular power spectrum. A study of these new higher order spectra is provided in App.~\ref{ap:sphten}. Despite many searches, we are unaware of any previous description of these tensors. The maximum likelihood analysis of the angular power spectrum of a Gaussian distribution on a sphere is reviewed in App.~\ref{ap:likelihood}, and the derivation of central moments of Gaussian fields is reviewed in App.~\ref{ap:moments}.

Some notations consistently appear in this paper. An unbiased estimator $\hat{X}$, dependent on random data, of some physical quantity $X$ is denoted with a `$\hat{\ \ }$'. Measurements of spherical-mean-normalized angular distributions are indicated throughout with a `$\tilde{\ \ }$'.

\section{Angular Power Spectra: From High to Low Statistics, From Sky Ensembles to a Single Sky}
\label{sec:cvar}
This section will briefly discuss the existing methods for power spectrum estimation, and preview the main results of this paper for comparison. The derivation of these results and detailed comparison with the existing methods are then carried out in the next section.

We will see that the current methods are suitable when data sets are very large, such as for galaxy surveys, where the goal is to ascertain the statistical correlations of the spatial distribution of the observations. Galaxy surveys are often analyzed to measure the parameters of a pre-stated cosmological model. As such, the power spectrum measurements in this context are estimators for the cosmic mean angular spectrum over an ensemble of skies that are realizations of the same distribution (i.e. produced by the fiducial cosmology). Because, of course, the experiments are limited to sample only the single sky they have access to, the analysis can never reach a statistical precision better than the cosmic variance of the measurement over the theoretical ensemble of possible skies. This cosmic variance represents a fundamental statistical limit to the precision that cosmological parameters can be determined for a given cosmological model.

In contrast, detection of high-energy cosmic radiation (such as $\gamma$-rays) have motivations of determining the sources of the radiation and learning about the high-energy processes that govern the Universe. Until a model of sources is established, experiments benefit by first focusing on measurements of the radiation distribution for our specific sky. In this context, the precision of these measurements is not bounded by cosmic variance; in fact, the distribution of a static sky can, in principle, be measured to any statistical precision, bounded only by the sky transience and irreducible experimental systematics. Matching the angular distribution of diffuse $\gamma$-rays to the distribution of potential sources is an important technique for resolving the $\gamma$-ray sky.

The additional precision achieved by measuring the power spectrum of the actual sky (rather than a cosmic mean power spectrum) is useful for searches of energy modulation features in the power spectrum from dark matter annihilation \citep{SiegalGaskins2009}. Not only is the power spectrum of dark matter annihilation predicted to be significantly different from other $\gamma$-ray sources \citep{Ando2006,Ando2007}, but it is sensitive to the abundance and profile of dark matter subhalos \citep{Campbell2013,Calore2014a}, as well as the velocity-dependence of extragalactic annihilations \citep{Campbell2010,Campbell2011}.

Now that the \emph{Fermi-LAT} has determined a diffuse component to the cosmic $\gamma$-rays and measured a positive power spectrum, this information is already being used to constrain models of the sources \citep*[e.g.,][]{FLAT2012,Cuoco2012,Harding2012,GomezVargas2012,Ando2013,Broderick2013b,Calore2014,DiMauro2014}. For example, blazars that are relatively sparse in the sky would generate a large intensity anisotropy if they were too bright, thus limiting their possible contribution to the net diffuse background \citep{Harding2012}. 

This pioneering work by the \emph{Fermi-LAT} collaboration had the foresight to estimate the uncertainty of the power spectrum using methods similar to the cosmological experiments---these correctly estimated the primary sources of uncertainty. The strategy applied by the \emph{Fermi-LAT} working group was to analyze the earliest data set with the assumption that it is Gaussian-distributed, and to test these assumptions with subsequent data (Komatsu, private communication).

However, we will see that neglected higher order terms due to finite count effects and non-Gaussianities can produce non-negligible contributions to the variance of the power spectrum, and can be determined reliably from the data, allowing the assumption of Gaussianity to be tested immediately. Neglecting the higher order terms can cause the power spectrum variance to be biased, though whether the bias is positive or negative depends on the properties of the skymap.

It was pointed out in \citet{Broderick2013a} that the power spectrum in the highest energy bin $\unit{10.4}{GeV}<E<\unit{50.0}{GeV}$ was measured to be inconsistently small because subsequent resolved point sources that were not masked in the original analysis appear to produce a larger intensity power spectrum than the original measurement (an estimated 3$\sigma$ discrepancy). This could indicate that the uncertainty in that energy bin was underestimated. Significant non-Gaussianities in the distribution of that energy bin would resolve this apparent discrepancy.

We now review the error analysis that is appropriate for the cosmological experiments measuring the statistics of large scale structure and formed the basis for the \emph{Fermi-LAT} analysis. We then compare this technique to the new exact methods derived in this paper for measuring the power spectrum of our particular sky, which we will preview in this section without derivation. The results reveal the conditions under which the original analysis applies. A full derivation is presented in \ssec{sec:clvar}, and the consequences for inferencing the power spectrum known to be sourced from a Gaussian sky is discussed in \ssec{ssec:gaussian}.

Perhaps the most famous angular power spectrum is of the temperature fluctuations of the CMB. The Gaussianity of these fluctuations implies that their statistical properties are fully specified by their power spectrum. The method of estimating the power spectrum of the temperature field $T(\bn)$ via the harmonic amplitudes
\begin{equation}
  c_{\ell m}=\int\der\bn\,Y^*_{\ell m}(\bn)\left[T(\bn)-T_0\right],
\end{equation}
where $T_0$ is the mean CMB temperature, is often called the pseudo-spectrum in the CMB literature \citep{Hivon2002}
\begin{equation}
  \label{eq:Clest}
  C_\ell=\frac{1}{2\ell+1}\sum_{m=-\ell}^\ell\left|c_{\ell m}\right|^2=\mathscr{N}_\ell+\hat{C}_\ell.
\end{equation}
The pseudo-spectrum estimator is biased by photon shot noise, sky pixelation noise, and instrument noise with spectrum $\mathscr{N}_\ell$, which is modeled and subtracted to determine the unbiased estimator $\hat{C}_\ell$ of the CMB power spectrum. For example, the instrument noise experienced by the High Frequency Instrument on the \emph{Planck} Satellite is due to its bolometers' detector noise (designed to be smaller than the photon noise produced from the cosmic background), thermal emission from the cooled telescope, and also from the cryogenic filter stages in the telescope \citep*{Holmes2008,Planck2014o}.

While early CMB experiments made use of the pseudo-spectrum estimator, realistic implementations of the pseudo-spectrum analysis must also consider effects due to sky pixelation, masking, non-uniform exposure, and instrument resolution \citep{Hivon2002}. Computational limitations require the sky to be pixelized as an effective way to reduce the large volume of data without loss of information \citep*{Tegmark1997b,Tegmark1997c}. Modern analyses tend to also use more general quadratic maximum likelihood estimator methods on large data sets, which can be made more computationally feasible \citep{Tegmark1997a} and more precise for certain applications \citep{Molinari2014}. Hybrid methods also exist \citep{Efstathiou2004,HAWC2014}. Maximum likelihood estimators generally assume the data are spatially Gaussian-distributed in order to write down a likelihood function.

It may be possible to generalize maximum-likelihood methods for non-Gaussian skymaps. In principle, it is possible to derive the probability density function (PDF) of any estimator of $C_\ell$ for any set of random event data of size $N$ from either a fixed skymap $S(\bn)$ or statistical ensemble of skymaps. This PDF can then be used to determine the likelihood function for the actual value for $C_\ell$, given a particular set of data, from which the maximum likelihood estimator and minimum variance estimate can be made. The dependence on $S(\bn)$, though, requires careful theoretical modeling of the skymap. When a reliable model of the skymap is unavailable, methods that do not use prior information about the spatial distribution of the skymap may be more desirable. Such a method is what is presented in this paper.

When the power spectrum of fluctuations is produced by a Gaussian random process, it will differ between different realizations of the random process. When $C_\ell$, derived with \eqn{eq:Clest}, is interpreted as an estimator of the cosmic mean power spectrum over all possible realizations of the Gaussian process, then the variance of $C_\ell$ from realization to realization, i.e., the \emph{cosmic variance}, is given by (see App.~\ref{ap:likelihood})
\begin{equation}
  \label{eq:cvarbare}
  \var{C_\ell}=\frac{2}{2\ell+1}C_\ell^2,
\end{equation}
explained in App.~\ref{ap:likelihood} to be minimal for the estimator in \eqn{eq:Clest}. Likewise, the variance of noise contributing to multipole $\ell$ is
\begin{equation}
  \var{\mathscr{N}_\ell}=\frac{2}{2\ell+1}\mathscr{N}_\ell^2.
\end{equation}
It was demonstrated in \citet{Knox1995} that if Gaussian fluctuations are pixelized, the determination of $C_\ell$ contains a pixel noise due to fluctuation correlations within a single pixel, and then the estimator $\hat{C_\ell}$ is $\chi^2_{2\ell+1}$ distributed with cosmic variance
\begin{equation}
  \label{eq:cvar}
  \var{\hat{C}_\ell}=\frac{2}{2\ell+1}\left(\mathscr{N}_\ell+C_\ell\right)^2.
\end{equation}
That is, it is a fundamental property of the statistical distribution of $C_\ell$ that the standard deviations of $C_\ell$ and $\mathscr{N}_\ell$ are added to each other, not the variances (see also App.~\ref{ap:likelihood} for justification). This simple prescription, properly applied, was shown to give a precise characterization of the variance of Boomerang CMB data \citep{Hivon2002}. We refer to this prescription as \emph{cosmic variance with noise}.

The \emph{WMAP} measurement of the CMB temperature power spectrum was presented with this uncertainty separated into two contributions \citep{WMAP2013}. The measured values of $\hat{C}_\ell$ had an uncertainty due to those terms containing the noise,
\begin{equation}
	\label{eq:noisevar}
	\var{\hat{C}_\ell}=\frac{2}{2\ell+1}\mathscr{N}_\ell\left(\mathscr{N}_\ell+2(C_{\text{th}})_\ell\right),
\end{equation}
with appropriate adjustments for binning in multipole and partial sky coverage, and where $(C_{\text{th}})_\ell$ is the best-fit theoretical power spectrum from a $\Lambda$CDM cosmology. For reasons that will become apparent, we refer to this variance as \emph{$C_\ell$-only}. It represents the statistical uncertainty of the measurement, accounting for differences of the measurement from the actual power spectrum of the sky due to randomness in the data. The remaining contribution is simply the cosmic variance. It gives the magnitude that the power spectrum of our sky's realization of the CMB fluctuations can randomly differ from the theoretical cosmic mean power spectrum. It is shown as an uncertainty band on $(C_{\text{th}})_\ell$, representing its status as a theoretical uncertainty.

The success of this analytic variance to capture so well the simulated variances of CMB experiments motivated the use of \eqn{eq:cvar} in the error estimate for an early measurement of the angular power spectrum of the diffuse gamma-ray background by the \emph{Fermi-LAT} collaboration \citep{FLAT2012}. In this application, the noise $\mathscr{N}_\ell$ corresponds to multipole-independent photon shot noise, which is the variance of the power spectrum for a Poisson point process where events are equally likely from each position in the sky. This approach was suitable for early $\gamma$-ray data sets for which the uncertainty is dominated by the shot noise. Since the shot noise term is well understood and was estimated to dominate the error bar, inaccuracies in the signal contribution to the variance were not a source of concern. Since the diffuse background appears isotropic on large angular scales, it was presumed to be dominantly from extragalactic sources. Thus, cosmic variance was included in the error estimate to aid the constraint of theoretical models of extragalactic sources on linear distance scales; however, its contribution to the error bars was negligible in that analysis.

The main goal of this paper is to carry out a detailed analytic study of the statistical uncertainty of measurements of angular power spectra derived from a finite number of independent random events, where the probability of each individual event observation is weighted according to the source's intrinsic skymap. We focus on the case where the observed events are high-energy $\gamma$-rays, with probability weighted by the apparent flux skymap of sources with intensity power spectrum $C_\ell$. Application to other experiments is discussed in \ssec{sec:exp}. We also focus in this paper on the measurement on our particular sky-realization's power spectrum, for which there is no cosmic variance. In \ssec{ssec:gaussian}, we will show how to apply our results to the case of measuring the cosmic mean power spectrum of a Gaussian distribution, for which cosmic variance is an important effect.

The photon shot noise has magnitude (see \ssec{ssec:shot})
\begin{equation}
  \mathscr{N}=\frac{4\pi N}{\varepsilon^2}
\end{equation}
after $N$ $\gamma$-rays are observed over the full sky with uniform exposure $\varepsilon$. Then \eqn{eq:Clest} predicts that the unbiased estimator $\hat{C}_{\ell,N}$ of the power spectrum of the signal source is related to the power spectrum $C_{\ell,N}$ of the $N$ observed events by
\begin{equation}
  C_{\ell,N}=\frac{4\pi N}{\varepsilon^2}+\hat{C}_{\ell,N}.
\end{equation}

In contrast, our analysis finds that \eqn{eq:Clest} is modified at low counts to
\begin{equation}
  \boxed{C_{\ell,N}=\frac{4\pi N}{\varepsilon^2}+\left(1-\frac{1}{N}\right)\hat{C}_{\ell,N}}
\end{equation}
and we find that $\hat{C}_{\ell,N}$ can be determined directly as a sum of Legendre polynomials of the `distances' between events
\begin{equation}
  \label{eq:Clestimator1}
  \hat{C}_{\ell,N}=\frac{4\pi N^2}{\varepsilon^2}\Bigg[\frac{1}{N(N-1)}\sum_{i=1}^N\sum_{\substack{j=1\\j\neq i}}^NP_\ell(\bn_i\cdot\bn_j)-\delta_{\ell,0}\Bigg].
\end{equation}
As usual, the inner product $\bn_1\cdot\bn_2$ is defined as the cosine of the angular distance between the positions. While this modification does not affect previous estimates of the power spectrum when $N\gg1$, the corrections to the measurement's variance are potentially more important.

The variance estimate that provided the starting point for \emph{Fermi-LAT}'s error analysis is the cosmic variance with shot noise variance from \eqn{eq:cvar},
\begin{equation}
  \label{eq:cvar2}
  \Big[\var{\hat{C}_{\ell,N}}\Big]_{\text{CV}}=\frac{2}{2\ell+1}\left(\frac{4\pi N}{\varepsilon^2}+C_\ell\right)^{\!\!2},
\end{equation}
in terms of the true power spectrum $C_\ell$. The strategy to estimate this variance is to substitute $C_\ell$ either with the measured spectrum $\hat{C}_{\ell,N}$, or with a modeled spectrum that best fits the data. Since, as we just explained, the measurement of our particular sky does not contain cosmic variance, we remove the cosmic variance for our analysis, as in \eqn{eq:noisevar}. Thus, the analytic estimate leaves us with the $C_\ell$-only variance
\begin{equation}
  \label{eq:cellonlyvar}
  \wvar{\hat{C}_{\ell,N}}=\frac{2}{2\ell+1}\frac{4\pi N}{\varepsilon^2}\left(\frac{4\pi N}{\varepsilon^2}+2\hat{C}_{\ell,N}\right).
\end{equation}
  
Our analysis finds that the exact variance of $\hat{C}_{\ell,N}$ is
\fbox{
\addtolength{\linewidth}{-4\fboxsep}%
\addtolength{\linewidth}{-2\fboxrule}%
\begin{minipage}{\linewidth}
\begin{align}
  \var{\hat{C}_{\ell,N}}&=\frac{1}{1-\frac{1}{N}}\Bigg[\!\frac{2}{2\ell+1}\left(\frac{4\pi N}{\varepsilon^2}\right)^{\!\!2}\label{eq:varprev}\\
  &\quad+\frac{4}{2\ell+1}\frac{4\pi(N-2)}{\varepsilon^2}C_\ell+\frac{32\pi^2}{\varepsilon^2}\,C_\ell^{(2)}\nonumber\\
  &\quad+\left(1-\frac{2}{N}\right)\frac{16\pi}{\varepsilon}\,C_\ell^{(3)}-\frac{4N-6}{N^2}\,C_\ell^2\Bigg],\nonumber
\end{align}
\end{minipage}
}
where $C_\ell^{(2)}$ is a new 2-point spectrum that can be written as a linear combination of components of $C_\ell$, and $C_\ell^{(3)}$ is a new 3-point spectrum that is a linear combination of components of the angular bispectrum (see App.~\ref{ap:sphten} for definitions and details).

The first two terms of \eqn{eq:varprev} are slight corrections of \eqn{eq:cellonlyvar}. The remaining three terms of \eqn{eq:varprev} are new corrections to the statistical variance of the power spectrum. $C_\ell^{(2)}$ provides a correction to the shot noise term of the spectrum (at $N^{-2}$), and contributes a dependence on the power spectrum's neighboring $\ell$ components. $C_\ell^{(3)}$ enters at the same order as the signal term (asymptotically $N^{-1}$), accounting for effects from the signal's bispectrum. The $C_\ell^2$ term provides trispectrum effects that appear to \emph{reduce} both the shot and signal terms of the variance. This reduction of the power spectrum variance by the presence of a significant trispectrum is not clearly understood.

\begin{figure*}
  \subfloat{\includegraphics[width=0.5\textwidth]{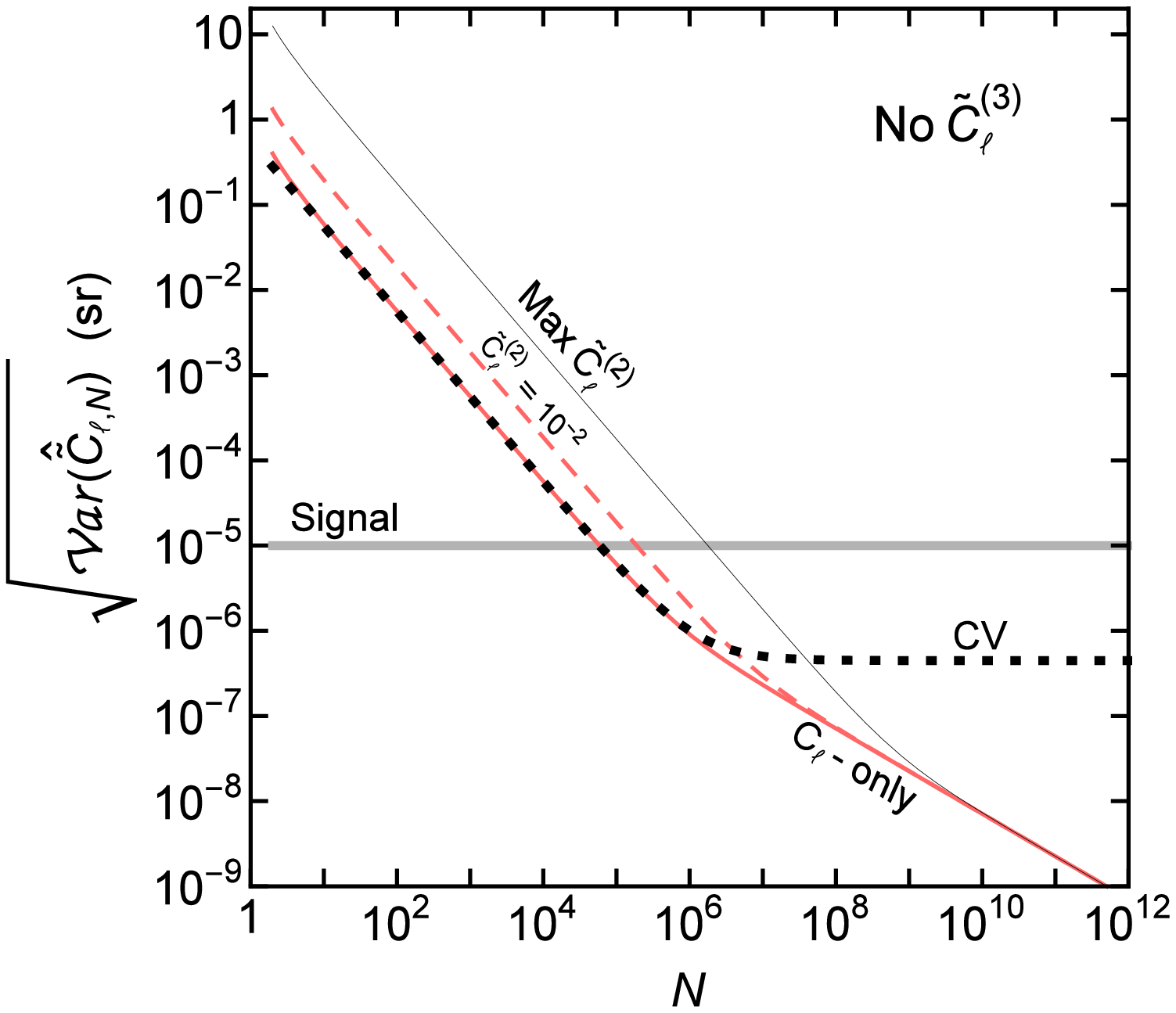}}
  \subfloat{\includegraphics[width=0.5\textwidth]{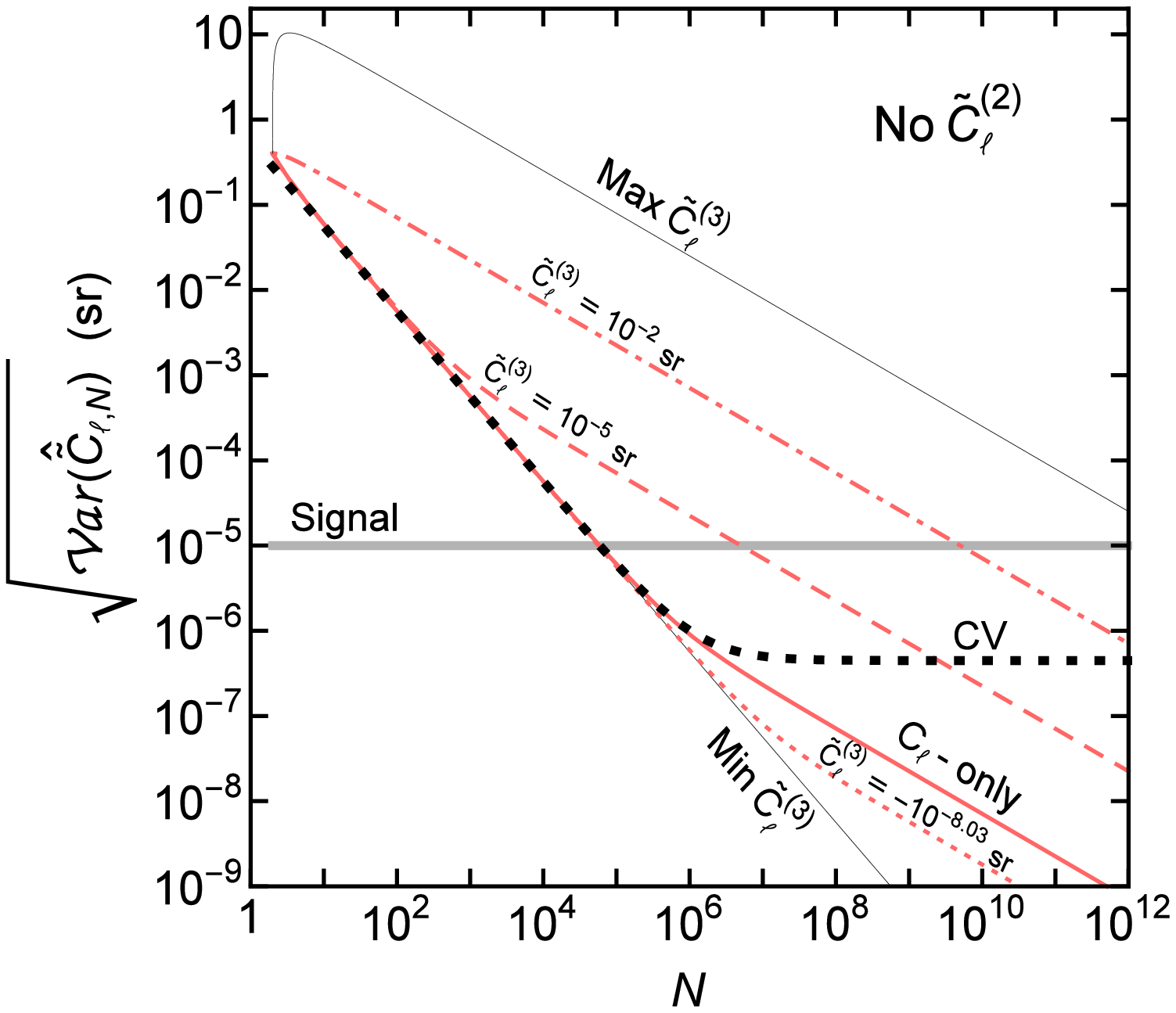}}
  \caption{\label{fig:clerror}Statistical uncertainty of the power spectrum of $N$ points at $\ell=500$ when $\unit{\tilde{C}_{500}=10^{-5}}{sr}$. The magnitude of $\tilde{C}_{500}$ is indicated by the horizontal gray band. The red solid line shows the `$C_\ell$-only' variance, when the new effects from $\tilde{C}_\ell^{(2)}$ and $\tilde{C}_\ell^{(3)}$ are negligible. For comparison, the cosmic variance with noise scenario---which is used for probing extragalactic source models---is shown with the black dotted curve labelled `CV'. Left: The effect of a non-negligible $\tilde{C}_\ell^{(2)}$ increases the shot-dominated portion of the uncertainty, as shown by the dashed red curve, though never enough to reach the `max' thin black line. Right: A significant positive bispectrum produces an early signal transition (red dashed curve) slowing the rate at which the uncertainty decreases with additional counts. This produces a much larger statistical error. The dot-dashed curve shows the effect for a very large $\tilde{C}_\ell^{(3)}$, becoming up to 1000 times larger than the $C_\ell$-only error, and showing that it may be possible for the signal term to always exceed the shot term. The signal term cannot be as large as the `max' thin black line. A negative bispectrum is possible (red dotted curve), and is the only way for this variance to be significantly below the $C_\ell$-only estimate for this small value of $\tilde{C}_\ell$, though not below the `min' thin black line. For comparison, the \emph{Fermi-LAT} analysis measured a similar power spectrum with $N\sim10^5-10^6$ events per energy bin over $\sim1/3$ of the sky \citep{FLAT2012}. See \ssec{ssec:compare} for a complete discussion of these results.}
\end{figure*}

The effects of the new terms in the variance are shown in Fig.~\ref{fig:clerror} in terms of the fluctuation power spectrum $\tilde{C}_\ell\equiv(\varepsilon/N)^2C_\ell$ for varying magnitudes of $\tilde{C}_\ell^{(2)}\equiv(\varepsilon/N)^2C_\ell^{(2)}$ and $\tilde{C}_\ell^{(3)}\equiv(\varepsilon/N)^3C_\ell^{(3)}$. For comparison, the cosmic variance with noise is also shown. A complete discussion of these effects is in \ssec{ssec:compare}.

In addition, we derived an unbiased estimator for the variance from the $N$ data points, presented in \ssec{ssec:varestimator}. Importantly, it allows the uncertainty to be estimated from the data, independent of any simulations or assumed models of the sources. There are simple estimators for the shot term and signal term separately, but their sum simplifies further to the total estimator
\begin{equation}
  \label{eq:Clvarest}
  \boxed{\wvar{\hat{C}_{\ell,N}}=\hat{C}_{\ell,N}^2-\hat{C}_{\ell,N}^{(4)}}
\end{equation}
using \eqn{eq:Clestimator1} and the estimator of the trispectrum term
\begin{align}
  &\hat{C}_{\ell,N}^{(4)}=\!\left(\!\frac{4\pi N^2}{\varepsilon^2}\!\right)^{\!\!2}\Bigg[\frac{1}{N(N-1)(N-2)(N-3)}\sum_{i_1}\sum_{i_2\neq i_1}\nonumber\\
  &\sumthree\sumfour\hspace{-3mm}P_\ell(\bn_1\cdot\bn_2)P_\ell(\bn_3\cdot\bn_4)-\delta_{\ell,0}\Bigg].
\end{align}
The simplicity of Eqns.~(\ref{eq:Clestimator1}) and (\ref{eq:Clvarest}) suggest that it is practical for sufficiently small counts to consider the events point-by-point as opposed to binned in pixels. This would have the additional benefit of not having to account for artificial pixel effects in the results.

The complete derivation of all these results is carried out in the following section.

\section{Angular Power Spectra of Cosmic Events}
\label{sec:clvar}
The new error analysis technique is now described. The problem statement for this analysis can be given in a general way. In this section, we will apply it specifically to high-energy $\gamma$-ray events, and will briefly discuss other experiments in \ssec{sec:exp}.

A position on the sky (i.e. on a sphere) is given by a unit vector $\bn$ pointing from the sphere's center. If a spherical coordinate system is defined on the sphere, then $\bn=(\theta,\phi)$ may be specified by polar angle $\theta$ and azimuthal angle $\phi$.

We identify a class of events to be analyzed, and identify their positions in the sky $\bn_1, \bn_2, \bn_3, \ldots$\,. For the case of $\gamma$-ray events, these could be a particular class of event reconstruction in some specified energy range.

Let $S(\bn)$ be the \emph{actual} angular distribution of the signal's intrinsic `skymap': the rate of events at each position, as would be observed in the limit of infinite $N$. For $\gamma$-rays, $S(\bn)$ is the apparent $\gamma$-ray flux over the energy range of interest.

Note that no effects due to the instrument's point spread function (PSF) are being applied to $S(\bn)$. However, we will explicitly derive in \ssec{ssec:estimator} that the only effect of the PSF is to convolve with $S(\bn)$, and that the measured spectrum is an estimator of the convolved skymap. We will thereafter assume that $S(\bn)$ refers to the intrinsic skymap convolved with the instrument PSF.

If $S(\bn)$ is non-negative and non-trivial (i.e., not zero everywhere), then a very useful quantity is the normalized skymap
\begin{equation}
  \tilde{S}(\bn)\equiv\frac{S(\bn)}{\int{\frac{\der\bn'}{4\pi}}\,S(\bn')},
\end{equation}
which fluctuates around 1 on the sphere. Note that $\der\bn=\der\phi\,\der(\cos\theta)$ denotes the usual sphere measure.

The power spectrum and fluctuation power spectrum of the skymap are
\begin{align}
  &C_\ell=\frac{1}{2\ell+1}\sum_{m=-\ell}^\ell\left|c_{\ell m}\right|^2,\\
  &\tilde{C}_\ell=\frac{1}{2\ell+1}\sum_{m=-\ell}^\ell\left|\tilde{c}_{\ell m}\right|^2,
\end{align}
where $c_{\ell m}$ is the spherical transform of $S-\int{S\,\der\bn/(4\pi)}$, and $\tilde{c}_{\ell m}$ is the spherical transform of $\tilde{S}-1$. The values of $c_{\ell m}$ are not invariant under rotations of the sphere and depend on the definition of the spherical coordinate system applied to the sphere, but $C_\ell$ is a rotation-invariant tensor.

The problem is to consider $N$ points received from random locations on the sphere with probability distribution specified by $\tilde{S}(\bn)$, and determine the mean and variance of the power spectrum of these events.

For the observation of radiation events, it is useful to determine the intensity power spectrum. Let $\mathcal{A}(\bn,t)$ be the effective area of the instrument to events from position $\bn$ at time $t$. Then the total exposure of the instrument is
\begin{equation}
  \varepsilon(t)=\int{\der\bn}\int_0^t{\der t'}\,\mathcal{A}(\bn,t')
\end{equation}
integrated over the field of observation in the sky, and instrument livetime. Let $N(t)$ be the number of detected events with this exposure. Then the total observed intensity is
\begin{equation}
  I(\bn,t)=\frac{\der N(\bn,t)}{\der\Omega\,\der\mathcal{A}\,\der t}=\frac{4\pi}{\varepsilon(t)}\sum_{i=1}^{N(t)}w_i(t)\,\delta(\bn-\bn_i),
\end{equation}
where $\Omega$ is the solid angle of observation, and
\begin{equation}
  w_i(t)\equiv\frac{\varepsilon(t)}{4\pi\int_0^t\der t'\,\mathcal{A}(\bn_i,t')}
\end{equation}
is the exposure weighting for event $i$.

The work in this paper will restrict to the simple experiment where the full sky is being observed with uniform exposure. In this case, $w_i=1$ for all events. These results can be generalized in future work to account for partial sky coverage with a non-uniform exposure map.

Denote $c_{\ell m,N}$ as the \emph{central} spherical transform of the intensity from $N$ events
\begin{align}
  c_{\ell m,N}&=\int\der\bn\,Y^*_{\ell m}(\bn)\left[I(\bn)-\int\frac{\der\bn'}{4\pi}I(\bn')\right]\nonumber\\
  &=\frac{4\pi}{\varepsilon}\sum_{i=1}^NY^*_{\ell m}(\bn_i)-\frac{\sqrt{4\pi}\,N}{\varepsilon}\delta_{\ell,0}\delta_{m,0},
\end{align}
with the intensity power spectrum of the detected events being
\begin{equation}
  C_{\ell,N}=\frac{1}{2\ell+1}\sum_{m=-\ell}^\ell\left|c_{\ell m,N}\right|^2.
\end{equation}
A useful alternative expression for the power spectrum can be found by using the spherical harmonic addition theorem in \eqn{eq:harmadd}, 
\begin{equation}
  C_{\ell,N}=\frac{4\pi}{\varepsilon^2}\sum_{i=1}^N\sum_{j=1}^NP_\ell(\bn_i\cdot\bn_j)-\frac{4\pi N^2}{\varepsilon^2}\delta_{\ell,0}.
\end{equation}

The fluctuation power spectrum, which also applies to counts of celestial objects, is found by normalizing the spherical transform by the mean intensity $N/\varepsilon$:
\begin{align}
  &\tilde{c}_{\ell m,N}(\bn_1,\ldots,\bn_N)=\frac{4\pi}{N}\sum_{i=1}^NY^*_{\ell m}(\bn_i)-\sqrt{4\pi}\delta_{\ell,0}\delta_{m,0},\\
  &\tilde{C}_{\ell,N}(\bn_1,\ldots,\bn_N)=\frac{4\pi}{N^2}\sum_{i=1}^N\sum_{j=1}^NP_\ell(\bn_i\cdot\bn_j)-4\pi\delta_{\ell,0}.\label{eq:ClinPl}
\end{align}
Statistical properties of these observables will now be considered. The results will be presented in terms of the fluctuation power spectrum, but can be converted to the intensity power spectrum by multiplying $\tilde{C}_\ell$ by the mean intensity squared $N^2/\varepsilon^2$.

In the limit that the uncertainty in $\varepsilon$ is negligible, the use of $C_\ell$ or $\tilde{C}_\ell$ are equivalent. Since the fluctuation spectra are independent of $\varepsilon$, they are not affected by any systematics present in the estimate of the exposure.

\subsection{Shot Noise and Point Spread: Angular Power Spectrum of Events from an Isotropic Sky}
\label{ssec:shot}
The analysis is begun by considering the case of an isotropic skymap, such that $\tilde{S}(\bn)=1$. This is both instructive as a review of results for a Poisson point process, and useful by providing some necessary intermediate formulae involving the effect of an instrument's PSF.

Define the point spread function as $F(\bn_s,\bn)$ as the probability density function (PDF) of observing (or reconstructing) an event at $\bn$ given that it was sourced from position $\bn_s$. In general, this function can depend on other parameters that categorize the events and can affect the angular precision of observation. As a PDF, $F$ is normalized over all possible reconstruction positions such that
\begin{equation}
  \int\frac{\der\bn}{4\pi}F(\bn_s,\bn)=1.
\end{equation}
Let $\psi=\cos^{-1}(\bn_s\cdot\bn)$ be the angle of misreconstruction. If the PSF angular dependence is isotropic (depending on $\psi$ alone), then it is a \emph{symmetric PSF} with $F(\bn_s,\bn)=F(\bn,\bn_s)$, and it follows that $F$ is also normalized to $1$ over the skymap positions $\bn_s$. However, this need not be the case in general.

For good angular resolution, the PSF is sometimes well-described in the flat-sky limit as a Gaussian beam, often expressed as
\begin{equation}
  \label{eq:gaussianbeam}
  F_{\text{flat}}(\psi)=\frac{1}{2\pi\sigma_b^2}\exp\left[-\frac{\psi^2}{2\sigma_b^2}\right],
\end{equation}
with width $\sigma_b$ expressed in radians. One finds that the fraction $f(p)$ of events observed within angles $\psi<p\sigma_b$ is $f(p)=1-e^{-p^2/2}$. The fraction of events within $\psi<\sigma_b$ is $f(1)\simeq0.393$, and the 68\% containment angle (i.e., the angular radius that contains 68\% of events from a point source) is $\psi_{68}\simeq1.51\sigma_b$. One simple extension of this to the sphere is the Fisher distribution \citep{Fisher1953}
\begin{equation}
  \label{eq:Fisherdist}
  F_F(\bn_s,\bn)=\sigma_b^{-2}\,\text{csch}\!\left(\sigma_b^{-2}\right)\exp\!\left[\frac{\bn_s\cdot\bn}{\sigma_b^2}\right].
\end{equation}

Armed with our two postulates of stationarity of $S$ and independence of the $N$ events, the determination of statistical moments of the power spectrum is divided into two parts. The first calculation is to determine, for fixed source positions $\bn_{s1},\bn_{s2},\ldots$, the moment averaged over the PSF, which will be denoted with an overbar
\begin{equation}
  \psfmean{X}(\bn_{s1},\ldots,\bn_{sN})\equiv\int\left[\prod_i\frac{\der\bn_i}{4\pi}F(\bn_{si},\bn_i)\right]X(\bn_1,\ldots\bn_N).
\end{equation}
With that result, the second step is to take an ensemble average of the $N$ observed event positions over the sphere, giving the full moment of the measurement denoted with angle brackets, 
\begin{equation}
  \label{eq:posmean}
  \mean{X}\equiv\int\left[\prod_i\frac{\der\bn_{si}}{4\pi}\tilde{S}(\bn_{si})\right]\psfmean{X}(\bn_{s1},\ldots,\bn_{sN}),
\end{equation}
written for an isotropic source as
\begin{equation}
  \label{eq:posmean0}
  \mean{X}_0\equiv\int\left[\prod_i\frac{\der\bn_{si}}{4\pi}\right]\psfmean{X}(\bn_{s1},\ldots,\bn_{sN}).
\end{equation}
This last step in \eqn{eq:posmean} is where the independence postulate of the received events is applied.

Consider the spherical transform $\tilde{c}_{\ell m,N}(\bn_1,\ldots,\bn_N)$ of the observed events, which is an interesting property of the distribution in its own right, and consider the mean of its measurement. Let $\bn_{si}$ be the actual skymap position of the observed/reconstructed event position $\bn_i$. Averaging over the PSF of each event gives
\begin{align}
  \psfmean{\tilde{c}_{\ell m,N}}&(\bn_{s1},\ldots,\bn_{sN})\nonumber\\
  &=\int\left[\prod_i\frac{\der\bn_i}{4\pi}F(\bn_{si},\bn_i)\right]\tilde{c}_{\ell m,N}(\bn_1,\ldots,\bn_N)\nonumber\\
  &=\frac{1}{N}\sum_{i=1}^NF_{\ell m}(\bn_{si}),\label{eq:psfclm}
\end{align}
where the spherical transform of the PSF is defined as
\begin{equation}
  F_{\ell m}(\bn_s)\equiv\int\der\bn\, Y_{\ell m}^*(\bn)\Big[F(\bn_s,\bn)-1\Big].
\end{equation}
Averaging over all source positions for an isotropic skymap, a symmetric PSF produces the usual result
\begin{equation}
  \label{eq:shotclm}
  \mean{\tilde{c}_{\ell m,N}}_0=\int\left[\prod_i\frac{\der\bn_{si}}{4\pi}\right]\psfmean{\tilde{c}_{\ell m,N}}(\bn_{s1},\ldots,\bn_{sN})=0.
\end{equation}
This tells us the mean of the spherical transform measurement of isotropy is 0, but does not tell us the level of statistical fluctuation we expect. To determine this variance, calculate
\begin{align}
  \psfmean{\left|\tilde{c}_{\ell m,N}\right|^2}&=\frac{4\pi}{N}\left(1-\delta_{\ell,0}\delta_{m,0}\right)+\frac{4\pi}{N^2}\Bigg[\sum_iF_{\ell m}^{\ \ \,m}(\bn_{si})\nonumber\\
  &\quad+\frac{1}{4\pi}\sum_i\sum_{j\neq i}F_{\ell m}(\bn_{si})F_{\ell}^{\,m}(\bn_{sj})\Bigg].\label{eq:psfclm2}
\end{align}
This result produces a rank 2 spherical transform (see Appendix~\ref{ap:sphten}), 
\begin{equation}
  F_{\ell m}^{\ \ \,m}(\bn_s)=\int\der\bn\left|Y_\ell^{\,m}(\bn)\right|^2\Big[F(\bn_s,\bn)-1\Big],
\end{equation}
from the 1-event contributions, the terms where the contribution from each factor of $\tilde{c}_{\ell m,N}$ is from the same event. The other `2-event' terms generate a product of rank~1 transforms. This leads for isotropic sky and symmetric PSF to a mean square of
\begin{equation}
  \label{eq:shotclmvar}
  \mean{\left|\tilde{c}_{\ell m,N}\right|^2}_0=\frac{4\pi}{N}\left(1-\delta_{\ell,0}\,\delta_{m,0}\right),
\end{equation}
which in this case is the expected variance $\var{\tilde{c}_{\ell m,N}}$. Thus, a lack of signal produces spherical transforms of the data that randomly distribute about zero with a variance of $4\pi/N$ (for $\ell>0$).

Carrying out this calculation with the power spectrum of the data produces a PSF average
\begin{align}
  &\psfmean{\tilde{C}_{\ell,N}}(\bn_{s1},\ldots,\bn_{sN})=\frac{1}{2\ell+1}\sum_m\psfmean{\left|\tilde{c}_{\ell m,N}\right|^2}\nonumber\\
  &=\frac{4\pi}{N}\left(1-\delta_{\ell,0}\right)+\frac{1}{N^2}\sum_i\sum_{j\neq i}\scw_{\ell}(\bn_{si},\bn_{sj}),\label{eq:psfCl}
\end{align}
in terms of the power spectrum of the PSF
\begin{align}
  \label{eq:windowfunction}
  &\scw_\ell(\bn_{s1},\bn_{s2})=\frac{1}{2\ell+1}\sum_mF_{\ell m}(\bn_{s1})F_\ell^{\,m}(\bn_{s2})\\
  &\!\!=\!4\pi\!\!\int\!\!\frac{\der\bn_1}{4\pi}\frac{\der\bn_2}{4\pi}\Big[F(\bn_{s1},\bn_1)\!-\!1\Big]P_\ell(\bn_1\!\cdot\!\bn_2)\Big[F(\bn_{s2},\bn_2)\!-\!1\Big]\nonumber
\end{align}

If the PSF is symmetric, then it depends only on $\mu\equiv\bn_s\cdot\bn$, and it has a corresponding \emph{instrument window function}
\begin{equation}
  W_\ell\equiv\int_{-1}^{1}\frac{\der\mu}{2}P_\ell(\mu)F(\mu).
\end{equation}
For the Fisher distribution in \eqn{eq:Fisherdist},
\begin{equation}
  (W_{F})_\ell=\sqrt{\frac{\pi}{2}}\sigma^{-1}\text{csch}(\sigma^{-2})I_{\ell+\frac{1}{2}}(\sigma^{-2}),
\end{equation}
where $I_\nu(x)$ is the $\nu^{\text{th}}$ modified Bessel function of the first kind. When $\sigma_b\ll1$, this is indistinguishable from
\begin{equation}
  (W_{\text{flat}})_\ell=e^{-\ell(\ell+1)\sigma_b^2/2},
\end{equation}
which at large $\ell$ is the Fourier transform of \eqn{eq:gaussianbeam}. With the window function, the power spectrum of a symmetric PSF is
\begin{equation}
  \label{eq:symmetricpsfpowspec}
  \scw_\ell(\bn_{s1},\bn_{s2})=4\pi W_\ell^2\left[P_\ell(\bn_{s1}\cdot\bn_{s2})-\delta_{\ell,0}\right].
\end{equation}

The spherical average of \eqn{eq:psfCl} is
\begin{equation}
  \label{eq:shotmean}
  \boxed{\mean{\tilde{C}_{\ell,N}}_0=\frac{4\pi}{N}\left(1-\delta_{\ell,0}\right).}
\end{equation}
This is the usual shot noise, or Poisson noise, of angular power spectra. This average is biased from the actual value of $0$ for the isotropic skymap. As is well known and will be verified in the next subsection, this shot noise also provides a statistical bias to the estimation of the power spectrum of an anisotropic skymap. 

The power spectrum squared of the $N$ events contains a quadruple sum over the $N$ positions
\begin{align}
  &\tilde{C}_{\ell,N}^2=\frac{(4\pi)^2}{N^4}\sum_{i_1=1}^N\sum_{i_2=1}^N\sum_{i_3=1}^N\sum_{i_4=1}^NP_\ell(\bn_{i_1}\cdot\bn_{i_2})P_\ell(\bn_{i_3}\cdot\bn_{i_4})\nonumber\\
  &\qquad\qquad-(4\pi)^2\delta_{\ell,0}\\
  &\equiv\left[\tilde{C}_{\ell,N}^2\right]_1+\left[\tilde{C}_{\ell,N}^2\right]_2+\left[\tilde{C}_{\ell,N}^2\right]_3+\left[\tilde{C}_{\ell,N}^2\right]_4-(4\pi)^2\delta_{\ell,0},\nonumber
\end{align}
which is convenient to separate into a 1-event term, 2-event term, and so on. The PSF mean of the 1-event term, i.e., the sum of those terms where all 4 events are the same, is
\begin{equation}
  \left[\psfmean{\tilde{C}_{\ell,N}^2}\right]_1=\frac{(4\pi)^2}{N^3}.
\end{equation}
The 2-event terms are categorized by
\begin{align}
  &\!\!\!\left[\psfmean{\tilde{C}_{\ell,N}^2}\right]_2=\frac{(4\pi)^2}{N^4}\sum_{i_1}\sum_{i_2\neq i_1}\int\frac{\der\bn_{i_1}}{4\pi}\frac{\der\bn_{i_2}}{4\pi}F(\bn_{si_1},\bn_{i_1})\nonumber\\
  &F(\bn_{si_2},\bn_{i_2})\Big[4P_\ell(\bn_{i_1}\cdot\bn_{i_2})+2P_\ell^2(\bn_{i_1}\cdot\bn_{i_2})+1\Big].
\end{align}
The result in the square brackets shows how the 7 ways to pair 4 position are partitioned. The 4 ways to split the events so three are the same (different from the fourth) give the first term in the square brackets. The second term, with $P_\ell^2$, is generated from pairings that split the azimuthal indices, i.e., with $i_1=i_3$ and $i_2=i_4$, or $i_1=i_4$ and $i_2=i_3$. The last term is from the pairing that respects the azimuthal indices, with $i_1=i_2$ and $i_3=i_4$. The Legendre polynomial term generates a power spectrum of the PSF, and the square polynomial term produces a composite power spectrum of $F$ (see Appendix~\ref{ap:sphten})
\begin{align}
  &\scw_\ell^{(2)}(\bn_{s1},\bn_{s2})\\
  &=\!\int\!\frac{\der\bn_1}{4\pi}\frac{\der\bn_2}{4\pi}F(\bn_1,\bn_{s1})P_\ell^2(\bn_1\!\cdot\!\bn_2)F(\bn_2,\bn_{s2})-\frac{1}{2\ell+1}.\nonumber
\end{align}
The 2-event term is thus
\begin{align}
  &\left[\psfmean{\tilde{C}_{\ell,N}^2}\right]_2=\frac{(4\pi)^2}{N^4}\Bigg\{\!\sum_{i_1}\sum_{i_2\neq i_1}\Bigg[4\Bigg(\frac{\scw_\ell(\bn_{si_1},\bn_{si_2})}{4\pi}+\delta_{\ell,0}\Bigg)\nonumber\\
  &\quad\ +2\Bigg(\scw_\ell^{(2)}(\bn_{si_1},\bn_{si_2})+\frac{1}{2\ell+1}\Bigg)\Bigg]+N(N-1)\Bigg\}\nonumber\\
  &=\frac{(4\pi)^2}{N^4}\Bigg[N(N-1)\left(1+\frac{2}{2\ell+1}+4\delta_{\ell,0}\right)\\
  &\quad\ +\sum_{i_1}\sum_{i_2\neq i_1}\Bigg(4\frac{\scw_\ell(\bn_{si_1},\bn_{si_2})}{4\pi}+2\scw_\ell^{(2)}(\bn_{si_1},\bn_{si_2})\Bigg)\Bigg]\nonumber
\end{align}

For the 3-event terms, there are 6 ways to choose 2 of the 4 positions to be the same event
\begin{align}
  &\left[\psfmean{\tilde{C}_{\ell,N}^2}\right]_3\!=\frac{(4\pi)^2}{N^4}\sum_{i_1}\sum_{i_2\neq i_1}\sumthree\int\frac{\der\bn_{i_1}}{4\pi}\frac{\der\bn_{i_2}}{4\pi}\frac{\der\bn_{i_3}}{4\pi}\nonumber\\
  &\qquad F(\bn_{si_1},\bn_{i_1})F(\bn_{si_2},\bn_{i_2})F(\bn_{si_3},\bn_{i_3})\left[2P_\ell(\bn_{i_1}\cdot\bn_{i_2})\right.\nonumber\\
  &\qquad+4\left.\!P_\ell(\bn_{i_1}\cdot\bn_{i_2})P_\ell(\bn_{i_2}\cdot\bn_{i_3})\right]
\end{align}
The first term in the square brackets comes from the 2 ways to pair events with the same azimuthal index, $i_1=i_2$ or $i_3=i_4$. The other pairings generate an open bispectrum of the PSF (see \eqn{eq:openbisp})
\begin{align}
  &\!\!\!\!\scw_\ell^{(3)}(\bn_{s1},\bn_{s2},\bn_{s3})=4\pi\int\frac{\der\bn_1}{4\pi}\frac{\der\bn_2}{4\pi}\frac{\der\bn_3}{4\pi}\nonumber\\
  &F(\bn_{s1},\bn_1)P_\ell(\bn_1\cdot\bn_3)F(\bn_{s3},\bn_3)P_\ell(\bn_3\cdot\bn_2)F(\bn_{s2},\bn_2)\nonumber\\
  &-\frac{\scw_\ell(\bn_{s1},\bn_{s2})}{2\ell+1}-4\pi\delta_{\ell,0}.
\end{align}
Then we find that the expression
\begin{align}
  &\!\!\left[\psfmean{\tilde{C}_{\ell,N}^2}\right]_3\!\!=\!\frac{(4\pi)^2}{N^4}\Bigg\{\!2(N-2)\!\sum_{i_1}\!\sum_{i_2\neq i_1}\!\!\Bigg[\!\frac{\scw_\ell(\bn_{si_1},\bn_{si_2})}{4\pi}+\delta_{\ell,0}\!\Bigg]\nonumber\\
  &\qquad+4\sum_{i_1}\!\sum_{i_2\neq i_1}\sumthree\!\!\Bigg[\!\frac{\scw_\ell^{(3)}(\bn_{si_1},\bn_{si_2},\bn_{si_3})}{4\pi}\nonumber\\
  &\qquad+\frac{\scw_\ell(\bn_{si_1},\bn_{si_2})}{4\pi(2\ell+1)}+\delta_{\ell,0}\!\Bigg]\Bigg\}\nonumber\\
  &=\frac{(4\pi)^2}{N^4}\Bigg[6N(N-1)(N-2)\delta_{\ell,0}\\
  &\qquad+2(N-2)\left(1+\frac{2}{2\ell+1}\right)\sum_{i_1}\!\sum_{i_2\neq i_1}\frac{\scw_\ell(\bn_{si_1},\bn_{si_2})}{4\pi}\nonumber\\
  &\qquad+4\sum_{i_1}\!\sum_{i_2\neq i_1}\sumthree\!\!\!\!\!\!\frac{\scw_\ell^{(3)}(\bn_{si_1},\bn_{si_2},\bn_{si_3})}{4\pi}\Bigg]\nonumber
\end{align}
constitutes all of the 3-event contributions.

Finally, the 4-event terms are simply
\begin{align}
  &\left[\psfmean{\tilde{C}_{\ell,N}^2}\right]_4=\frac{(4\pi)^2}{N^4}\!\sum_{i_1}\!\sum_{i_2\neq i_1}\Bigg[\frac{\scw_\ell(\bn_{si_1},\bn_{si_2})}{4\pi}+\delta_{\ell,0}\Bigg]\nonumber\\
  &\qquad\sumthree\sumfour\Bigg[\frac{\scw_\ell(\bn_{si_3},\bn_{si_4})}{4\pi}+\delta_{\ell,0}\Bigg]\nonumber\\
  &=\frac{(4\pi)^2}{N^4}\Bigg[N(N-1)(N-2)(N-3)\delta_{\ell,0}\\
  &+\!\!\sum_{i_1}\!\!\sum_{i_2\neq i_1}\sumthree\!\!\sumfour\!\!\!\!\!\!\!\!\!\!\!\!\!\!\!\frac{\scw_\ell(\bn_{si_1},\bn_{si_2})}{4\pi}\frac{\scw_\ell(\bn_{si_3},\bn_{si_4})}{4\pi}\!\Bigg].\nonumber
\end{align}

The combination of the terms has cancellations, leaving us with the major result
\begin{align}
  &\psfmean{\tilde{C}_{\ell,N}^2}=\frac{(4\pi)^2}{N^4}\Bigg\{N^2\left[1+\left(1-\frac{1}{N}\right)\frac{2}{2\ell+1}\right](1-\delta_{\ell,0})\nonumber\\
  &+\sum_{i_1}\sum_{i_2\neq i_1}\Bigg[2N\left[1+\left(1-\frac{2}{N}\right)\frac{2}{2\ell+1}\right]\frac{\scw_{\ell}(\bn_{si_1},\bn_{si_2})}{4\pi}\nonumber\\
  &+2\scw_\ell^{(2)}(\bn_{si_1},\bn_{si_2})+\sumthree\!\!\Bigg(4\frac{\scw_\ell^{(3)}(\bn_{si_1},\bn_{si_2},\bn_{si_3})}{4\pi}\nonumber\\
  &+\!\sumfour\!\!\!\!\frac{\scw_\ell(\bn_{si_1}\bn_{si_2})}{4\pi}\frac{\scw_\ell(\bn_{si_3},\bn_{si_4})}{4\pi}\Bigg)\Bigg]\Bigg\},\label{eq:psfCl2}
\end{align}
from which all other conclusions are easily derived. For instance, we have for an isotropic background that the measurement of shot noise with a symmetric PSF has second moment
\begin{equation}
  \mean{\tilde{C}_{\ell,N}^2}_{\!0}=\left(\frac{4\pi}{N}\right)^{\!\!2}\!\left[1+\left(1-\frac{1}{N}\right)\frac{2}{2\ell+1}\right](1-\delta_{\ell,0}),
\end{equation}
and the variance of shot noise is
\begin{equation}
  \label{eq:shotvar}
  \boxed{\var{\tilde{C}_{\ell,N}}_0=\left(\frac{4\pi}{N}\right)^{\!\!2}\!\left(1-\frac{1}{N}\right)\frac{2}{2\ell+1}(1-\delta_{\ell,0}).}
\end{equation}

It is worth noting that Eqns.~(\ref{eq:shotclm}), (\ref{eq:shotclmvar}), (\ref{eq:shotmean}), and (\ref{eq:shotvar}) for the moments of an isotropic sky (shot noise) are the results one gets without a PSF, with $F(\bn_s,\bn)=1$. This demonstrates that a symmetric PSF does not affect the statistical properties of measurements of the spherical transform or power spectrum of an isotropic skymap.

\subsection{An Unbiased Estimator of the Source Angular Power Spectrum}
\label{ssec:estimator}
Now consider an intrinsic skymap that is anisotropic with normalized distribution $\tilde{S}(\bn)$. The spectra moments must now be calculated as in \eqn{eq:posmean} instead of \eqn{eq:posmean0}.

When taking the position ensemble mean of a transform or spectrum of the PSF, each skymap-weighted integration that acts on the PSF has the effect of convolving the skymap over the PSF
\begin{equation}
  \label{eq:psfconvolution}
  \tilde{S}_F(\bn)=\int\frac{\der\bn_s}{4\pi}\tilde{S}(\bn_s)F(\bn_s,\bn).
\end{equation}
This PSF-convolved skymap contains the information that the instrument is capable of probing. Information that is lost in the convolution is inaccessible because of the finite angular resolution of the instrument. Therefore, the power spectrum that is being probed by cosmic events is not the power spectrum of the complete skymap, but the power spectrum of the convolved skymap.

Using \eqn{eq:symmetricpsfpowspec}, the power spectrum $\tilde{C}_{F\ell}$ of the skymap convolved by a symmetric PSF is related to $\tilde{C}_\ell$ by
\begin{equation}
  \tilde{C}_{F\ell}=W_\ell^2\tilde{C}_\ell.
\end{equation}
We will not de-convolve the skymap by dividing out the window function, but will express results more simply in terms of the directly-observed PSF-convolved skymap.

For the rest of this article, it will be implicitly understood that the spherical transforms $\tilde{c}_{\ell m}$ and angular spectra $\tilde{C}^{(k)}_\ell$ of the skymap are of the PSF-convolved skymap $\tilde{S}_F(\bn)$. These results also explain how to determine the effects of an arbitrary, asymmetric PSF when measuring an isotropic sky.

The ensemble averages of angular spectra of the PSF are then just the angular spectra of the PSF-convolved skymap,
\begin{align}
  &\mean{\tilde{F}_{\ell m}(\bn_s)}=\tilde{c}_{\ell m},\nonumber\\
  &\mean{\tilde{F}_{\ell m}^{\ \ \,m}(\bn_s)}=\tilde{c}_{\ell m}^{\ \ \ m},\nonumber\\
  &\mean{\scw_\ell(\bn_{s1},\bn_{s2})}=\tilde{C}_\ell\text{ when }\bn_{s1}\neq\bn_{s2},\nonumber
\end{align}
and so on.

Then it follows from Eqns.~(\ref{eq:psfclm}) and (\ref{eq:psfclm2}) that the mean and variance of the spherical transform of $N$ events are
\begin{align}
  &\mean{\tilde{c}_{\ell m,N}}=\tilde{c}_{\ell m},\\
  &\var{\tilde{c}_{\ell m,N}}=\frac{4\pi}{N}\left(1+\tilde{c}_{\ell m}^{\ \ \ m}\right)(1-\delta_{\ell,0}\delta_{m,0}).
\end{align}
Thus, $\tilde{c}_{\ell m,N}$ is an unbiased estimator of the spherical transform of the source skymap, convolved with the PSF. The variance of this measurement is due to the finite statistics of the sampling of the skymap, and decreases indefinitely as the observed number of events increases. Note that this variance does not account for cosmic variance of the sources, but is due to sampling of a fixed sky. The mean spherical transform after marginalizing over cosmic variance is zero for Gaussian-distributed sources.

The event-ensemble average of the power spectrum of $N$ events follows from \eqn{eq:psfCl}
\begin{equation}
  \label{eq:aveClN}
  \boxed{\mean{\tilde{C}_{\ell,N}}=\frac{4\pi}{N}(1-\delta_{\ell,0})+\left(1-\frac{1}{N}\right)\tilde{C}_\ell.}
\end{equation}
This result verifies previous estimates, as in \eqn{eq:Clest}, that the shot noise of \eqn{eq:shotmean} biases the power spectrum of the $N$ events as an estimator of the power spectrum of $\tilde{S}_F(\bn)$. Additionally, this calculation reveals that for low counts, there is a normalizing coefficient in front of the signal.

An unbiased estimator, $\hat{\tilde{C}}_{\ell,N}$, for $\tilde{C}_\ell$ from $N$ events is therefore found by subtracting the shot noise and re-normalizing $\tilde{C}_{\ell,N}$
\begin{equation}
  \label{eq:Chat}
  \hat{\tilde{C}}_{\ell,N}\equiv\frac{1}{1-\frac{1}{N}}\left[\tilde{C}_{\ell,N}-\frac{4\pi}{N}(1-\delta_{\ell,0})\right],
\end{equation}
so that $\mean{\hat{\tilde{C}}_{\ell,N}}=\tilde{C}_\ell$. Applying \eqn{eq:ClinPl}, we find this estimator is simply
\begin{equation}
  \label{eq:ChatD}
  \boxed{\hat{\tilde{C}}_{\ell,N}=4\pi(\hat{\Delta}_{\ell,N}-\delta_{\ell,0}),}
\end{equation}
where we introduce the irreducible angular power spectrum estimator
\begin{equation}
  \label{eq:irred1}
  \hat{\Delta}_{\ell,N}(\bn_1,\ldots,\bn_N)=\frac{1}{N(N-1)}\sum_i\sum_{j\neq i}P_\ell(\bn_i\cdot\bn_j),
\end{equation}
an unbiased estimator for the dimensionless, irreducible power spectrum
\begin{equation}
  \Delta_\ell\equiv\frac{\tilde{C}_\ell}{4\pi}+\delta_{\ell,0}.
\end{equation}
Eqn.~(\ref{eq:ChatD}) expresses the correct frequentist estimate for the angular power spectrum from $N$ events.

\subsection{The Statistical Variance of the Angular Power Estimator}
\label{ssec:apsvar}
The variance of $\hat{\tilde{C}}_{\ell,N}$ is determined from the mean of the square. First, from \eqn{eq:psfCl2},
\begin{align}
  \label{eq:meansq}
  &\mean{\tilde{C}_{\ell,N}^2}=\left(\frac{4\pi}{N}\right)^{\!\!2}\!\!\left(1-\frac{1}{N}\right)\!\Bigg[\!\left(\frac{N}{N-1}+\frac{2}{2\ell+1}\right)+2\tilde{C}_\ell^{(2)}\nonumber\\
  &\qquad+2\left(N+(N-2)\frac{2}{2\ell+1}\right)\frac{\tilde{C}_\ell}{4\pi}+4(N-2)\frac{\tilde{C}_\ell^{(3)}}{4\pi}\nonumber\\
  &\qquad+(N-2)(N-3)\left(\frac{\tilde{C}_\ell}{4\pi}\right)^{\!\!2}\Bigg](1-\delta_{\ell,0}),
\end{align}
then noting from \eqn{eq:Chat} that
\begin{equation}
  \var{\hat{\tilde{C}}_{\ell,N}}=\frac{1}{\left(1-\frac{1}{N}\right)^2}\,\var{\tilde{C}_{\ell,N}}=\frac{\mean{\tilde{C}_{\ell,N}^2}-\mean{\tilde{C}_{\ell,N}}^{\!2}}{\left(1-\frac{1}{N}\right)^2}
\end{equation}
and applying \eqn{eq:aveClN} in the form
\begin{align}
  \mean{\tilde{C}_{\ell,N}}^{\!2}=&\ \left(\frac{4\pi}{N}\right)^{\!\!2}\!\left(1-\frac{1}{N}\right)\!(1-\delta_{\ell,0})\\
  &\left[\frac{N}{N-1}+2N\frac{\tilde{C}_\ell}{4\pi}+N(N-1)\left(\frac{\tilde{C}_\ell}{4\pi}\right)^{\!\!2}\right],\nonumber
\end{align}
we have the variance of the unbiased estimator of $\tilde{C}_\ell$ from $N$ random events to be
\begin{align}
  \label{eq:Chatvar}
  &\var{\hat{\tilde{C}}_{\ell,N}}=\frac{(4\pi)^2}{N(N-1)}\Bigg\{\frac{2}{2\ell+1}(1-\delta_{\ell,0})+2\tilde{C}_\ell^{(2)}\\
  &+4(N-2)\!\left[\frac{1}{2\ell+1}\frac{\tilde{C}_\ell}{4\pi}+\frac{\tilde{C}_\ell^{(3)}}{4\pi}\right]-(4N-6)\!\left(\frac{\tilde{C}_\ell}{4\pi}\right)^{\!\!2}\Bigg\}.\nonumber
\end{align}
Converting the fluctuation spectra to intensity spectra recovers \eqn{eq:varprev}, where it is assumed that any uncertainty in the exposure $\varepsilon$ can be neglected. It's useful to consider the last term as from two separate contributions
\fbox{
\addtolength{\linewidth}{-4\fboxsep}%
\addtolength{\linewidth}{-2\fboxrule}%
\begin{minipage}{\linewidth}
\begin{align}
  \label{eq:Chatvar2}
  &\!\!\!\!\!\var{\hat{\tilde{C}}_{\ell,N}}=\frac{(4\pi)^2}{N(N-1)}\Bigg\{2\Bigg[\frac{1-\delta_{\ell,0}}{2\ell+1}+\tilde{C}_\ell^{(2)}-\left(\frac{\tilde{C}_\ell}{4\pi}\right)^{\!\!2}\Bigg]\nonumber\\
  &\!\!\!\!\!\!\qquad+4(N-2)\!\left[\frac{1}{2\ell+1}\frac{\tilde{C}_\ell}{4\pi}+\frac{\tilde{C}_\ell^{(3)}}{4\pi}-\left(\frac{\tilde{C}_\ell}{4\pi}\right)^{\!\!2}\right]\Bigg\}.
\end{align}
\end{minipage}
}
The first collection of terms, which we will refer to as the \textbf{shot} term, contains the shot noise contribution and is modified by 2-point and 4-point correlations. The shot term of the variance evolves as $N^{-2}$ and is important during the early stages of the experiment. The second set of terms, collectively referred here as the \textbf{signal} term of the variance, evolves as $N^{-1}$ and has 3-point and 4-point corrections. It is intriguing that the 4-point influence is to always decrease the signal variance. Since the central angular power spectrum and open angular bispectrum may be negative, it seems plausible at present that the signal term could be negative. However, the positivity of the variance as $N\rightarrow\infty$ must prevent this, suggesting the existence of non-trivial bounds for the spherical tensors. A detailed discussion of the new higher-order effects due to $C_\ell^{(2)}$, $C_\ell^{(3)}$, and $C_\ell^{(4)}$ is carried out in \ssec{ssec:compare}.

\subsection{Unbiased Estimators for the Statistical Variance}
\label{ssec:varestimator}
Due to the fact that the experiment has no knowledge of the intrinsic skymap, the rotation-invariant angular spectra are unknown and the variance of the power spectrum needs to be estimated from the data. Just as we determined an unbiased estimator of the power spectrum from the observed event positions in \eqn{eq:Chat}, unbiased estimators of the higher order spectra can also be determined, and these estimators will allow us to estimate the statistical uncertainty of the measurement of $\tilde{C}_\ell$.

The central rank-2 spherical transform of $N$ point events
\begin{equation}
  \tilde{c}_{\ell m_1\hphantom{m_2}\!,N}^{\hphantom{\ell m_1}\!m_2}(\bn_1,\ldots,\bn_N)=\frac{4\pi}{N}\sum_{i=1}^NY^*_{\ell m_1}(\bn_i)Y_{\ell}^{\,m_2}(\bn_i)-\delta_{m_1}^{\ \ \ m_2}
\end{equation}
produces a composite angular power spectrum
\begin{align}
  &\!\!\tilde{C}_{\ell,N}^{(2)}(\bn_1,\ldots,\bn_N)=\frac{1}{(2\ell+1)^2}\!\!\sum_{m_1=-\ell}^\ell\sum_{m_2=-\ell}^\ell\!\Bigg[\delta_{m_1}^{\hphantom{m_1}\!m_2}\,\delta^{m_1}_{\hphantom{m_1}m_2}\nonumber\\
  &\qquad-2\frac{4\pi}{N}\,\delta^{m_1}_{\hphantom{m_1}m_2}\sum_iY^*_{\ell m_1}(\bn_i)Y_{\ell}^{\,m_2}(\bn_i)\nonumber\\
  &\qquad+\left(\frac{4\pi}{N}\right)^{\!\!2}\!\sum_i\left|Y_\ell^{\,m_1}(\bn_i)\right|^2\left|Y_\ell^{\,m_2}(\bn_i)\right|^2\nonumber\\
  &+\left(\frac{4\pi}{N}\right)^{\!\!2}\!\sum_i\sum_{j\neq i}Y^*_{\ell m_1}(\bn_i)Y_\ell^{\,m_1}(\bn_j)Y^*_{\ell m_2}(\bn_j)Y_\ell^{\,m_2}(\bn_i)\Bigg]\nonumber\\
  &=\frac{1}{2\ell+1}-\frac{2}{2\ell+1}+\frac{1}{N}+\frac{1}{N^2}\!\sum_i\sum_{j\neq i}P_\ell^2(\bn_i\cdot\bn_j)\nonumber\\
  &=\frac{1}{N^2}\sum_{i=1}^N\sum_{j=1}^NP_\ell^2(\bn_i\cdot\bn_j)-\frac{1}{2\ell+1}.
\end{align}
The PSF-mean is
\begin{align}
  &\psfmean{\tilde{C}_{\ell,N}^{(2)}}(\bn_{s1},\ldots,\bn_{sN})=\frac{1}{N}+\frac{1}{N^2}\!\sum_i\sum_{j\neq i}\bigg[\scw_{\ell}^{(2)}(\bn_{si},\bn_{sj})\nonumber\\
  &\qquad+\frac{1}{2\ell+1}\bigg]-\frac{1}{2\ell+1}\nonumber\\
  &=\frac{1}{N^2}\!\sum_i\sum_{j\neq i}\scw_{\ell}^{(2)}(\bn_{si},\bn_{sj})+\frac{1}{N}\left(1-\frac{1}{2\ell+1}\right),
\end{align}
and the source ensemble average is found to be
\begin{equation}
  \mean{\tilde{C}_{\ell,N}^{(2)}}=\left(1-\frac{1}{N}\right)\tilde{C}_\ell^{(2)}+\frac{1}{N}\left(1-\frac{1}{2\ell+1}\right).
\end{equation}
Thus, an unbiased estimator for the composite power spectrum of the projected skymap is
\begin{align}
  \hat{\tilde{C}}_{\ell,N}^{(2)}&\equiv\frac{1}{1-\frac{1}{N}}\left[\tilde{C}_{\ell,N}^{(2)}-\frac{1}{N}\left(1-\frac{1}{2\ell+1}\right)\right]\\
  &=\hat{\Delta}_{\ell,N}^{\!(2)}-\frac{1}{2\ell+1},
\end{align}
where the irreducible composite power spectrum is defined as
\begin{equation}
  \label{eq:irred2}
  \hat{\Delta}_{\ell,N}^{\!(2)}(\bn_1,\ldots,\bn_N)\equiv\frac{1}{N(N-1)}\sum_{i_1}\sum_{i_2\neq i_1}P_\ell^2(\bn_{i_1}\cdot\bn_{i_2}).
\end{equation}

Similarly, the open bispectrum of $N$ events is
\begin{align}
  &\tilde{C}_{\ell,N}^{(3)}=4\pi\bigg[\frac{1}{N^3}\sum_{i_1,i_2,i_3}P_\ell(\bn_{i_1}\cdot\bn_{i_3})P_\ell(\bn_{i_3}\cdot\bn_{i_2})\nonumber\\
  &\hspace{17mm}-\frac{1}{2\ell+1}\frac{1}{N^2}\sum_{i_1,i_2}P_\ell(\bn_{i_1}\cdot\bn_{i_2})\bigg]\\
  &\qquad=4\pi\Bigg\{\frac{1}{N}\left(\frac{1}{N}-\frac{1}{2\ell+1}\right)+\frac{1}{N^3}\sum_{i_1}\sum_{i_2\neq i_1}\Bigg[\nonumber\\
  &\qquad\quad\left(2-\frac{N}{2\ell+1}\right)P_\ell(\bn_{i_1}\cdot\bn_{i_2})+P_\ell^2(\bn_{i_1}\cdot\bn_{i_2})\nonumber\\
  &\qquad\quad+\sumthree \!\!P_\ell(\bn_{i_1}\cdot\bn_{i_3})P_\ell(\bn_{i_3}\cdot\bn_{i_2})\Bigg]\Bigg\}\nonumber,
\end{align}
with PSF-mean
\begin{align}
  &\psfmean{\tilde{C}_{\ell,N}^{(3)}}(\bn_{si_1},\ldots,\bn_{si_N})=4\pi\Bigg\{\frac{1}{N}\left(\frac{1}{N}-\frac{1}{2\ell+1}\right)\nonumber\\
  &+\frac{1}{N^3}\sum_{i_1}\sum_{i_2\neq i_1}\Bigg[\left(2-\frac{N}{2\ell+1}\right)\left(\frac{\scw_\ell(\bn_{si_1},\bn_{si_2})}{4\pi}+\delta_{\ell,0}\right)\nonumber\\
  &+\left(\scw_\ell^{(2)}(\bn_{si_1},\bn_{si_2})+\frac{1}{2\ell+1}\right)+\sumthree\!\Bigg(\delta_{\ell,0}\nonumber\\
  &+\frac{1}{2\ell+1}\frac{\scw_\ell(\bn_{si_1},\bn_{si_2})}{4\pi}+\frac{\scw_\ell^{(3)}(\bn_{si_1},\bn_{si_2},\bn_{si_3})}{4\pi}\Bigg)\Bigg]\Bigg\}\nonumber\\
  &=\frac{4\pi}{N^2}\Bigg\{1-\frac{1}{2\ell+1}+\frac{1}{N}\sum_{i_1}\sum_{i_2\neq i_1}\Bigg[\nonumber\\
  &\qquad2\left(1-\frac{1}{2\ell+1}\right)\frac{\scw_\ell(\bn_{si_1},\bn_{si_2})}{4\pi}+\scw_\ell^{(2)}(\bn_{si_1},\bn_{si_2})\nonumber\\
  &\qquad+\sumthree\frac{\scw_\ell^{(3)}(\bn_{si_1},\bn_{si_2},\bn_{si_3})}{4\pi}\Bigg]\Bigg\},
\end{align}
and source ensemble average
\begin{align}
  &\mean{\tilde{C}_{\ell,N}^{(3)}}=4\pi\Bigg\{\!\left(1-\frac{1}{2\ell+1}\right)\!\left[\frac{1}{N^2}+\!\frac{2}{N}\!\left(1-\frac{1}{N}\right)\!\frac{\tilde{C}_\ell}{4\pi}\right]\nonumber\\
  &\qquad+\left(1-\frac{1}{N}\right)\Bigg[\frac{\tilde{C}_\ell^{(2)}}{N}+\left(1-\frac{2}{N}\right)\frac{\tilde{C}_\ell^{(3)}}{4\pi}\Bigg]\Bigg\}.
\end{align}
Replacing the 2-point spectra with their unbiased estimators, we find an unbiased estimator for the open bi\-spectrum to be

\begin{align}
  \hat{\tilde{C}}_{\ell,N}^{(3)}&\equiv\frac{1}{\left(1-\frac{1}{N}\right)\left(1-\frac{2}{N}\right)}\bigg[\tilde{C}_{\ell,N}^{(3)}-\frac{4\pi}{N}\tilde{C}_{\ell,N}^{(2)}\nonumber\\
  &\quad-\frac{2}{N}\left(1-\frac{1}{2\ell+1}\right)\left(\tilde{C}_{\ell,N}-\frac{4\pi}{N}\right)\bigg]\\
  &=4\pi\left(\hat{\Delta}_{\ell,N}^{\!(3)}-\frac{\hat{\Delta}_{\ell,N}}{2\ell+1}\right),
\end{align}
in terms of the irreducible spectrum in \eqn{eq:irred1} and
\begin{align}
  \label{eq:irred3}
  &\hat{\Delta}_{\ell,N}^{\!(3)}(\bn_1,\ldots,\bn_N)\equiv\frac{1}{N(N-1)(N-2)}\\
  &\qquad\sum_{i_1}\sum_{i_2\neq i_1}\sumthree\!\!\!P_\ell(\bn_{i_1}\cdot\bn_{i_3})P_\ell(\bn_{i_3}\cdot\bn_{i_2}).\nonumber
\end{align}

Finally, from \eqn{eq:meansq}, the unbiased estimator for the 4-point spectrum is
\begin{align}
  &\hat{\tilde{C}}_{\ell,N}^{(4)}=\tilde{C}_{\ell,N}^2\hspace{-7mm}\widehat{\phantom{\tilde{C}_{\ell,N}^2}}\ \ \equiv\frac{(4\pi)^2}{\left(1-\frac{1}{N}\right)\left(1-\frac{2}{N}\right)\left(1-\frac{3}{N}\right)}\Bigg\{\left(\frac{\tilde{C}_{\ell,N}}{4\pi}\right)^{\!\!2}\nonumber\\
  &\quad-\frac{2}{N}\left[2\frac{\tilde{C}_{\ell,N}^{(3)}}{4\pi}-\frac{\tilde{C}_{\ell,N}^{(2)}}{N}+\left(1-\frac{4}{N}+\frac{2}{2\ell+1}\right)\frac{\tilde{C}_{\ell,N}}{4\pi}\right]\nonumber\\
  &\quad+\frac{1-\delta_{\ell,0}}{N^2}\left(1-\frac{6}{N}+\frac{2}{2\ell+1}\right)\Bigg\}\\
  &=(4\pi)^{2}\!\left(\hat{\Delta}_{\ell,N}^{\!(4)}-\delta_{\ell,0}\right),
\end{align}
where
\begin{align}
  \label{eq:irred4}
  &\hat{\Delta}_{\ell,N}^{\!(4)}(\bn_1,\ldots,\bn_N)\equiv\frac{1}{N(N-1)(N-2)(N-3)}\sum_{i_1}\sum_{i_2\neq i_1}\nonumber\\
  &\qquad\sumthree\sumfour\!\!\!\!P_\ell(\bn_{i_1}\cdot\bn_{i_2})P_\ell(\bn_{i_3}\cdot\bn_{i_4}).
\end{align}

With these new estimators, we find that the simplest expression for the unbiased estimator of the power spectrum's variance is in terms of the irreducible spectra
\fbox{
\addtolength{\linewidth}{-4\fboxsep}%
\addtolength{\linewidth}{-2\fboxrule}%
\begin{minipage}{\linewidth}
\begin{align}
  \label{eq:varhatchat}
  \wvar{\hat{\tilde{C}}_{\ell,N}}=\frac{(4\pi)^2}{N(N-1)}\Big[2\Big(&\hat{\Delta}_{\ell,N}^{\!(2)}-\hat{\Delta}_{\ell,N}^{\!(4)}\Big)\\
  +4(N-2)\Big(&\hat{\Delta}_{\ell,N}^{\!(3)}-\hat{\Delta}_{\ell,N}^{\!(4)}\Big)\Big].\nonumber
\end{align}
\end{minipage}
}
At low $N$, the shot term determines the variance, determined from the irreducible composite power spectrum with correction from the disjoint trispectrum. At high $N$, the signal term dominates, which is generated by the irreducible open bispectrum, and negative trispectrum correction. Positivity of the variance at small $N$ requires $\hat{\Delta}_{\ell,N}^{\!(2)}\geq\hat{\Delta}_{\ell,N}^{\!(4)}$, and at large $N$ demands $\hat{\Delta}_{\ell,N}^{\!(3)}\geq\hat{\Delta}_{\ell,N}^{\!(4)}$.

Simulations presented in \ssec{ssec:sims} demonstrate that at low values of $N$, it is possible for the $N$ data points to produce negative estimates of the variance. In fact, this estimator doesn't produce a useful error estimate until enough data is received that the width of the distribution of $\wvar{\hat{\tilde{C}}_{\ell,N}}$ is smaller than an order of magnitude. Determining an estimate for the minimum number of counts required for a robust error estimate of the power spectrum is an important subject for future work.

Decomposing the squared irreducible power spectrum into 2, 3, and 4-event terms
\begin{align}
  \label{eq:specidentity}
  &\hat{\Delta}_{\ell,N}^{\!2}=\frac{1}{N(N-1)}\Big[2\hat{\Delta}_{\ell,N}^{\!(2)}+4(N-2)\hat{\Delta}_{\ell,N}^{\!(3)}\nonumber\\
  &\hspace{27mm}+(N-2)(N-3)\hat{\Delta}_{\ell,N}^{\!(4)}\Big],
\end{align}
reveals the variance estimator to simply be
\begin{equation}
  \label{eq:varhatchat2}
  \boxed{\wvar{\hat{\tilde{C}}_{\ell,N}}=(4\pi)^2\Big(\hat{\Delta}_{\ell,N}^{\!2}-\hat{\Delta}_{\ell,N}^{\!(4)}\Big)=\hat{\tilde{C}}_{\ell,N}^2-\hat{\tilde{C}}_{\ell,N}^{(4)}.}
\end{equation}
This is the simplest form for the estimator of the power spectrum variance from data. One important consequence is that the condition for $\hat{\tilde{C}}_{\ell,N}$ to be non-zero to high confidence (i.e., for it's uncertainty to be much smaller than its measured magnitude) is for $\hat{\tilde{C}}_{\ell,N}^{(4)}$ to be positive and a significant fraction of the magnitude of $\hat{\tilde{C}}_{\ell,N}^2$.

\subsection{The Statistical Variance for a Gaussian-Distributed Skymap}
\label{ssec:gaussian}
In this subsection, we show how one can return to an ensemble of skymaps by applying the new formalism to the example of a Gaussian skymap.

The results derived in this section so far are quite general, applicable to any spatial distribution of source for the observed events $\bn_i$. Additional sensitivity should be achievable with an accurate model of the sources. 

If a statistical model of the distribution of a hypothetical source is to be tested, it is important to do so in a statistical framework that limits the ensemble of realizations to those consistent with the proposed model. Once a model is established in terms of some parameters to be determined from the measurement, then a maximum likelihood analysis may be applied and a minimum variance achieved for the class of distributions considered.

Alternatively, in the case where the statistical model can be expressed in terms of correlation functions of the skymap, then ensemble averages of $\tilde{C}_\ell$, $\tilde{C}_\ell^{(2)}$, $\tilde{C}_\ell^{(3)}$, and $\tilde{C}_\ell^2$ can be determined and used to calculate the variance of the mean power spectrum, averaged over the ensemble defined by the statistical model.

To demonstrate this point, the variance of a Gauss\-i\-an-distributed source is determined. Assume that the normalized skymap $\tilde{S}(\bn)$ is Gaussian-distributed (statistically stationary and isotropic) with fixed covariance specified as the Legendre polynomial transform of the mean power spectrum $\meangb{\tilde{C}_\ell}$, as in Appendix~\ref{ap:likelihood}, where the average $\meang{X}$ denotes the ensemble average over all skymap realizations of the Gaussian distribution $\mathcal{G}$. Recall that $\tilde{S}$ was normalized by the spherical average of measurements, and as such $\meangb{\tilde{S}(\bn)}$ is not necessarily 1. However, the results expressed here make the further assumption that any deviation of $\meangb{\tilde{S}(\bn)}$ from 1 is negligible for our purposes.

Since the effects of the PSF are more complicated in this section than a simple convolution, this subsection will indicate explicitly those quantities for which the PSF affect is applied with a subscript `$F$', as in Eqn.~(\ref{eq:psfconvolution}). Initially, the PSF is neglected.

The assumption of Gaussianity allows all central moments of $\tilde{S}$ to be expressed in terms of the covariance $\mathcal{C}(\bn,\bn')$ (or equivalently $\meangb{\tilde{C}_\ell}$) only, as explained in Appendix~\ref{ap:moments}. In this case, the cosmic mean of the $\hat{\tilde{C}}_{\ell,N}$-variance in \eqn{eq:Chatvar} can be expressed solely in terms of $\meangb{\tilde{C}_\ell}$. That is, it provides a Gaussian variance estimate that depends only on the estimate of $\tilde{C}_\ell$.

The expression for $\meangb{\tilde{C}_\ell^{(2)}}$ follows from \eqn{eq:cl2fromcl}, and a vanishing central 3-moment implies that $\tilde{C}_\ell^{(3)}=0$.

For the mean quadratic term of the variance, apply \eqn{eq:4moment} to the ensemble average of \eqn{eq:clfrompl} squared,
\begin{align}
  &\!\!\meang{\tilde{C}_\ell^2}\!=(4\pi)^2\!\!\int\!\frac{\der\bn_1}{4\pi}\frac{\der\bn_2}{4\pi}\frac{\der\bn_3}{4\pi}\frac{\der\bn_4}{4\pi}P_\ell(\bn_1\cdot\bn_2)P_\ell(\bn_3\cdot\bn_4)\!\big[\nonumber\\
  &\qquad\mathcal{C}(\bn_1\cdot\bn_2)\mathcal{C}(\bn_3\cdot\bn_4)+\mathcal{C}(\bn_1\cdot\bn_3)\mathcal{C}(\bn_2\cdot\bn_4)+\nonumber\\
  &\qquad\mathcal{C}(\bn_1\cdot\bn_4)\mathcal{C}(\bn_2\cdot\bn_3)\big].
\end{align}
Then substitute \eqn{eq:covarfromcl} for the covariance and apply Legendre polynomial contractions \eqn{eq:sumleg} to find
\begin{align}
  \meang{\tilde{C}_\ell^2}&=\left[1+2\int\frac{\der\bn}{4\pi}\frac{\der\bn'}{4\pi}P_\ell^2(\bn\cdot\bn')\right]\meang{\tilde{C}_\ell}^{\!\!2}\nonumber\\
  &=\left(1+\frac{2}{2\ell+1}\right)\meang{\tilde{C}_\ell}^{\!\!2},\label{eq:meangcellsq}
\end{align}
providing yet another derivation of the Gaussian cosmic variance of $\tilde{C}_\ell$, Eqns.~(\ref{eq:cvarbare}) and (\ref{eq:cosmicvar}); doing so, we demonstrated how to determine any higher moments of $\tilde{C}_\ell$.

Thus, for a theoretically Gaussian-distributed sky\-map, the cosmic mean variance of the power spectrum estimator is
\begin{align}
  &\meang{\var{\hat{\tilde{C}}_{\ell,N}}}\!\!=\!\frac{(4\pi)^2}{N(N-1)}\Bigg[\frac{2(1-\delta_{\ell,0})}{2\ell+1}+\!\frac{4(N-2)}{2\ell+1}\frac{\meangb{\tilde{C}_\ell\!}}{4\pi}\nonumber\\
  &\qquad\qquad-(4N-6)\left(1+\frac{2}{2\ell+1}\right)\left(\frac{\meangb{\tilde{C}_\ell}}{4\pi}\right)^{\!\!2}\nonumber\\
  &\qquad\qquad+2\sum_{\ell'=0}^{2\ell}(2\ell'+1)\wignerthreej{\ell}{\ell}{\ell'}{0}{0}{0}^2\frac{\meangb{\tilde{C}_{\ell'}}}{4\pi}\Bigg].
\end{align}
In this setup, $\hat{\tilde{C}}_{\ell,N}$ is not just an estimator for the $\tilde{C}_\ell$ of the particular sky being sampled, but an estimator of the cosmic mean $\meangb{\tilde{C}_\ell}$:
\begin{equation}
  \left.\left\langle\tilde{C}_\ell\right\rangle\hspace{-7.6mm}\widehat{\phantom{\left\langle\tilde{C}_\ell\right\rangle}}\ \right._{\mathcal{G},N}=4\pi(\hat{\Delta}_{\ell,N}-\delta_{\ell,0}).
\end{equation}
That is, a physical theory that predicts the Gaussianity of $S(\bn)$ must also predict values of $\meangb{\tilde{C}_\ell}$ that are consistent with the measurement of $\hat{\tilde{C}}_\ell$. In this case, cosmic variance is an important consideration when comparing the theory to the measurement since the sky being sampled could be any random realization. We conjecture that the variance of the ensemble-averaged power spectrum estimator is well-described by the sum of the cosmic variance of the power spectrum and the ensemble average of the statistical variance of the fixed-sky's power spectrum estimator.

To justify the use of the Gaussian-source model, the experiment needs to verify that $\hat{\tilde{C}}_{\ell,N}^{(3)}=0$ to within uncertainties. The measured value of $\hat{\tilde{C}}_{\ell,N}$ has variance
\fbox{
\addtolength{\linewidth}{-4\fboxsep}%
\addtolength{\linewidth}{-2\fboxrule}%
\begin{minipage}{\linewidth}
\begin{align}
  &\varb{\left.\left\langle\tilde{C}_\ell\right\rangle\hspace{-7.6mm}\widehat{\phantom{\left\langle\tilde{C}_\ell\right\rangle}}\ \right._{\mathcal{G},N}}=\meang{\var{\hat{\tilde{C}}_\ell}}+\meang{\tilde{C}_\ell^2}-\meang{\tilde{C}_\ell}^{\!\!2}\\
  &=\frac{1}{1-\frac{1}{N}}\frac{2}{2\ell+1}\Bigg[\left(\frac{4\pi}{N}\right)^{\!\!2}+2\left(1-\frac{2}{N}\right)\frac{4\pi}{N}\meang{\tilde{C}_\ell}\nonumber\\
  &\qquad\qquad+\left(1-\frac{2}{N}\right)\left(1-\frac{3}{N}\right)\meang{\tilde{C}_\ell}^{\!2}\Bigg]\nonumber\\
  &\qquad+\frac{(4\pi)^2}{N(N-1)}\Bigg[-(4N-6)\left(\frac{\meangb{\tilde{C}_\ell}}{4\pi}\right)^{\!\!2}\nonumber\\
  &\qquad\qquad+2\sum_{\ell'=0}^{2\ell}(2\ell'+1)\wignerthreej{\ell}{\ell}{\ell'}{0}{0}{0}^2\frac{\meangb{\tilde{C}_{\ell'}}}{4\pi}\Bigg].\label{eq:clvargauss}
\end{align}
\end{minipage}
}
This equation represents an update on the standard \eqn{eq:cvar}, which corresponds to the first group of terms. Since these terms are suppressed at high $\ell$ by a factor of $(2\ell+1)^{-1}$, then for any finite value of $N$, there must be a (possibly large) multipole above which the new second group of terms is dominant. These new terms take into account the `statistical wandering' of the measured value of the angular power spectrum of our particular skymap that occurs with the addition of more detected events. Further comparison to \eqn{eq:cvar} is carried out in \ssec{ssec:compare}.

The effect of the PSF in this context will now be considered. With our model for $\meang{\tilde{C}_\ell}$ and our knowledge of the PSF $F(\bn_s,\bn)$, the cosmic mean power spectrum of skies that have been convolved by $F$ can be expressed as
\begin{equation}
  \label{eq:PSFClGauss}
  \meang{\tilde{C}_{F\ell}}=\sum_{\ell'}\scw_{G\ell\ell'}\meang{\tilde{C}_{\ell'}}+R_\ell,
\end{equation}
with
\begin{equation}
  \scw_{G\ell\ell'}\equiv\int\frac{\der\bn_{s1}}{4\pi}\frac{\der\bn_{s2}}{4\pi}(2\ell'+1)P_{\ell'}(\bn_{s1}\cdot\bn_{s2})\frac{\scw_{\ell}(\bn_{s1},\bn_{s2})}{4\pi}
\end{equation}
and
\begin{equation}
  R_\ell\equiv\int\frac{\der\bn_{s1}}{4\pi}\frac{\der\bn_{s2}}{4\pi}\scw_\ell(\bn_{s1},\bn_{s2}),
\end{equation}
where $\scw_\ell$ is defined in \eqn{eq:windowfunction}. Note that $R_\ell$ vanishes in the usual scenario where $F$ is symmetric. Let us assume this standard scenario. Then, using \eqn{eq:symmetricpsfpowspec},
\begin{equation}
  W_{G\ell\ell'}=W_\ell^2(\delta_{\ell\ell'}-\delta_{\ell,0}\delta_{\ell',0}),
\end{equation}
and we recover the usual result that 
\begin{equation}
  \meang{\tilde{C}_{F\ell}}=W_\ell^2\meang{\tilde{C}_{\ell}}
\end{equation}
for symmetric PSFs.

As before, the mean composite spectrum $\meangb{\tilde{C}_{F\ell}^{(2)}}$ is the same linear combination of mean power spectra. Also, as long as $F$ is symmetric, the convolution of $F$ over a Gaussian sky still has no bispectrum. The mean disjoint trispectrum can be expressed with symmetric PSF as in \eqn{eq:meangcellsq}. We then find that \eqn{eq:clvargauss} does indeed still hold for the PSF-convolved sky, for a symmetric PSF.

\section{Analytical and Numerical Comparisons of Methods}
\label{sec:comp}
Now that we have succeeded in determining an analytic treatment of the variance of the angular power spectrum of events on a sphere, it is helpful to compare our results with the previous analytical estimates and identify the new effects. We then present some simple simulations that verify these effects and demonstrate the effectiveness of the variance estimates of the power spectrum.

\subsection{Discussion of the New Contributions to the Power Spectrum Variance}
\label{ssec:compare}
As discussed in \ssec{sec:cvar}, previous analytic estimates of the variance of the angular power spectrum were based on Gaussian cosmic variance with noise. The statistical variance of the estimator of the power spectrum \eqn{eq:Chatvar2} is a generalization of the $C_\ell$-only variance of \eqn{eq:cellonlyvar}, or
\begin{equation}
  \label{eq:Clonly}
  \var{\hat{\tilde{C}}_{\ell,N}}\simeq\frac{2}{2\ell+1}\frac{4\pi}{N}\left(\frac{4\pi}{N}(1-\delta_{\ell,0})+2\tilde{C}_\ell\right).\ \ \text{($C_\ell$-only)}
\end{equation}
It is convenient to re-express this as
\begin{equation}
  \label{eq:Clonly2}
  \var{\hat{\tilde{C}}_{\ell,N}}\simeq\left(\frac{4\pi}{N}\right)^{\!\!2}\!\!\Bigg[\frac{2(1-\delta_{\ell,0})}{2\ell+1}+\frac{4N}{2\ell+1}\frac{\tilde{C}_\ell}{4\pi}\Bigg].
\end{equation}
for direct comparison with \eqn{eq:Chatvar2}. We see that \eqn{eq:Clonly} is indeed valid under certain conditions. When $N\gg1$, the first term reproduces the first-order shot effect, and the second term corresponds to the first-order estimate of the signal term of the variance. However, newly identified second order effects may provide important corrections. The multipole-dependent factor of the shot term is corrected by $C_\ell^{(2)}-C_\ell^{(4)}/(4\pi)^2$, the difference of the composite power spectrum with the disjoint trispectrum. Similarly, the signal term's multipole dependence is corrected by $C_\ell^{(3)}/(4\pi)-C_\ell^{(4)}/(4\pi)^2$, the difference of the open bispectrum with the disjoint trispectrum.

Examples of statistical uncertainties of $\hat{C}_{500,N}$ when $\hat{C}_{500,N}=10^{-5}$ sr are shown by the red curves in Fig.~\ref{fig:clerror}. In all cases, the uncertainty separates into two regimes. At small $N$, the shot term is dominant and the uncertainty decreases as $N^{-1}$. At large $N$, the statistical uncertainty is now dominated by the signal term and continues to shrink, albeit at the reduced rate of $N^{-1/2}$.

For a power spectrum as small in magnitude as in this example, the disjoint trispectrum $C_\ell^{(4)}$ does not contribute. If $C_\ell^{(2)}$ and $C_\ell^{(3)}$ are also of negligible magnitude, then the variance reduces back to the $C_\ell$-only variance.

The necessary and sufficient conditions for the power spectrum variance to be $C_\ell$-only depend on the magnitude of $N$, and are given as follows.
\begin{enumerate}
\item $N\gg1$. When $N$ is $\mathcal{O}(1)$, the variance has a small-counts correction, but this is usually not that significant.
\item Second-order shot term effects are negligible when the error is not signal-dominated,
\begin{equation}
  \left|\tilde{C}_\ell^{(2)}-\left(\frac{\tilde{C}_\ell}{4\pi}\right)^{\!\!2}\right|\ll\frac{1}{2\ell+1}\text{ if }N\not\gg2\pi/\tilde{C}_\ell.
\end{equation}
\item Second-order signal term effects are negligible when the error is shot-dominated,
\begin{equation}
  \left|\frac{\tilde{C}_\ell^{(3)}}{4\pi}-\left(\frac{\tilde{C}_\ell}{4\pi}\right)^{\!\!2}\right|\ll\frac{1}{2N(2\ell+1)}\text{ if }N\ll2\pi/\tilde{C}_\ell.
\end{equation}
\item Second-order signal term effects are negligible when the error is not shot-dominated,
\begin{equation}
  \left|\frac{\tilde{C}_\ell^{(3)}}{4\pi}-\left(\frac{\tilde{C}_\ell}{4\pi}\right)^{\!\!2}\right|\ll\frac{1}{2\ell+1}\frac{\tilde{C}_\ell}{4\pi}\text{ if }N\not\ll2\pi/\tilde{C}_\ell.
\end{equation}
\item Second-order shot term effects are negligible when the error is signal-dominated,
\begin{equation}
  \left|\tilde{C}_\ell^{(2)}-\left(\frac{\tilde{C}_\ell}{4\pi}\right)^{\!\!2}\right|\ll\frac{2N}{2\ell+1}\tilde{C}_\ell\text{ if }N\gg2\pi/\tilde{C}_\ell.
\end{equation}
\end{enumerate}
The range of counts for which the variance remains $C_\ell$-only depends on the multipole being analyzed, and in principle it is possible for some or all multipoles' power spectrum to never have $C_\ell$-only variances.

The approximate number $N_*$ of events received when the $C_\ell$-only variance transitions from shot-dominated to signal-dominated is
\begin{equation}
  N_*=\frac{4\pi}{\tilde{C}_\ell},
\end{equation}
at which time the magnitude of the variance is
\begin{equation}
  \var{\hat{C}_{\ell,N_*}}=\frac{4}{2\ell+1}\tilde{C}_\ell^2.
\end{equation}
Thus, for large multipoles, the magnitude of the statistical error bar is already much less than the magnitude of the power spectrum at the time the uncertainty becomes signal-dominated. The power spectrum is therefore resolved while its error is still shot-dominated.

These results can be modified by the presence of significant higher-order spectra $\tilde{C}_\ell^{(2)}$ and $\tilde{C}_\ell^{(3)}$ in the skymap. These effects make the uncertainty larger--potentially much larger; the only exception being when $\tilde{C}_\ell^{(3)}<0$ and $N>N_*$, where the uncertainty becomes smaller than $C_\ell$-only, but by this time, the power spectrum is usually already detected to high significance. 

A significant composite power spectrum $\tilde{C}_\ell^{(2)}$ can occur at multipole $\ell$ if power spectrum amplitudes $\tilde{C}_{\ell'}$ at neighboring multipoles $\ell'$ are much larger than $\tilde{C}_\ell$, and are also significant compared to $1/(2\ell+1)$. The composite power spectrum increases the magnitude of the shot term, and the variance is no longer $C_\ell$-only. It will become $C_\ell$-only at large $N$ if the open bispectrum is negligible. A significant $\tilde{C}_\ell^{(2)}$ increases the number of events needed to resolve $\tilde{C}_\ell$, and also increases the number of events where the transition to signal domination occurs.

In the case that there are significant three-point correlations, the open bispectrum $\tilde{C}_\ell^{(3)}$ can be significant. A non-negligibile $\tilde{C}_\ell^{(3)}$ causes the signal transition of the uncertainty to occur at lower $N$ for positive open bispectra, and at higher $N$ for negative open bispectra. If this happens, the spectrum is not $C_\ell$-only at large $N$ when the variance is signal-dominated.

In all cases except for negative open bispectrum, the second-order effects cause the variance to be \emph{larger} than the $C_\ell$-only variance. However, for negative open bispectrum, the signal-dominated variance will be \emph{smaller} than the $C_\ell$-only variance; that is, the presence of a negative open bispectrum decreases the statistical variation of measurement of the sky's power spectrum, allowing it to be determined more precisely than for a sky without an open bispectrum, or with positive open bispectrum. This will have little effect on the number of events required to initially resolve the power spectrum.

If the open bispectrum is significant and positive, it causes an early transition to a signal-term-dominated error bar, which shrinks as $N^{-1/2}$ instead of $N^{-1}$. If $\tilde{C}_\ell^{(3)}$ is so large that this transition occurs before the power spectrum is resolved, the time required to resolve the spectrum is significantly increased. Thus, the presence of angular bispectrum complicates the measurement of the angular power spectrum. If, in this case, the bispectrum effect were neglected, a spurious detection of the power spectrum could be made.

Now we will discuss the measurement of the cosmic mean power spectrum for the case of a Gaussian angular distribution. This measurement requires the cosmic variance of the observed sky-realization to be taken into acount. The new results of \ssec{ssec:gaussian} should be compared with the Gaussian cosmic variance with noise estimate \eqn{eq:cvar2}.
For a more direct comparison, it is convenient to re-express \eqn{eq:cvar2} as
\begin{align}
  \label{eq:oldvar2}
  \!\!\!\!\!\!\!\!\left[\var{\hat{\tilde{C}}_{\ell,N}}\right]_{\text{CV}}\!\!=\left(\frac{4\pi}{N}\right)^{\!\!2}\!\!\Bigg[&\frac{2(1-\delta_{\ell,0})}{2\ell+1}+\frac{4N}{2\ell+1}\frac{\tilde{C}_\ell}{4\pi}\nonumber\\
  &+\frac{2N^2}{2\ell+1}\left(\frac{\tilde{C}_\ell}{4\pi}\right)^{\!\!2}\Bigg].
\end{align}

\begin{figure*}
  \subfloat[$\ell=10$]{\includegraphics[width=0.5\textwidth]{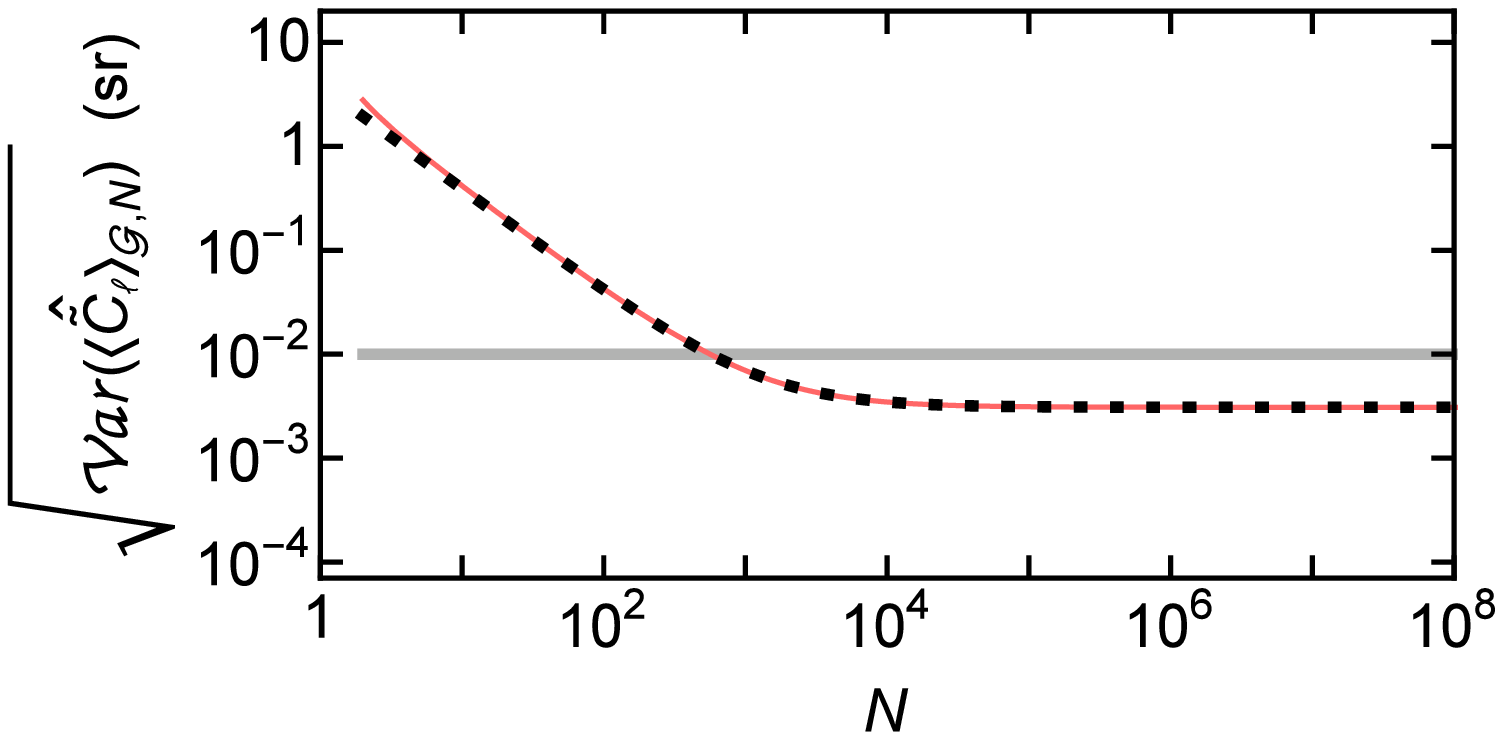}}
  \subfloat[$\ell=100$]{\includegraphics[width=0.5\textwidth]{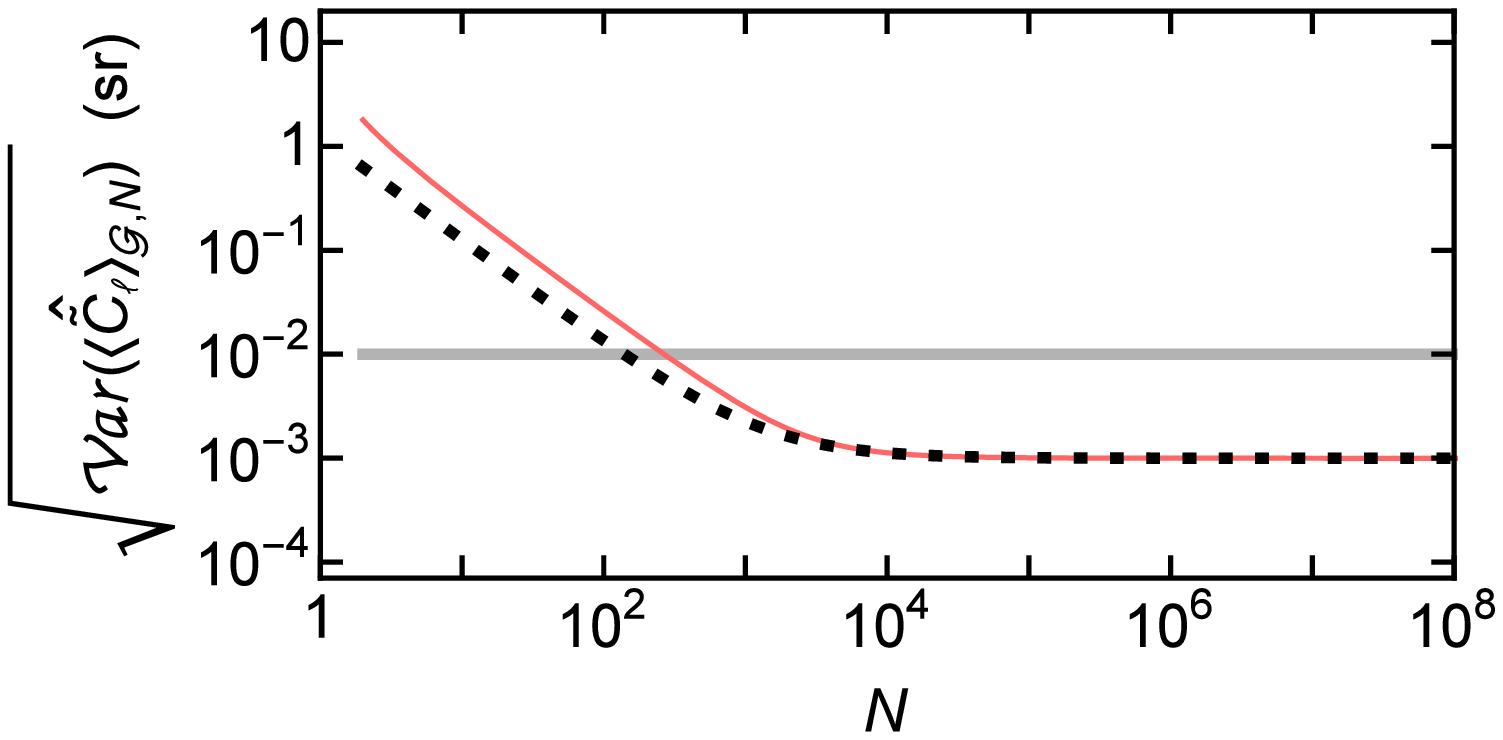}}\\\vspace{-5mm}
  \subfloat[$\ell=1000$]{\includegraphics[width=0.5\textwidth]{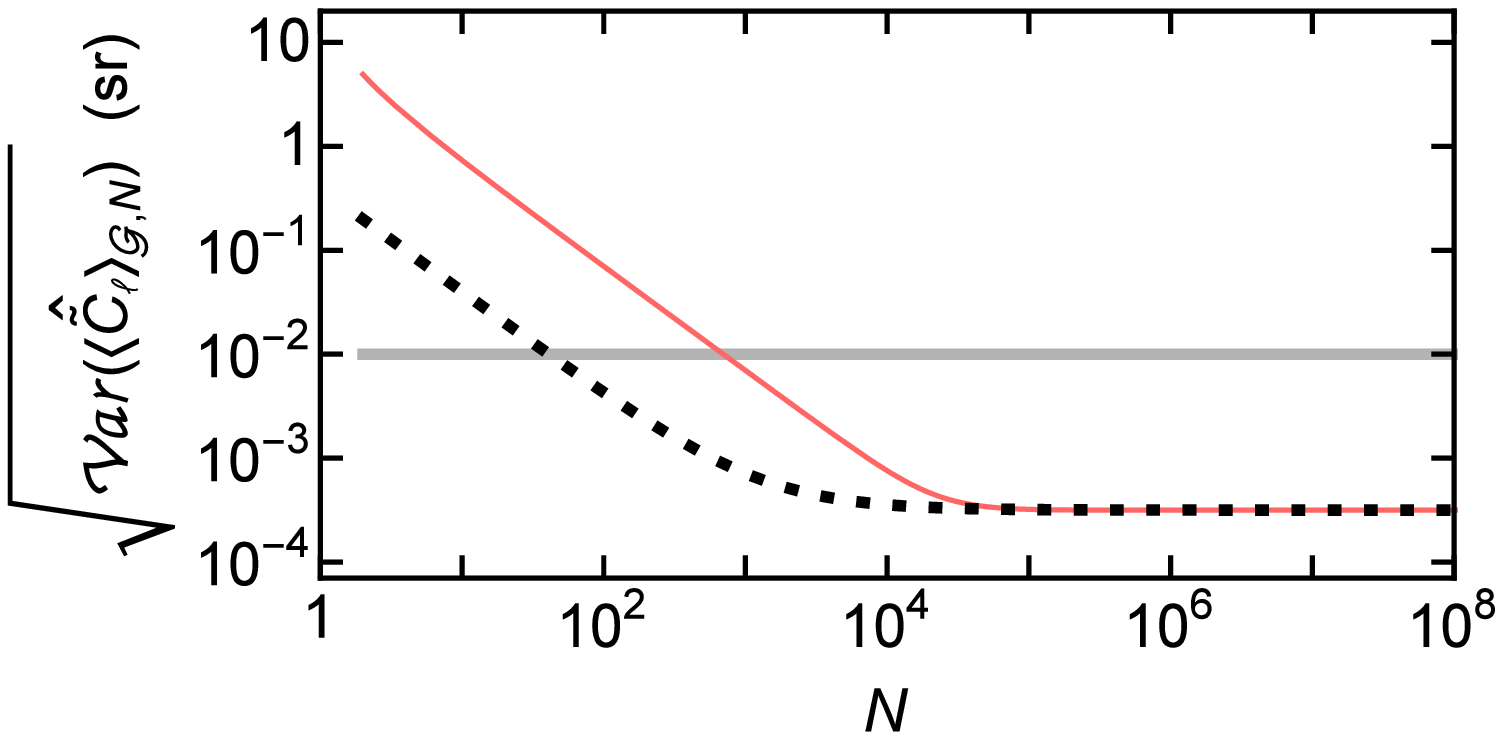}}
  \subfloat[$\ell=10\ 000$]{\includegraphics[width=0.5\textwidth]{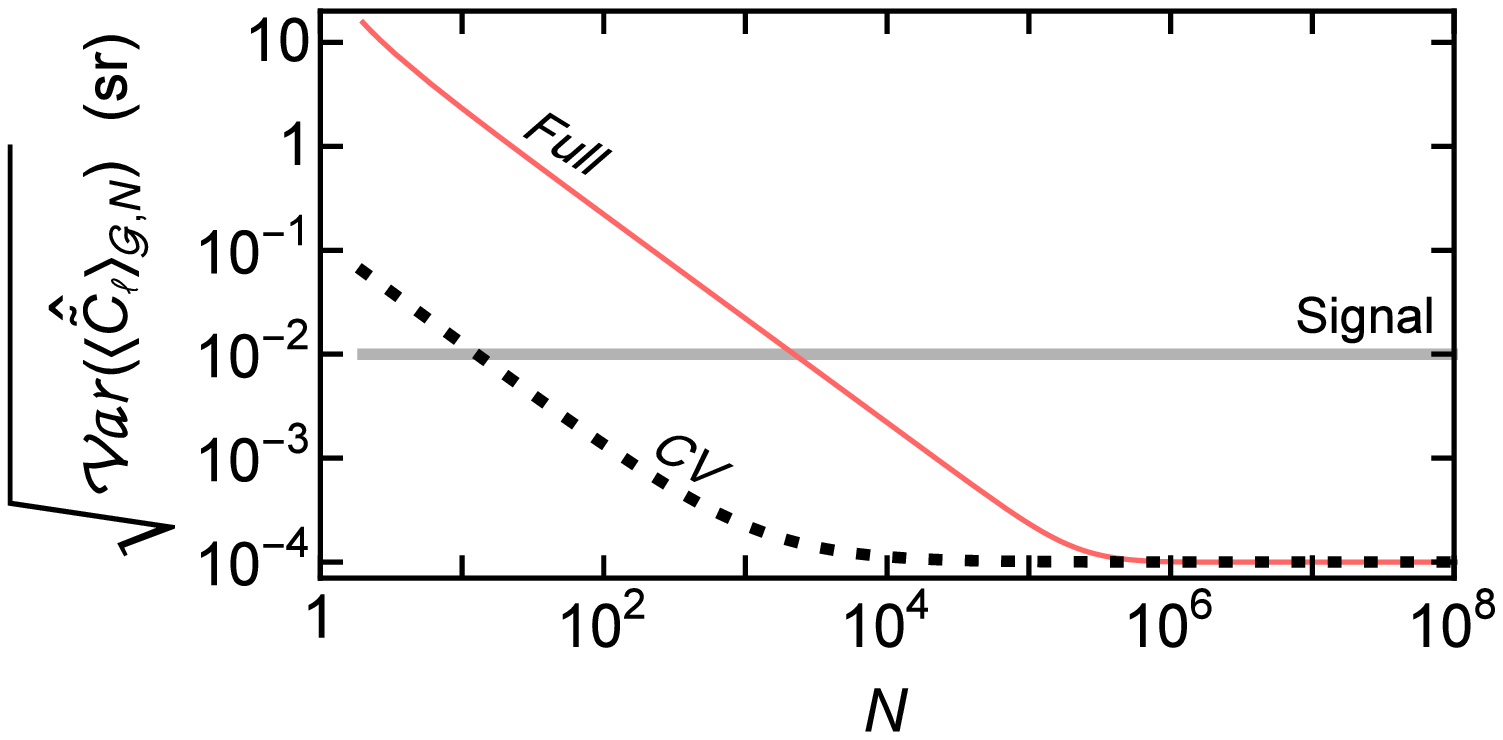}}
  \caption{\label{fig:clgerror}Examples of the statistical uncertainty of the cosmic mean power spectrum (i.e., ensemble-averaged over all possible Universes) of a Gaussian signal, sampled with $N$ random points from a single random position within a single random Universe. This is opposed to Fig.~\ref{fig:clerror} which shows the uncertainty of the power spectrum estimate of one particular skymap without prior assumptions about the signal's spatial distribution. For demonstration purposes, each example here models $\meangb{\tilde{C}_{\ell'}\!}\propto4\pi/[\ell'(\ell'+1)]$ with normalization $\meangb{\tilde{C}_\ell}=\unit{10^{-2}}{sr}$ for the multipole being considered. The black dotted lines show the cosmic variance (CV) with noise using \eqn{eq:cvar2}, while the red solid lines show the full statistical error using \eqn{eq:clvargauss}.}
\end{figure*}

The updated variance in \eqn{eq:clvargauss} contains both low-count corrections and new terms. The new terms are not diminished by $(2\ell+1)^{-1}$ like the previously known terms are. Thus, the new effects become more important at large $\ell$. Fig.~\ref{fig:clgerror} shows examples of the total variance of the estimator for the cosmic mean power spectrum for a Gaussian-distributed signal and scale-invariant Sachs-Wolfe spectrum (see \eqn{eq:swcl2})
\begin{equation}
  \meangb{\tilde{C}_\ell}\propto\frac{4\pi}{\ell(\ell+1)}.
\end{equation}
For each example shown, we fixed $\meangb{\tilde{C}_\ell}=\unit{10^{-2}}{sr}$. For such a large amplitude, the new effects are seen to increase the finite-count variance significantly at high multipoles, until the cosmic variance dominates the statistical uncertainty. For $\ell\ga100$ in these examples, there is additional shot error induced by $\meangb{\tilde{C}_\ell^{(2)}}$ until the cosmic variance dominates, which happens when
\begin{equation}
  N\ga\frac{4\pi}{\ \meangb{\tilde{C}_\ell}}\sqrt{(2\ell+1)\meangb{\tilde{C}_\ell^{(2)}}}.
\end{equation}

\subsection{Comparison with Simulations}
\label{ssec:sims}

We now compare the discussed formalisms with the results of Monte-Carlo simulations. These results address the single-sky measurement of power spectrum. The purpose is to simulate skymaps that demonstrate the new effects in this paper, not to simulate physical sources. We will describe two skymaps chosen for our comparison, discuss the simulations and implementation of the power spectra evaluations, and present the results.

For simple first tests, we simulate skymaps that contain anisotropy of only one multipole $\ell_0$. The strategy is to simulate observations of events from two different skymaps with the same power spectrum, but with different open bispectrum, and determine if the variance of the power spectrum estimate is affected.

We consider single-multipole, real-valued, normalized skymaps with multipole $\ell_0$ of the form
\begin{align}
  \label{eq:S1ell}
  \tilde{S}&(\bn;\ell_0;\eta_0,\eta_1,\dotsc,\eta_{\ell_0})=1+\eta_0Y_{\ell_0}^{\ 0}(\bn)\\
  &\qquad\qquad+\sum_{m=1}^{\ell_0}\eta_m\left[Y_{\ell_0}^{\ m}(\bn)+(-1)^mY_{\ell_0}^{\ -m}(\bn)\right],\nonumber
\end{align}
where the real constants $\eta_0,\dotsc,\eta_{\ell_0}$ are constrained only by the requirement that $\tilde{S}(\bn)\geq0$ at all positions $\bn$. Note that the imaginary parts of the spherical harmonics can also be added here with imaginary coefficients. 

These functions have power spectrum
\begin{equation}
  \tilde{C}_\ell=\frac{\delta_{\ell,\ell_0}}{2\ell_0+1}\left(\eta_0^2+2\sum_{m=1}^{\ell_0}\eta_m^2\right),
\end{equation}
composite power spectrum
\begin{equation}
  \tilde{C}_\ell^{(2)}=
  \begin{cases}
    0,&\ell<\left\lceil\frac{\ell_0}{2}\right\rceil,\\
    \displaystyle\frac{1}{4\pi}\left(\eta_0^2+2\sum_{m=1}^{\ell_0}\eta_m^2\right)\wignerthreej{\ell}{\ell}{\ell_0}{0}{0}{0}^{\!\!2},&\ell\geq\left\lceil\frac{\ell_0}{2}\right\rceil,
  \end{cases}
\end{equation}
and open bispectrum
\begin{align}
  \tilde{C}_\ell^{(3)}=&\ \delta_{\ell,\ell_0}\frac{1}{\sqrt{4\pi(2\ell_0+1)}}\wignerthreej{\ell_0}{\ell_0}{\ell_0}{0}{0}{0}\Bigg[\eta_0^3\wignerthreej{\ell_0}{\ell_0}{\ell_0}{0}{0}{0}\nonumber\\
  &+3\!\left[1+(-1)^{\ell_0}\right]\!\!\sum_{m_1=1}^{\ell_0}\sum_{m_2=0}^{m_1-1}(-1)^{m_1}\eta_{m_1}\eta_{m_2}\eta_{m_1-m_2}\nonumber\\
  &\qquad\wignerthreej{\ell_0}{\ell_0}{\ell_0}{-m_1}{m_2}{m_1-m_2}\Bigg].
\end{align}
For odd multipoles, only the azimuthally symmetric mode $\eta_0$ contributes to the open bispectrum. Our numerical experiments chose $\ell_0=12$ in order to have enough symmetries and degrees of freedom to produce interesting effects with not too large of a parameter landscape.

In order to resolve the bispectrum effect, we want a significant magnitude of $(2\ell+1)\tilde{C}_\ell^{(3)}/\tilde{C}_\ell$. One function that achieves a ratio of $-0.19$ is, using \eqn{eq:S1ell},
\begin{align}
  \tilde{S}_B(\bn)&\equiv\tilde{S}(\bn;12;0,0,0.3,0,-0.4,0,-0.4,0,\nonumber\\
  &\qquad-0.3,0,0.3,0,-0.3).
\end{align}
It has a power spectrum of $\tilde{C}_\ell=0.0544\,\delta_{\ell,12}$ sr, and open bispectrum $\tilde{C}_\ell^{(3)}=-0.000413\,\delta_{\ell,12}$ sr. This distribution's bispectrum constitutes an expected 17\% decrease in the signal term of the power spctrum's variance.

The distribution without bispectrum that was simulated for comparison was the one where $\eta_6$ is the only non-zero component,
\begin{equation}
	\tilde{S}_{N\!B}(\bn)\equiv1+\frac{\sqrt{17}}{5}[Y_{12}^{\ 6}(\bn)+Y_{12}^{\ -6}(\bn)].
\end{equation}

\begin{figure*}
  \subfloat[$\tilde{S}_{N\!B}(\bn)$]{\includegraphics[width=0.5\textwidth]{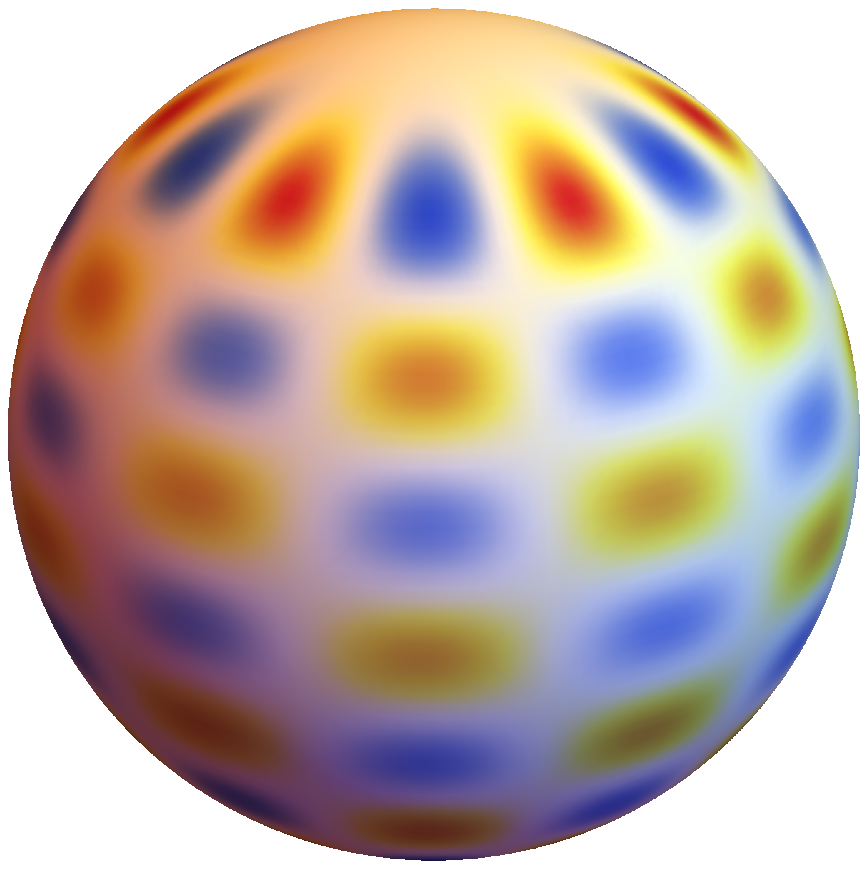}}
  \subfloat[$\tilde{S}_B(\bn)$]{\includegraphics[width=0.5\textwidth]{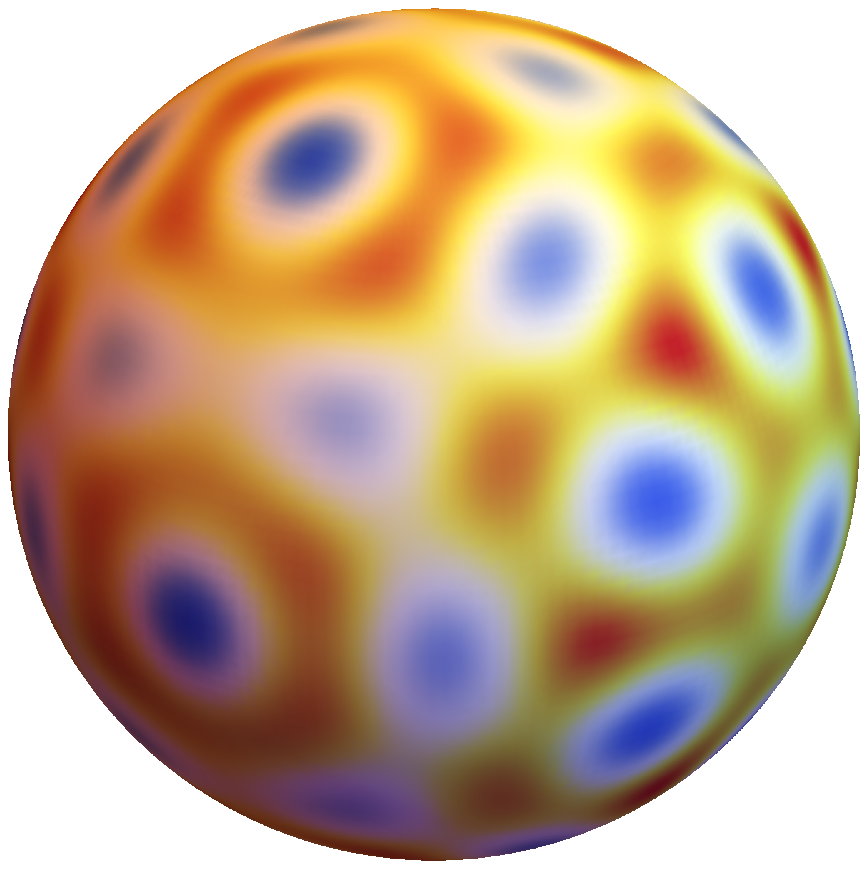}}
  \caption{\label{fig:spheres}Two distributions on the sphere with only $\ell=12$ harmonic modes. These distributions have equivalent power spectra. $\tilde{S}_{N\!B}$ has no open bispectrum, while $\tilde{S}_B$ has a relatively strong and negative open bispectrum.}
\end{figure*}

A density map of these two distributions, viewed with the same projection, is given in Fig.~\ref{fig:spheres}.

For each skymap, points on the sphere were randomly generated \citep{Knuth1997} with probability density function given by the skymap. For each trial, $10^5$ points were generated, and $10^5$ trials were performed for each skymap. For each trial, the irreducible power spectrum, composite spectrum, and open bispectrum were evaluated event-by-event for the first 1000 events, and only the power spectrum for the remaining 9000 events. The same set of random number generator seeds were used for the two skymaps.

It is efficient to calculate the irreducible power spectra iteratively as each event is added. A very efficient algorithm for the evaluation of Legendre polynomials with double precision operations that is also accurate to very high mulitpoles was implemented \citep{Bogaert2012,Harrison2009}. The irreducible power spectrum in \eqn{eq:irred1} was then determined as
\begin{equation}
  \hat{\Delta}_{\ell,N}=\frac{N-2}{N}\hat{\Delta}_{\ell,N-1}+\frac{2}{N(N-1)}\sum_{i=1}^{N-1}P_\ell(\bn_i\cdot\bn_N),
\end{equation}
and similarly for the irreducible composite power spectrum $\hat{\Delta}_{\ell,N}^{(2)}$ in \eqn{eq:irred2}, using $P_\ell^2$ instead of $P_\ell$. The irreducible open bispectrum in \eqn{eq:irred3} can be determined event-by-event with
\begin{align}
  \hat{\Delta}_{\ell,N}^{(3)}=&\ \frac{N-3}{N}\hat{\Delta}_{\ell,N-1}^{(3)}+\frac{2}{N(N-1)(N-2)}\sum_{i=1}^{N-1}\Bigg[\nonumber\\
  &\sum_{j=1}^{i-1}\Big[P_\ell(\bn_i\cdot\bn_N)+P_\ell(\bn_i\cdot\bn_j)\Big]P_\ell(\bn_j\cdot\bn_N)\nonumber\\
  &+\sum_{j=i+1}^{N-1}P_\ell(\bn_i\cdot\bn_j)P_\ell(\bn_j\cdot\bn_N)\Bigg].
\end{align}

Thus, the addition of the $N$'th event involves $N-1$ evaluations of the Legendre polynomial to determine the new power spectrum. Saving these evaluations in a look-up table, determination of the new composite spectrum and open bispectrum require $\mathcal{O}(N)$ and $\mathcal{O}(N^2)$ table look-ups, respectively. The irreducible disjoint trispectrum $\hat{\Delta}_{\ell,N}^{(4)}$ in \eqn{eq:irred4} can be determined from the other spectra using the spectrum identity in \eqn{eq:specidentity}.

These evaluations for $\ell=12$ took approximately three days to complete on a dual-core desktop computer for all trials of both skymaps.

\begin{figure*}
  \subfloat[$N=2$]{\includegraphics[width=0.3\textwidth]{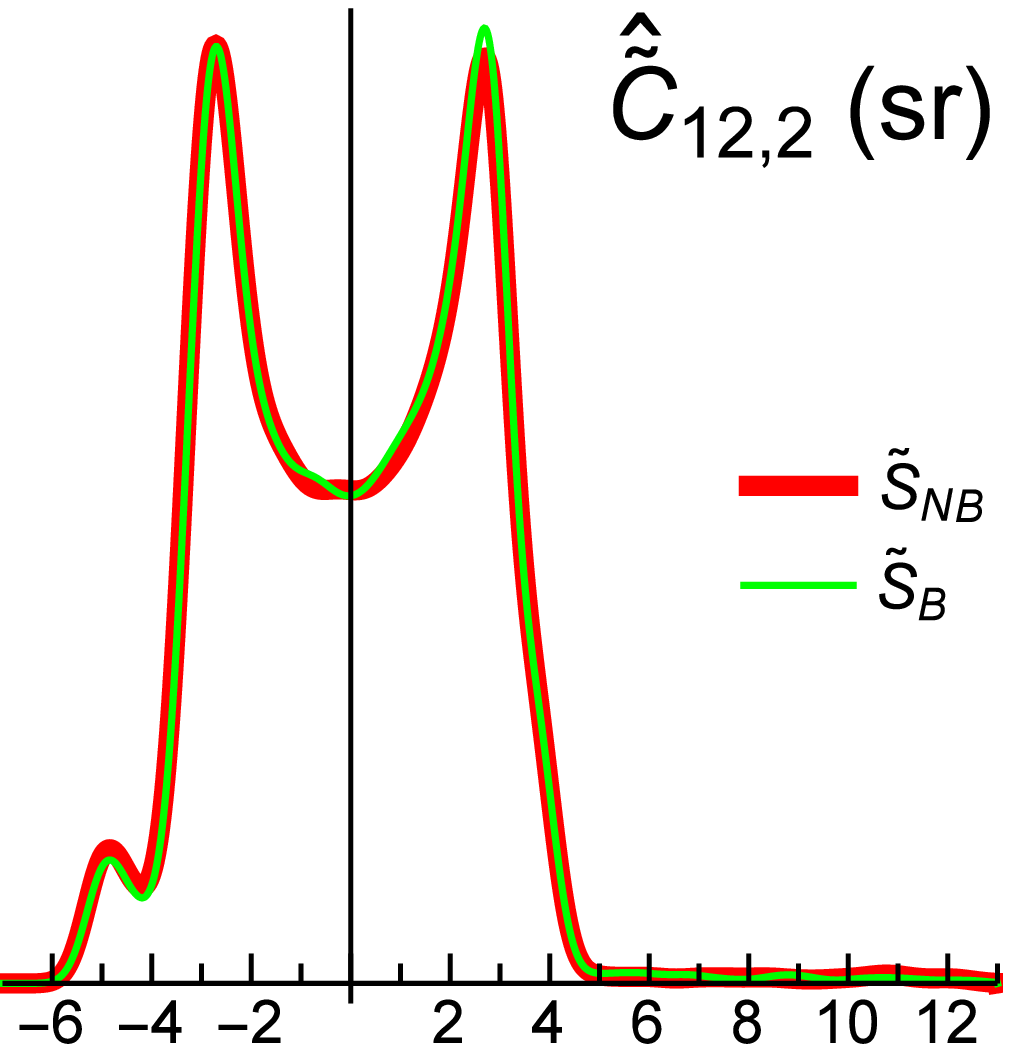}}\qquad
  \subfloat[$N=3$]{\includegraphics[width=0.3\textwidth]{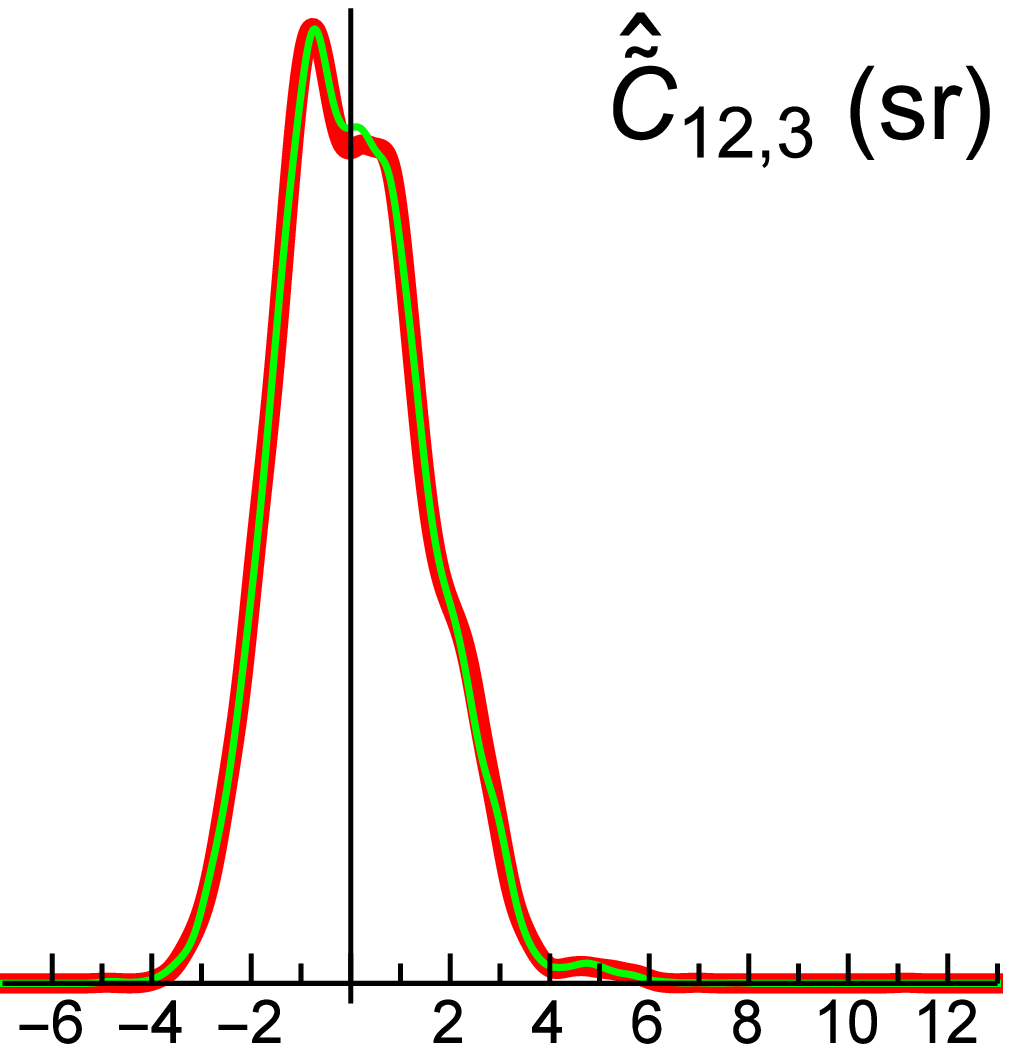}}\qquad
  \subfloat[$N=4$]{\includegraphics[width=0.3\textwidth]{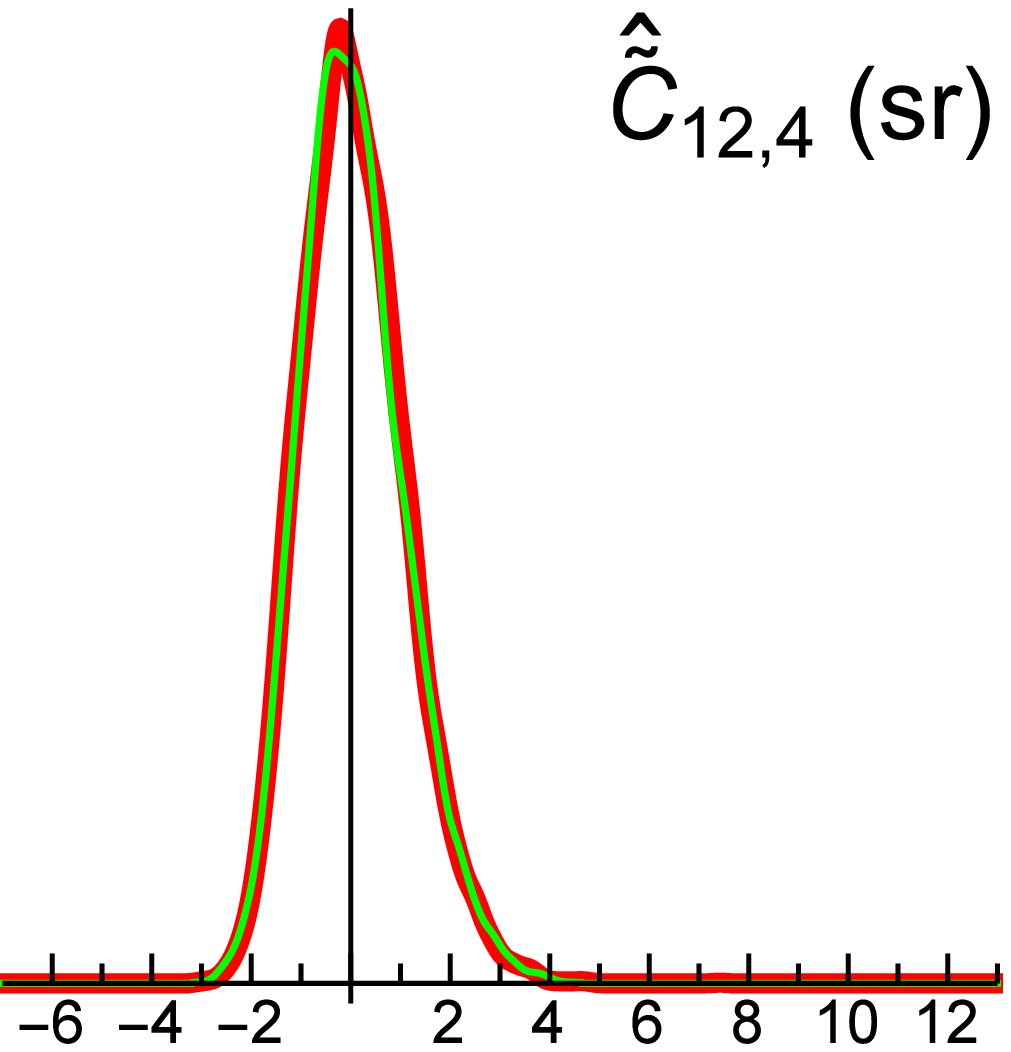}}
  \caption{\label{fig:powspecfewevent}Distribution of the angular power spectrum estimator \eqn{eq:ChatD} of the first $N$ events received. At just a few events, the distributions are wide, and there is no significant difference between the two different skymaps.}
\end{figure*}

We begin analyzing the results by comparing the distribution of power spectrum estimates at fixed $N$ for each sky. Fig.~\ref{fig:powspecfewevent} shows the distributions for the first few events. Interestingly, the estimator at 2 events shows a strong bimodal distribution, but this quickly collapses to a more typical bell curve by $N=4$. Initially, the distributions are very wide compared to the expectation value, and the distributions from the two skymaps cannot be distinguished at this number of trials.

\begin{figure*}
  \subfloat[$N=100$\newline\mbox{}\hspace{5mm}$\hat{\sigma}_{N\!B}=(4.77\pm0.60)\times10^{-2}$ sr\newline\mbox{}\hspace{5mm}$\hat{\sigma}_B=(4.47\pm0.55)\times10^{-2}$ sr\newline\mbox{}\hspace{5mm}$\sigma_{N\!B}=4.74\times10^{-2}$ sr\newline\mbox{}\hspace{5mm}$\sigma_B=4.52\times10^{-2}$ sr]{\includegraphics[width=0.3\textwidth]{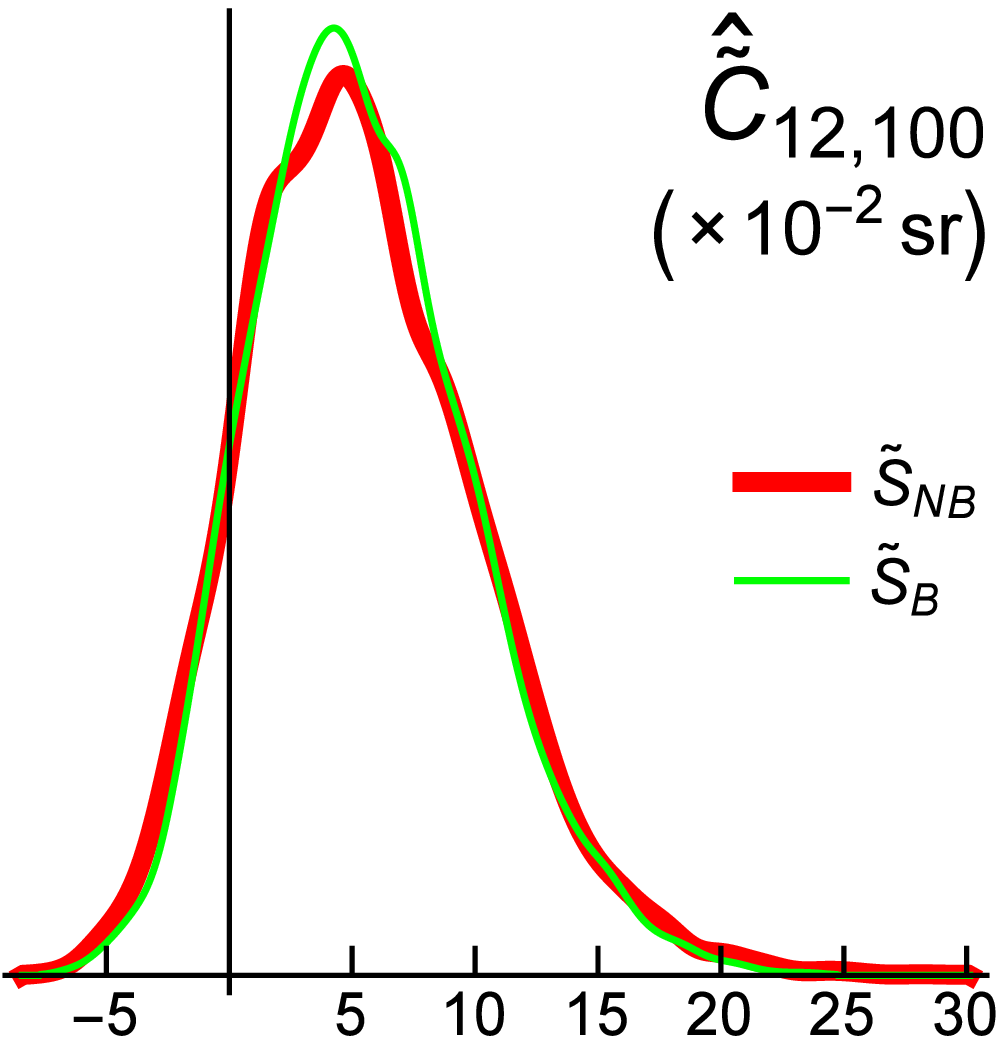}}\qquad
  \subfloat[$N=1000$\newline\mbox{}\hspace{5mm}$\hat{\sigma}_{N\!B}=(1.03\pm0.12)\times10^{-2}$ sr\newline\mbox{}\hspace{5mm}$\hat{\sigma}_B=(0.96\pm0.11)\times10^{-2}$ sr\newline\mbox{}\hspace{5mm}$\sigma_{N\!B}=1.05\times10^{-2}$ sr\newline\mbox{}\hspace{5mm}$\sigma_B=0.95\times10^{-2}$ sr]{\includegraphics[width=0.3\textwidth]{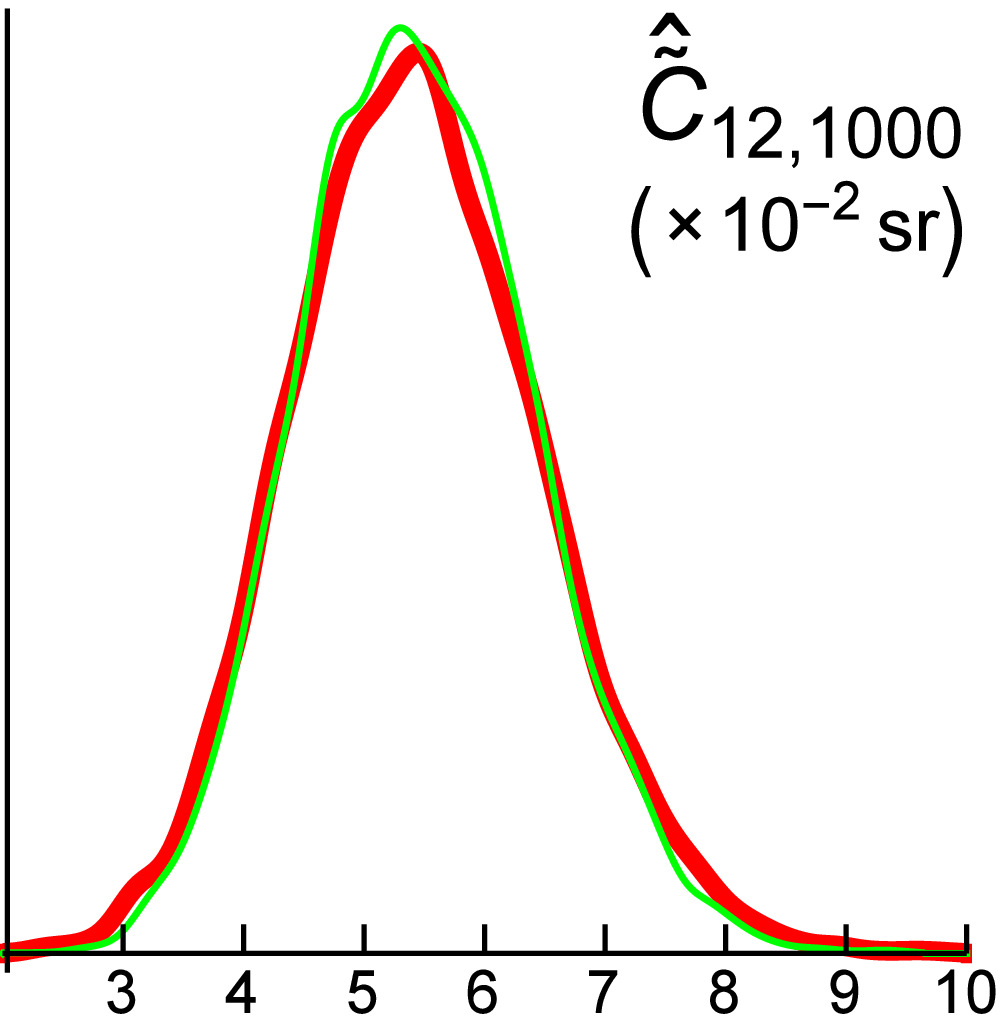}}\qquad
  \subfloat[$N=10000$\newline\mbox{}\hspace{5mm}$\hat{\sigma}_{N\!B}=(0.312\pm0.037)\times10^{-2}$ sr\newline\mbox{}\hspace{5mm}$\hat{\sigma}_B=(0.282\pm0.033)\times10^{-2}$ sr\newline\mbox{}\hspace{5mm}$\sigma_{N\!B}=0.314\times10^{-2}$ sr\newline\mbox{}\hspace{5mm}$\sigma_B=0.279\times10^{-2}$ sr]{\includegraphics[width=0.3\textwidth]{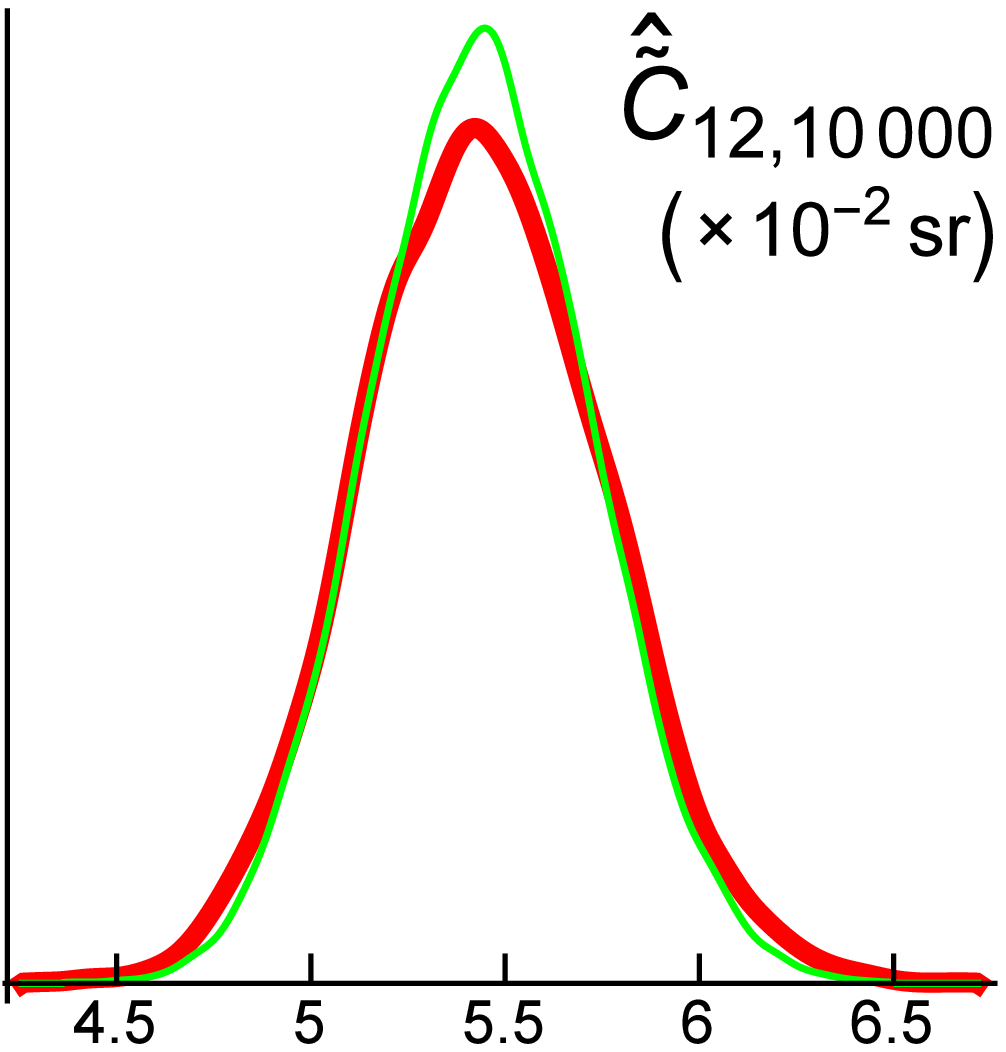}}
  \caption{\label{fig:powspecdist}Distribution of the angular power spectrum estimator \eqn{eq:ChatD} after observation of 100, 1000, and 10 000 events. Note the changing scale of the axis. As $N$ increases, the distributions become narrower, centered on the correct power spectrum of $\tilde{C}_{12}=0.0544$ sr. The standard deviations $\hat{\sigma}$ of these distributions are specified in the subcaptions, estimated with Eqns.~(\ref{eq:CtrVar})--(\ref{eq:CtrVarEst}). The skymap $\tilde{S}_B$ with the negative open bispectrum is consistently producing a slightly narrower distribution, though the difference is not statistically significant for each case individually. However, the standard deviations consistently agree with the expected values $\sigma=(\var{\hat{\tilde{C}}_{12,N}})^{1/2}$ from \eqn{eq:Chatvar2} to a better precision than their uncertainties indicate.}
\end{figure*}

It is demonstrated in Fig.~\ref{fig:powspecdist} that, as $N$ increases, the probability distribution of the power spectrum narrows, converging to the actual power spectrum. Comparing the $\hat{\tilde{C}}_{\ell,N}$ distributions of the two skymaps, the distribution from $\tilde{S}_B$ is narrower at higher $N$. The unbiased estimator of the variance $\var{\hat{\tilde{C}}^{\text{tr}}}$ of the power spectra $\tilde{C}_1^{\text{tr}}, \tilde{C}_2^{\text{tr}}, \dotsc, \tilde{C}_{N_{\text{tr}}}^{\text{tr}}$ observed after $N_{\text{tr}}$ trials is the usual result
\begin{equation}
  \label{eq:CtrVar}
  \wvar{\hat{\tilde{C}}^{\text{tr}}}=\frac{1}{N_{\text{tr}}-1}\sum_{p=1}^{N_\text{tr}}\left(\tilde{C}_p^{\text{tr}}-\hat{\tilde{C}}^{\text{tr}}\right)^2,
\end{equation}
where the estimated power spectrum from trials is
\begin{equation}
  \hat{\tilde{C}}^{\text{tr}}=\frac{1}{N_{\text{tr}}}\sum_{p=1}^{N_\text{tr}}\tilde{C}_p^{\text{tr}}.
\end{equation}
To determine whether a measured difference in the distribution width from the two skymaps is significant, we estimate the variance of our variance estimator to see how precisely $\var{\hat{\tilde{C}}^{\text{tr}}}$ was resolved at the present number of trials. It is a straightforward exercise to show that
\begin{align}
  \wvarb{\wvar{\hat{\tilde{C}}^{\text{tr}}}}=&\ \frac{1}{(N_{\text{tr}}-2)(N_{\text{tr}}-3)}\Bigg[\sum_{p=1}^{N_{\text{tr}}}\left(\tilde{C}_p^{\text{tr}}-\hat{\tilde{C}}^{\text{tr}}\right)^4\nonumber\\
  &-\left(N_{\text{tr}}-\frac{3}{N_{\text{tr}}}\right)\left(\wvar{\hat{\tilde{C}}^{\text{tr}}}\right)^2\Bigg].
\end{align}
Then the standard deviation and its uncertainty are estimated as
\begin{equation}
	\label{eq:CtrVarEst}
  \hat{\sigma}_{C^{\text{tr}}}=\left(\wvar{\hat{\tilde{C}}^{\text{tr}}}\right)^{1/2}\pm\left(\wvarb{\wvar{\hat{\tilde{C}}^{\text{tr}}}}\right)^{1/4}.
\end{equation}
These estimates are presented in Fig.~\ref{fig:powspecdist} for the distributions shown there. The standard deviation from the skymap $\tilde{S}_{N\!B}$ is denoted $\hat{\sigma}_{N\!B}$, and $\hat{\sigma}_B$ is the standard deviation of the estimates from $\tilde{S}_B$. The measured standard deviations are in excellent agreement with the expected values determined with \eqn{eq:Chatvar2}. Additionally, for $N=10000$, $\sigma_B$ is at 1$\sigma$ tension with $\hat{\sigma}_{N\!B}$, and $\sigma_{N\!B}$ is at 1$\sigma$ tension with $\hat{\sigma}_B$.

\begin{figure*}
  \subfloat[$N=10$]{\includegraphics[width=0.3\textwidth]{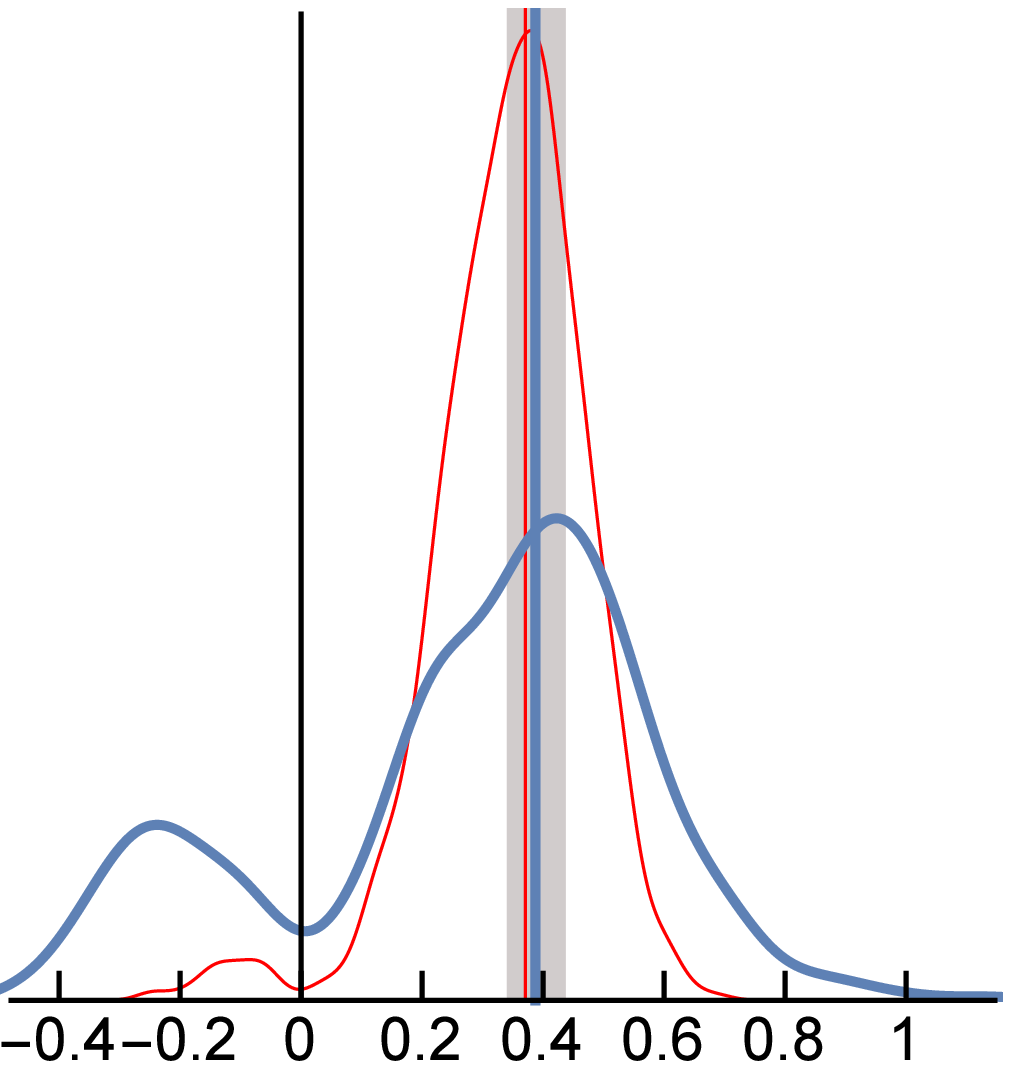}}\qquad
  \subfloat[$N=100$]{\includegraphics[width=0.3\textwidth]{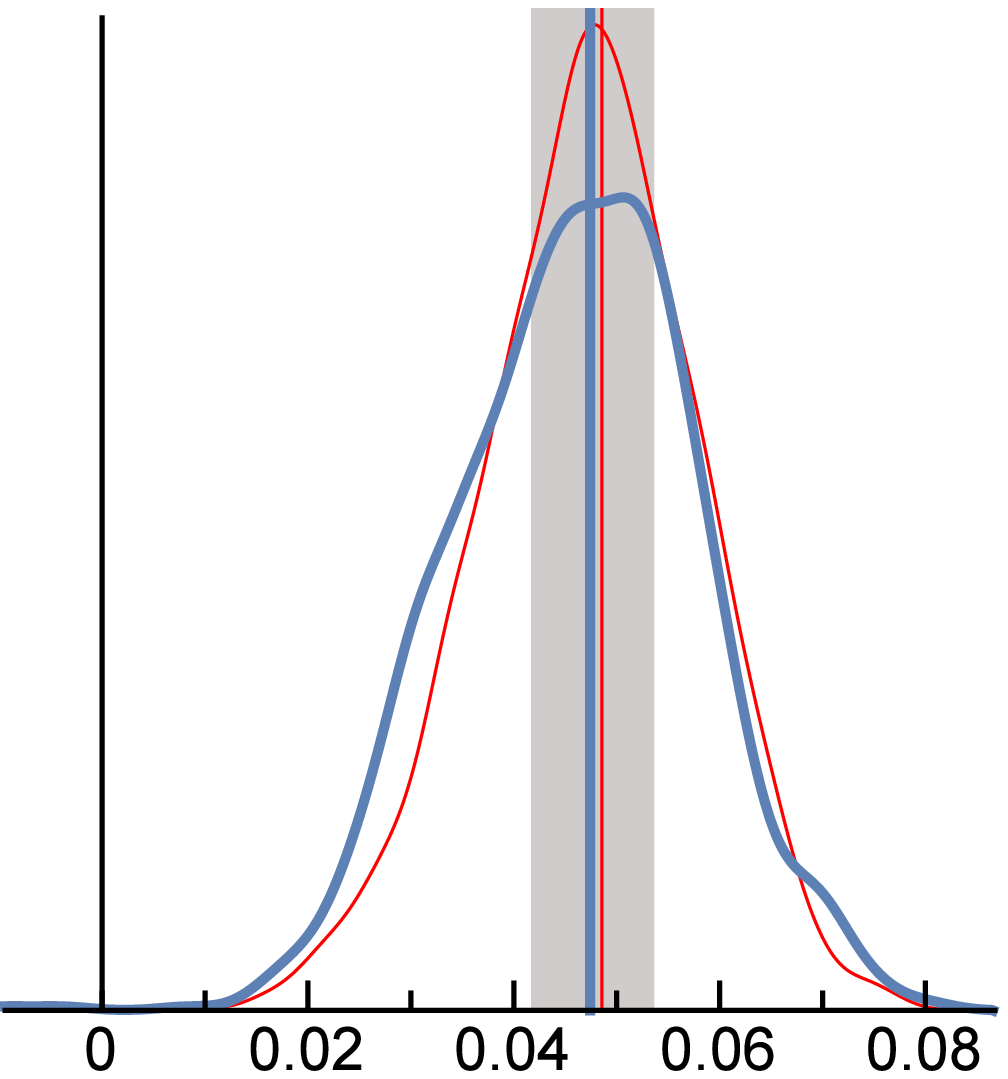}}\qquad
  \subfloat[$N=1000$]{\includegraphics[width=0.3\textwidth]{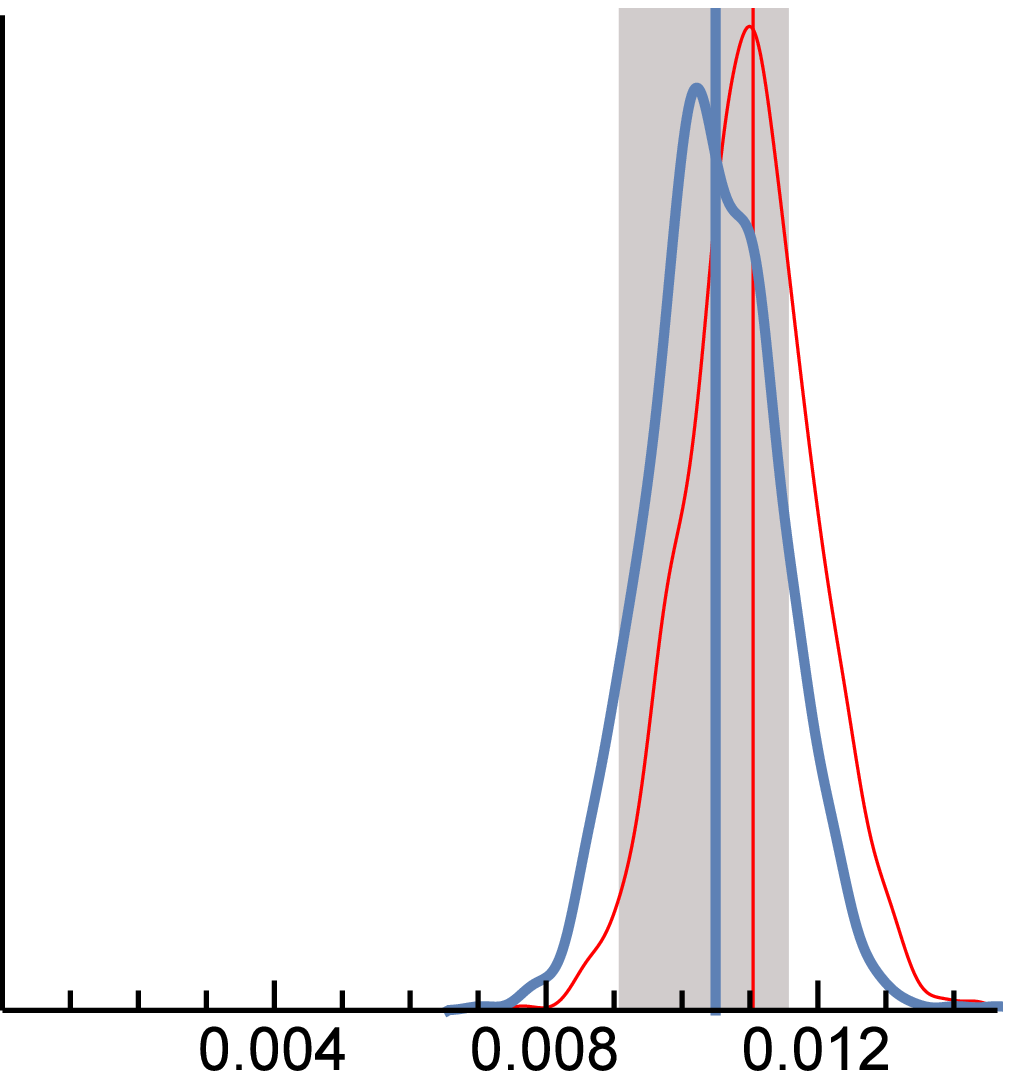}}
  \caption{\label{fig:varestnb}Uncertainty estimates of the power spectrum for the $\tilde{S}_{N\!B}$ skymap. Thick curve: distribution of the angular power spectrum's full uncertainty estimator $s\left|\wvar{\hat{\tilde{C}}_{12,N}}\right|^{1/2}$ with $s$ the sign of the variance estimate. Thick vertical line: expected full uncertainty determined from the known skymap. Thin curve: distribution of the $C_\ell$-only uncertainty estimator. Thin vertical line: actual value of the $C_\ell$-only uncertainty. Light rectangle: measured uncertainty from the simulated trials.}
\end{figure*}

\begin{figure*}
  \subfloat[$N=10$]{\includegraphics[width=0.3\textwidth]{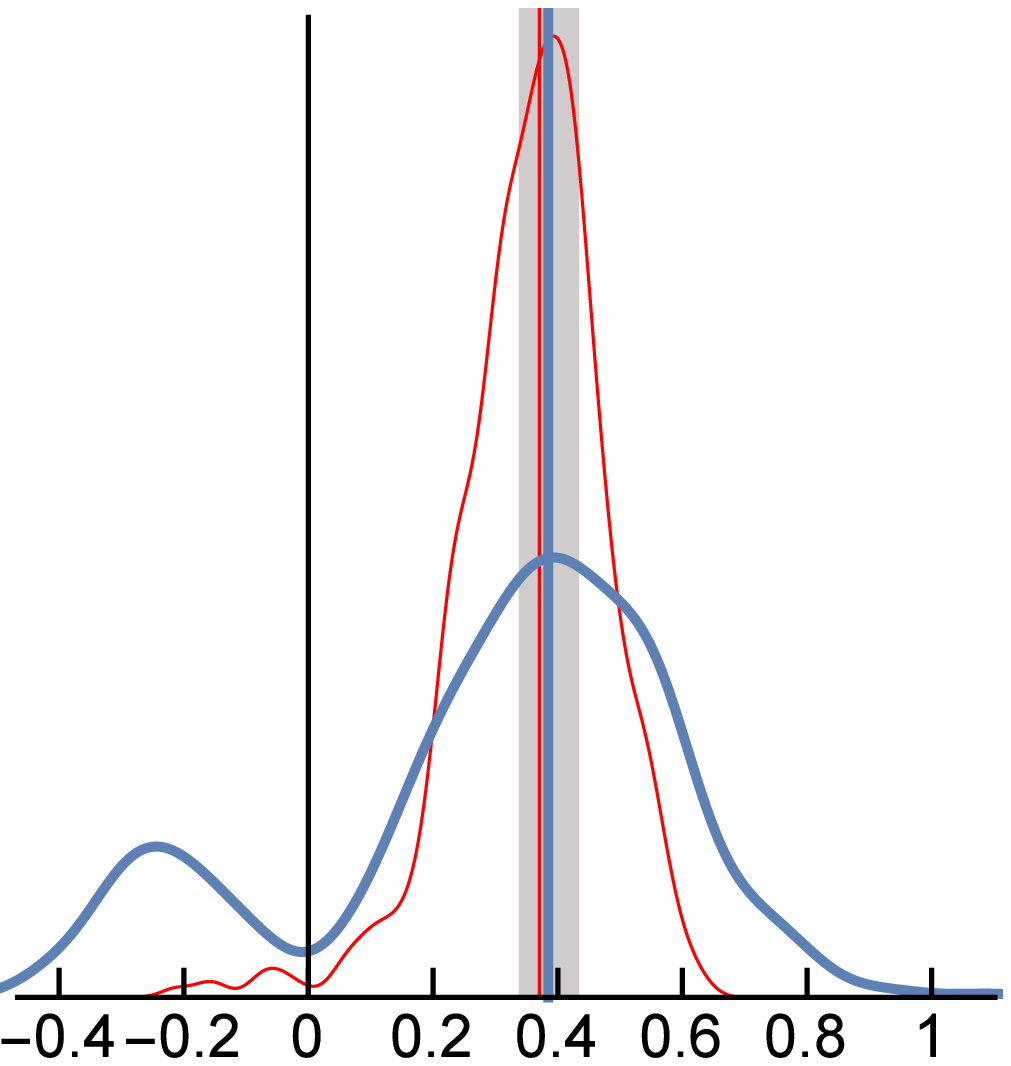}}\qquad
  \subfloat[$N=100$]{\includegraphics[width=0.3\textwidth]{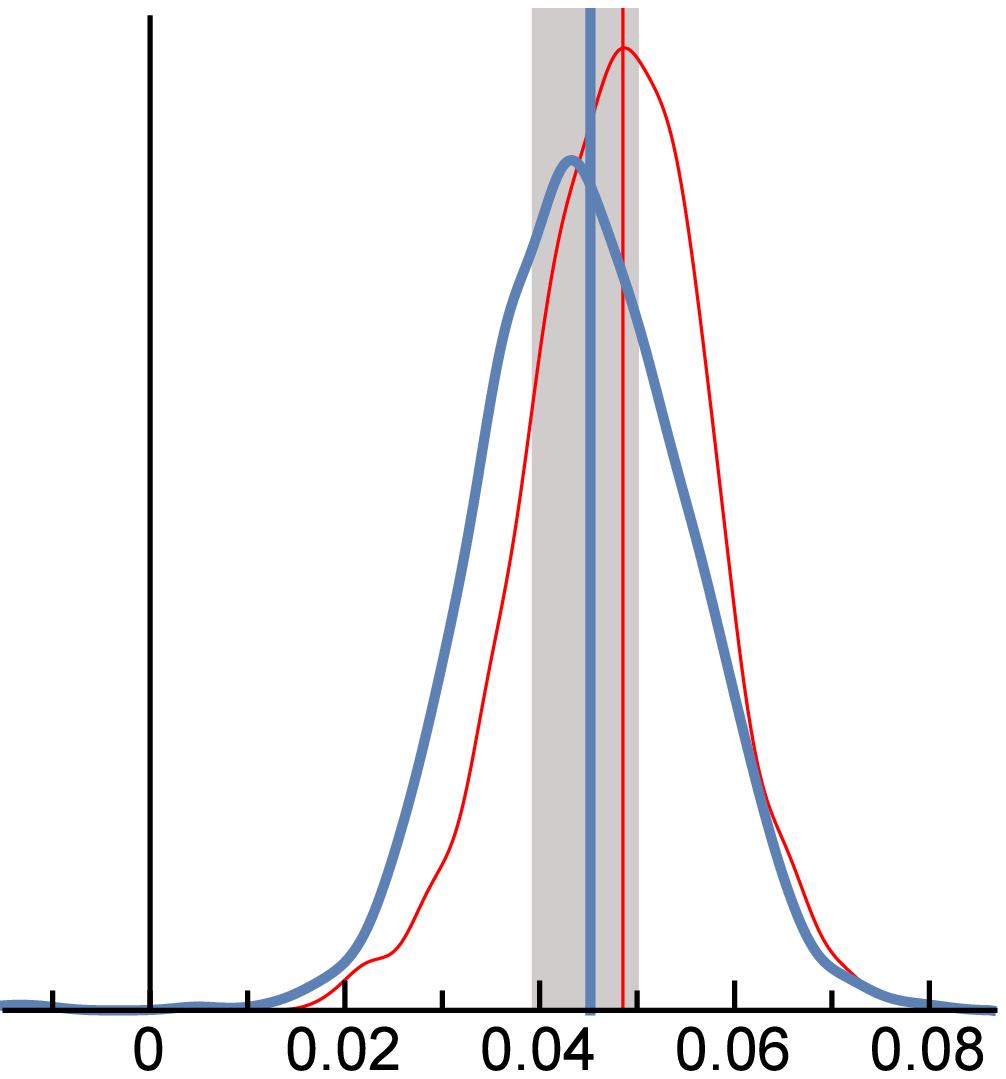}}\qquad
  \subfloat[$N=1000$]{\includegraphics[width=0.3\textwidth]{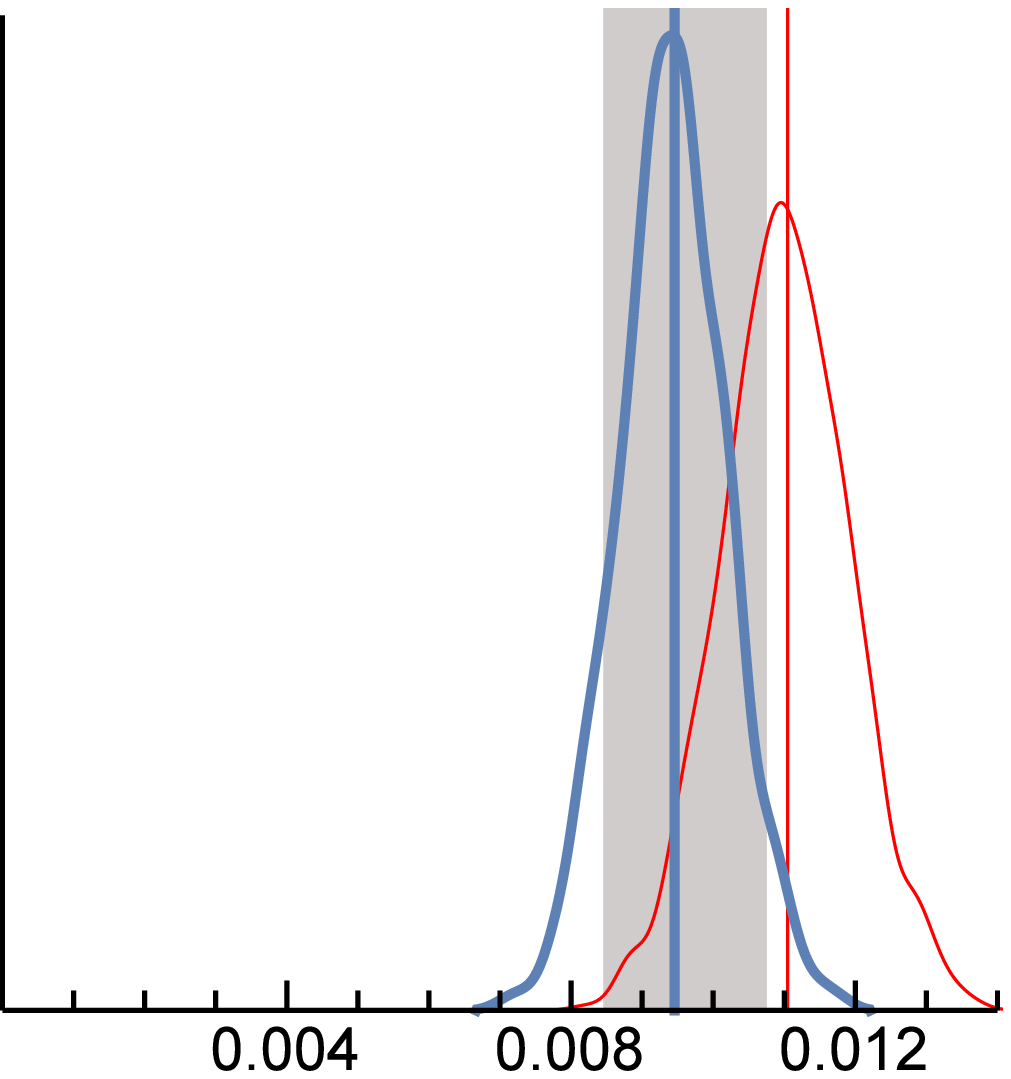}}
  \caption{\label{fig:varestb}Same as Fig.~\ref{fig:varestnb} for the $\tilde{S}_{B}$ skymap. At $N=1000$, the $C_\ell$-only variance is no longer consistent with the measured standard deviation of the distribution, making the secondary bispectrum and trispectrum effects statistically interesting.}
\end{figure*}

Having determined estimators for all the spectra up to $N=1000$, we can test the effectiveness of the full variance estimator \eqn{eq:varhatchat} by considering its statistical distribution, compared to the $C_\ell$-only estimator from substituting $\hat{\tilde{C}}_{\ell,N}$ for $\tilde{C}_\ell$ in \eqn{eq:Clonly}. The uncertainty of the power spectrum is to be given by the square root of the variance. Though the variance estimator is unbiased, at low event counts the distribution of the estimator is so wide as to allow negative estimates of the variance. For visualization purposes, we will plot the root of negative variances as negative (rather than imaginary) uncertainties. The results for the full estimator are shown in the thick curves of Figs.~\ref{fig:varestnb} and \ref{fig:varestb}. Shown in the thin curves are the $C_\ell$-only estimator. The actual values of these uncertainties in terms of the known sky distributions are shown by vertical lines. The shaded region is the 1$\sigma$ measurement of the uncertainty $\hat{\sigma}_{C^{\text{tr}}}$ from the variation in the measured power spectra of the simulated trials.

The $N=10$ plots reveal how unreliable the variance estimators are at low event counts. A significant fraction of trials produce negative variances, but the $C_\ell$-only variance is more reliable and more accurate, even though it is biased. At $N=100$, negative variances are very rare, and the width of the error's distribution is within an order of magnitude, making it suitable for use in error estimation. Both variance estimators are consistent with the simulation. At $N=1000$, the width of the error estimators is around 10\%. For the $\tilde{S}_{N\!B}$ skymap, both estimators agree with the simulation. Since there is no bispectrum, only the small trispectrum effect causes the small separation of the full error from the $C_\ell$-only error. In the $\tilde{S}_B$ skymap, the significant negative bispectrum causes the full error estimate to be reduced even further from the $C_\ell$-only estimate. The simulation exhibits a similar downward shift in its standard deviation $\hat{\sigma}_{C^{\text{tr}}}$, such that the $C_\ell$-only estimate is at 1$\sigma$ tension with the simulation. This supports the validity of the work in this paper.

These experiments suggest that the unbiased variance estimator may only be reliable at estimating the actual variance when a power spectrum signal has been resolved. It is recommended that early analyses use the shot noise variance \eqn{eq:shotvar} until $\tilde{C}_\ell=0$ is no longer consistent with that uncertainty estimate. At this point, the $C_\ell$-only uncertainty estimate will probably make the uncertainty large enough so that zero power is consistent. Once signal is again appearing, then at this stage the full unbiased uncertainty estimator should be reliable and most desirable. 

Additional simulations will have the ability to produce more significant evidence supporting the new theoretical results, and determine more precise protocols for their safe usage. 

\section{Discussion: Implications for Experiments}
\label{sec:exp}
We now discuss implications of the new results on experiments.

In order to understand the effect that these results will have on the \emph{Fermi-LAT} $\gamma$-ray power spectrum measurement, one needs to determine estimates of the composite power spectrum $\hat{\tilde{C}}_{\ell,N}^{(2)}$ and open bispectrum $\hat{\tilde{C}}_{\ell,N}^{(3)}$ of the events in each energy bin. If these are negligible, then the results are well described with a $C_\ell$-only variance if making a skymap measurement, or with the cosmic variance with noise estimate if carrying out a cosmic mean analysis for sources on linear cosmological distance scales (that are Gaussian-distributed). Given their observed small variation in $\hat{\tilde{C}}_{\ell,N}\sim10^{-5}$ sr, then one would expect $\hat{\tilde{C}}_{\ell,N}^{(2)}$ to be of a similar magnitude, and therefore negligible compared to $(2\ell+1)^{-1}$ of the shot noise term.

The \emph{Fermi-LAT}'s measurement of $\hat{C}_{\ell,N}$ in the highest energy bin from 10.4 GeV to 50.0 GeV was shown to have been a bit underestimated when compared to the power spectrum of point sources subsequently detected after the anisotropy analysis was completed \citep{Broderick2013a}. It is intriguing this tension could be explained if that energy bin had a significant, positive open bispectrum. However, before estimates of the higher-order uncertainty contributions can be made using the formulas derived in this paper, they must be modified to additionally take into account the masking of parts of the sky and the non-uniform exposure map.

If a significant open bispectrum is detected in the diffuse $\gamma$-ray background, there will be an unfortunate decrease in significance of the measured power spectrum, but this would be in exchange for the exciting detection of higher-order angular correlations that would be very constraining for source population models. Such an observation would indicate significant departure from the galaxy distribution on the scales of structure which is Gaussian-distributed, suggesting a source of gamma-rays from a more local distribution. However, if it is found that $\tilde{C}_\ell$ has been correctly measured, this immediately puts new bounds on the magnitude of the open bispectrum $\tilde{C}_\ell^{(3)}$.

Extragalactic sources correlated with large scale structure on linear cosmological distance scales would be expected to be Gaussian-distributed to within Poisson noise if sparse (such as active galactic nuclei), or Gaussian cosmic variance if dense (for instance, star-forming galaxies). The presence of a large bispectrum would be an indication of a population of sources that is either extragalactic at sub-Mpc scales (from galaxies in the local Universe), or Galactic. Example Galactic sources may be a population of unresolved pulsars \citep{Calore2014}, or annihilation in dark matter subhalos \citep{Campbell2013}. 

A way to probe the presence of an angular bispectrum affecting the power spectrum measurement is to analyze the time-sequence $\hat{\tilde{C}}_{\ell,N}$ as a function of $N$. This value of the estimated power spectrum wanders randomly around the actual skymap's power spectrum, with an amplitude that is indicative of the magnitude of the statistical uncertainty. With the collection of additional data, this amplitude diminishes in magnitude roughly as $N^{-1}$ and the estimator converges to the skymap's power spectrum. However, an early transition to a signal-dominated error that shrinks as $N^{-1/2}$, before the power spectrum has been resolved, would be evidence for the presence of a significant open bispectrum. The robustness of detecting the bispectrum using a single time-sequence of data can be tested with simulations in future work.

Now consider other experiments interested in measuring angular power spectra. For the angular distribution probes of cosmology, such as galaxy surveys, the amount of data is usually so high that the statistical uncertainty is dominated by cosmic variance and the new findings in this paper have no new effect. However, if galaxy distances can be made precise enough to allow accurate separation into such thin redshift slices that the number of galaxies per redshift slice is $\la10^6$, then cosmic variance of each redshift slice may no longer be the dominant statistical uncertainty and the new methods in this paper would be of use.

When counting objects such as galaxies, or blazars, etc., the role of the point spread function in \ssec{sec:clvar} is replaced by a selection function that specifies the completeness of observations. The `skymap' refers in this case to the complete map of the objects of interest.

For high-energy radiation events, the skymap is the complete apparent flux map of all radiation sources convolved by propagation effects. The propagation effects are most severe for charged cosmic rays which are deflected and hard-scattered within the medium. Since $\gamma$-rays and neutrinos interact less, the measurement of their high-energy angular power spectrum in different energy bins has the potential to identify the sources. One can use this information to separate distinct components of radiation from different sources \citep{Hensley2013}. Even hard spectral features too faint to be seen in the flux spectrum could still be observed in anisotropies \citep{Zhang2004,Campbell2013}. If the $\gamma$-rays or neutrinos have a particular angular power spectrum that matches the distribution of a potential source, this would be strong evidence for the origin of the high-energy radiation, motivating a cross-correlation analysis to be carried out between the radiation and that particular point source catalog \citep[e.g.,][]{Xia2011,Ando2014}.

A natural application of these methods is to \emph{IceCube} neutrino data. The astrophysical component in the energy range 30 TeV--2 PeV has a flux that produces roughly 10 events per year. If the source is stationary with a large fluctuation power spectrum $\sim 10^{-2}$ sr and small or negative open bispectrum, then the power spectrum could be begin to be resolved within 10 years of data. Another interesting analysis is determination of the power spectrum of the lower energy neutrinos from \emph{IceCube}; the presence of an anisotropic astrophysical component can be searched for among the highly numerous and isotropic atmospheric neutrinos.

Significant power spectra in the arrival direction of charged cosmic rays would indicate angular clustering that potentially contains information about the distribution of nearby sources and the structure of fields and material in the rays' path to us \citep{Sutherland2012,Keivani2015}. A new way to test for the presence of a significant dipole distribution is to determine if the $\ell=1$ component of the power spectrum is significant. If this amplitude is significant, the phase of the dipole can be ascertained from the values of the three $\ell=1$ spherical harmonic transforms.

\section{Conclusions}
\label{sec:conc}
To summarize, we have carried out an error analysis of the angular power spectrum of a finite set of points on a sphere. The statistical framework is modeled with only two postulates: a stationary `skymap' distribution of sources $S(\bn)$ and the assumption of independence of the detected events. For this analysis, the observation of the points is made with uniform efficiency and exposure over the entire sphere. 

We have presented an estimator of the skymap's power spectrum in \eqn{eq:Clestimator1} or \eqn{eq:ChatD} that is unbiased with respect to a fixed number of observed events $N$, and we determined the variance of the measured power spectrum as a function only of $N$ and $S(\bn)$ in \eqn{eq:varprev} or \eqn{eq:Chatvar2}. The power spectrum estimator is consistent with usual pseudo-spectrum techniques, and its statistical variance is found to agree at leading order with previous analytic estimates.

New, previously unquantified effects of the power spectrum variance have been discovered that are generated by three effects: the power spectrum at neighboring multipoles (the composite power spectrum $C_\ell^{(2)}$), the bispectrum of the skymap (in a form called the open bispectrum $C_\ell^{(3)}$), and the skymap's trispectrum (given by the disjoint trispectrum $C_\ell^{(4)}$). A study of these new higher-order spectra is presented in Appendix~\ref{ap:sphten}. They are the only possible `two-correlator' spectra, and are also the only spherical tensors that can be derived from rank 2 spherical transforms of the skymap.

These new higher order effects will be negligible under certain conditions that have been precisely identified. In this work, the power spectrum variance is called $C_\ell$-only at those values of $N$ where these new higher-order effects are negligible, corresponding to error estimates used in previous works. The new terms, with the open bispectrum in particular, have the potential to add orders of magnitude to the variance. Thus, the neglect of these new terms may result in spurious detections of the power spectrum. A negative open bispectrum causes the signal term of the variance to be smaller than $C_\ell$-only.

The behavior of the power spectrum variance is classified by two regimes in $N$: in the small-$N$ regime, the variance is dominated by a shot term that goes as $N^{-2}$ (for $N\gg1$); when $N$ is large, a signal term dominates and the variance goes as $N^{-1}$. The composite power spectrum minus disjoint trispectrum is a new contribution to the shot term, added to the shot noise variance. The signal term, which in previous work was the variance due to the power spectrum itself, is found to have additional terms from the open bispectrum minus disjoint trispectrum.

An unbiased estimator of the power spectrum's variance is identified in \eqn{eq:varhatchat} or \eqn{eq:varhatchat2}, calculated from the data. Remarkably, this estimate is made without any prior assumptions about properties of the skymap spatial distribution. Efficient, highly parallelizable codes have been shown to be very efficient at calculating this variance from position data. This makes these exact methods viable for use (at least for the full sky, uniform exposure limit) to quite large counts, relaxing the necessity of binning the data into pixels on the sphere, which either introduces the need to correct for `pixel' features in the power spectrum, or uses a large number of tiny pixels that requires more computation to determine the spectrum than the exact method.

These estimators are very important, providing a means for experiments to test for the presence of higher-order contributions to the power spectrum uncertainty. The presence of a large open bispectrum in the diffuse $\gamma$-ray background measured by \emph{Fermi-LAT} would decrease the significance of the measured power spectrum, but would provide exciting new information about the distribution of the source populations. On the other hand, verification of the existing power spectrum measurements would provide stringent bounds on the magnitude of non-Gaussianities in the distribution. Those $\gamma$-ray distributions consistent with being Gaussian to within cosmic variance are consistent with being produced by extragalactic sources over linear cosmological distances. The presence of non-Gaussianities would suggest the presence of a population of more local sources, either near-extragalactic, or Galactic sources.

We showed how these results can be applied to a statistical model of the spatial distribution of the skymap by considering an example analysis of a Gauss\-i\-an-distributed skymap. In this context, the prediction of the power spectrum is subject to cosmic variance from statistical variations between different skymap realizations of the Gaussian process. The open bispectrum vanishes for Gaussian distributions. Otherwise, the previous analytic estimates for the uncertainty of the power spectrum in these measurements was reproduced to first order, with additional higher order contributions from the composite power spectrum and disjoint trispectrum as expressed in \eqn{eq:clvargauss}. These new contributions become important at large $\ell$ since these new terms are not suppressed by $(2\ell+1)^{-1}$, as is the case for the standard terms. The usual uncertainty is restored when enough data is collected for the statistical uncertainty to be dominated by the cosmic variance. Thus, power spectrum measurements of large data sets that are cosmic variance limited remain unaffected by the new findings.

The effect of the instrument point spread function (PSF) on the power spectrum measurement is found to simply be that the measured spectra are of the skymap convolved with the PSF. However, further effects may be induced by asymmetric PSFs on the inference of the cosmic mean power spectrum of a modeled ensemble of Universes.

Simulations were carried out of two skymaps of equivalent power spectrum, but one without open bispectrum and the other with significant open bispectrum. The standard deviation of power spectrum estimates using the simulated data were shown to prefer the variance results of this paper with the bispectrum effect over the $C_\ell$-only variance.

The new formalism in this paper is important for smaller datasets, such as for events of incident high-energy radiation, where the power spectrum provides information about the distribution of sources and propagation effects. After around $10^6$ detected events, it is likely more efficient to use pixeled-sky estimates for which the uncertainty results presented here can be adapted.

\section*{Acknowledgments}
For informative discussions, suggestions, or comments on an early version of this manuscript, I thank Shin'ichiro Ando, Aur\'elien Benoit-L\'evy, Jonathan Blazek, Alessandro Cuoco, Mattia Fornasa, Nicolao Fornengo, Chris Gordon, Eric Huff, Kevin Huffenberger, Boris Leistedt, Peter Melchior, Ashley Ross, Bj\"orn Malte Sch\"afer, Hee-Jong Seo, Paul Sutter, David Weinberg, and especially thank John Beacom, Chris Hirata, and Eiichiro Komatsu for their additional time and particularly insightful feedback. The program to generate the sphere distribution plots was inspired by `Nodal Domains of Spherical Harmonics' from the Wolfram Demonstrations Project, contributed by Michael Trott. This research is partly funded by NSF Grant PHY-1101216 awarded to John Beacom.

\appendix

\section{Rotationally Invariant Spherical Tensors}
\label{ap:sphten}
This appendix introduces the notations and formal mathematical structures that appear in the results of this paper. For further background on spherical tensors, their products, and their rotations, we recommend \citet{Edmonds1957} and \citet{Hu2001}.

\hspace{-\parindent}\textbf{Spherical Tensors:} Let $\bn$ denote any unit position vector in 3 dimensional Euclidean space. In the context of this paper, spherical tensors are those generated by the spherical harmonic functions and any square-integrable scalar function $S(\bn)$ defined on the unit sphere. This is described below. Thus, these tensors inherit their coordinate transformation properties and symmetries from the spherical harmonics $Y_\ell^{\,m}(\bn)$.

Common conventions for the spherical harmonics are taken. They are normalized according to the orthonormality relation
\begin{equation}
  \int\der\bn[Y_{\ell_1}^{\ m_1}(\bn)]^*Y_{\ell_2}^{\ m_2}(\bn)=\delta_{\ell_1 \ell_2}\,\delta_{m_1m_2},
\end{equation}
where * denotes complex conjugation, and $\delta_{x_1x_2}$ is the Kronecker delta. It is important in this application to note that with this standard normalization, the spherical harmonics have physical units of inverse angle, which is equivalent to inverse length on the sphere.

The index $\ell$ is often called the multipole. It specifies the angular diameter $\psi\sim\pi/\ell$ of positive or negative regions of the spherical harmonics. Since the sphere is topologically compact, continuity of the spherical harmonics requires the spectrum of multipoles be discrete, and they are defined to take values $\ell=0,1,2,\ldots.$

The index $m$ is sometimes called the azimuthal index. An azimuthal index of a spherical tensor must always be associated with one multipole index. However, note that a multipole index can generally have any number of azimuthal indices associated with it. For a given value of its associated multipole $\ell$, the azimuthal index spans the integer values $-\ell\leq m\leq\ell$, delineating the $2\ell+1$ orthogonal harmonic modes with angular scale $\pi/\ell$.

To generate spherical tensors that are invariant under rotations, it is convenient to introduce tensors with covariant azimuthal indices $S_{\ell m}$ and contravariant azimuthal indices $S_\ell^{\ m}$, related by
\begin{equation}
  S_{\ell m}\equiv\sum_{m'=-\ell}^{\ell}g^{(\ell)}_{m,m'}S_\ell^{\ m'}=(-1)^mS_\ell^{\ -m},
\end{equation}
where the $\ell$-dimensional spherical metric tensor has components
\begin{equation}
  g^{(\ell)}_{m,m'}\equiv(-1)^m\delta_{m,-m'},\quad |m|\leq\ell.
\end{equation}

The rank of a spherical tensor is the number of multipole indices it contains, and the number of azimuthal indices for each multipole index. The full rank can be specified by a list of azimuthal ranks. For example, a rank (2,1,0) spherical tensor $S_{\ell_1,m_{11},m_{12},\ell_2,m_2,\ell_3}$ has 3 multipole indices of rank 2, 1, and 0, respectively. The rotation-invariant spherical tensors have only rank 0 multipole indices.

New spherical tensors can also be constructed by contracting a covariant and contravariant azimuthal index with the same multipole. For example, a rank 0 tensor $T_\ell$ can be constructed from a rank 2 spherical tensor $S_{\ell m_1}^{\ \ \ \ m_2}$ by contraction of the azimuthal indices,
\begin{equation}
  T_\ell=\frac{1}{2\ell+1}\sum_{m=-\ell}^\ell S_{\ell m}^{\ \ \ m}.
\end{equation}

\hspace{-\parindent}\textbf{Spherical Transforms:} Let $S(\bn)$ be a non-negative, real-valued, integrable function on the sphere. Then we may normalize $S$ without loss of generality such that
\begin{equation}
  \label{eq:snorm}
  \int\frac{\der\bn}{4\pi}\tilde{S}(\bn)=1,
\end{equation}
where
\begin{equation}
  \tilde{S}(\bn)\equiv \frac{S(\bn)}{\int\frac{\der\bn'}{4\pi}S(\bn')}.
\end{equation}
Spherical tensors of any combination of non-zero rank multipoles can be generated from $\tilde{S}(\bn)-1$ by spherical harmonic transforms, with the $Y_\ell^{\,m}$ generating contravariant indices, and the complex conjugates $(-1)^mY_\ell^{\,-m}=(Y_\ell^{\,m})^*\equiv Y^*_{\ell m}$ generating covariant indices.

The central spherical harmonic transform of $\tilde{S}(\bn)$ is a rank 1 spherical tensor.
\begin{align}
  \label{eq:transform1}
  \tilde{S}_{\ell m}&=\int\der\bn\ Y^*_{\ell m}(\bn)\left[\tilde{S}(\bn)-1\right]\\
  \tilde{S}_\ell^{\ m}&=\int\der\bn\ Y_\ell^{\,m}(\bn)\left[\tilde{S}(\bn)-1\right]
\end{align}
These tensors have units of angle. Since $\int\der\bn\,Y_\ell^{\,m}(\bn)=\delta_{\ell,0}\sqrt{4\pi}$, then taking the central transform only adjusts the $\ell=0$ component as

\begin{equation}
  \tilde{S}_{\ell m}=\int\der\bn\ Y^*_{\ell m}(\bn)\tilde{S}(\bn)-\delta_{\ell,0}\sqrt{4\pi}.
\end{equation}

Similarly, there are four different rank 2 spherical transforms of $S$, which in this case are unitless.
\begin{align}
  \tilde{S}_{\ell m_1 m_2}&=\int\der\bn\ Y^*_{\ell m_1}(\bn)Y^*_{\ell m_2}(\bn)\left[\tilde{S}(\bn)-1\right]\\
  \tilde{S}_{\ell m_1}^{\ \ \ \ m_2}&=\int\der\bn\ Y^*_{\ell m_1}(\bn)Y_\ell^{\,m_2}(\bn)\left[\tilde{S}(\bn)-1\right]\\
  \tilde{S}_{\ell\ \ \ \,m_2}^{\ m_1}&=\int\der\bn\ Y_\ell^{\,m_1}(\bn)Y^*_{\ell m_2}(\bn)\left[\tilde{S}(\bn)-1\right]\\
  \tilde{S}_\ell^{\ m_1 m_2}&=\int\der\bn\ Y_\ell^{\,m_1}(\bn)Y_\ell^{\,m_2}(\bn)\left[\tilde{S}(\bn)-1\right]\label{eq:transform2}
\end{align}
Taking the central moment adjusted all of the multipoles on the diagonal of the azimuth indices,
\begin{equation}
  \tilde{S}_{\ell m_1}^{\ \ \ \ m_2}=\int\der\bn\ Y^*_{\ell m_1}(\bn)Y_\ell^{\,m_2}(\bn)\tilde{S}(\bn)-\delta_{m_1}^{\ \ \ m_2}.
\end{equation}

It is worth noting that the rank~2 transforms can be expressed as sums of the rank~1 transforms by applying the result for a product of spherical harmonics,
\begin{align}
  &Y_{\ell_1}^{\,m_1}(\bn)Y_{\ell_2}^{\,m_2}(\bn)=\sum_{\ell=|\ell_2-\ell_1|}^{\ell_2+\ell_1}\sum_{m=-\ell}^{\ell}Y^*_{\ell m}(\bn)\\
  &\sqrt{\frac{(2\ell_1+1)(2\ell_2+1)(2\ell+1)}{4\pi}}\wignerthreej{\ell_1}{\ell_2}{\ell}{0}{0}{0}\!\!\wignerthreej{\ell_1}{\ell_2}{\ell}{m_1}{m_2}{m},\nonumber
\end{align}
where $\wignerthreej{\ell_1}{\ell_2}{\ell_3}{m_1}{m_2}{m_3}$ is the Wigner 3-j symbol. Then \eqn{eq:transform2} gives
\begin{align}
  \tilde{S}_\ell^{\ m_1m_2}=\ &(2\ell+1)\sum_{\ell'=0}^{2\ell}\sum_{m'=-\ell'}^{\ell'}\sqrt{\frac{2\ell'+1}{4\pi}}\nonumber\\
  &\wignerthreej{\ell}{\ell}{\ell'}{0}{0}{0}\!\!\wignerthreej{\ell}{\ell}{\ell'}{m_1}{m_2}{m'}\tilde{S}_{\ell'm'}.\label{eq:transform1totransform2}
\end{align}
This is likely a general method to generate a rank~2 contravariant spherical tensor from any rank~1 covariant tensor.

From the spherical transforms, we can generate the low-order rotationally invariant spherical tensors associated with $\tilde{S}(\bn)$. Note that the contraction of the rank 2 transform is trivial with
\begin{equation}
  \frac{1}{2\ell+1}\sum_m\tilde{S}_{\ell m}^{\ \ \,m}=0.
\end{equation}

\hspace{-\parindent}\textbf{Angular Power Spectrum:} In the special case where $\tilde{S}(\bn)$ is normalized as in \eqn{eq:snorm}, its angular power spectrum is sometimes called the \emph{fluctuation} or \emph{dimensionless} angular power spectrum, and its symbols are marked with a $\tilde{\ \ }$ to differentiate from other normalizations.

The lowest order rank~0 spherical tensor is the angular power spectrum, generated by contracting two rank~1 transforms with equivalent multipoles
\begin{equation}
  \tilde{C}_{\ell}=\frac{1}{2\ell+1}\sum_m\tilde{S}_{\ell m}\tilde{S}_\ell^{\ m}=\frac{1}{2\ell+1}\sum_m\left|\tilde{S}_{\ell m}\right|^2.
\end{equation}
This has units of solid angle. Consequently, $\tilde{C}_\ell$ will always be normalized by a factor of $4\pi$ in expressions, which can be traced to the choice of normalization condition for the spherical harmonic functions. It is instructive to substitute \eqn{eq:transform1} and apply the spherical harmonic addition theorem
\begin{equation}
  \label{eq:harmadd}
  \frac{1}{2\ell+1}\sum_mY^*_{\ell m}(\bn_1)Y_\ell^{\,m}(\bn_2)=\frac{1}{4\pi}P_\ell(\bn_1\cdot\bn_2),
\end{equation}
in terms of Legendre polynomials $P_\ell(x)$, to find
\begin{align}
  \label{eq:clfrompl}
  \tilde{C}_\ell&=4\pi\int\frac{\der\bn_1}{4\pi}\frac{\der\bn_2}{4\pi}[\tilde{S}(\bn_1)-1]P_\ell(\bn_1\cdot\bn_2)[\tilde{S}(\bn_2)-1]\\
  &=4\pi\left[\int\frac{\der\bn_1}{4\pi}\frac{\der\bn_2}{4\pi}\tilde{S}(\bn_1)P_\ell(\bn_1\cdot\bn_2)\tilde{S}(\bn_2)-\delta_{\ell,0}\right].
\end{align}
In this form, we can make the field-theoretic interpretation of the power spectrum as two field configurations $\tilde{S}(\bn_1)-1$ and $\tilde{S}(\bn_2)-1$ connected by a propagator, or in this application called a correlator, $P_\ell(\bn_1\cdot\bn_2)$. Taking the central transforms only shifts the monopole to $\tilde{C}_0=0$.

One interesting consequence of \eqn{eq:clfrompl} is found using the fact that
\begin{align}
  \sum_\ell(2\ell+1)P_\ell(\bn_i\cdot\bn_j)&=4\pi\sum_\ell\sum_mY^*_{\ell m}(\bn_i)Y_\ell^{\,m}(\bn_j)\nonumber\\
  &=4\pi\,\delta(\bn_i-\bn_j)\label{eq:Plclosure}
\end{align}
on the sphere to find
\begin{equation}
  \sum_{\ell=0}^{\infty}(2\ell+1)\tilde{C}_\ell=\int\!\der\bn\,[\tilde{S}(\bn)-1]^2.
\end{equation}
The left-hand sum is finite if and only if $\tilde{C}_\ell$ asymptotically vanishes faster than $\ell^{-2}$. This must be the case if $\tilde{S}(\bn)-1$ is square-integrable on the sphere. However, for a truly discrete source with $\tilde{S}(\bn)$ a sum of $\delta$-functions, then $\tilde{S}(\bn)-1$ is not square-integrable and the $\ell$-sum must diverge.

\begin{figure}
  \subfloat[$\tilde{C}_\ell$]{\quad\includegraphics[width=0.06\textwidth]{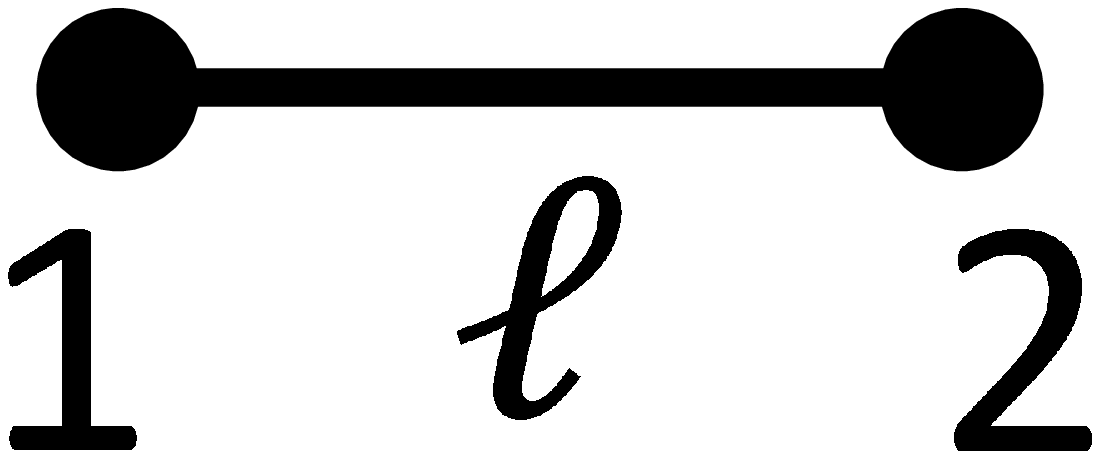}\quad}
  \subfloat[$\tilde{B}_{\ell_1\ell_2\ell_3}$]{\quad\includegraphics[width=0.06\textwidth]{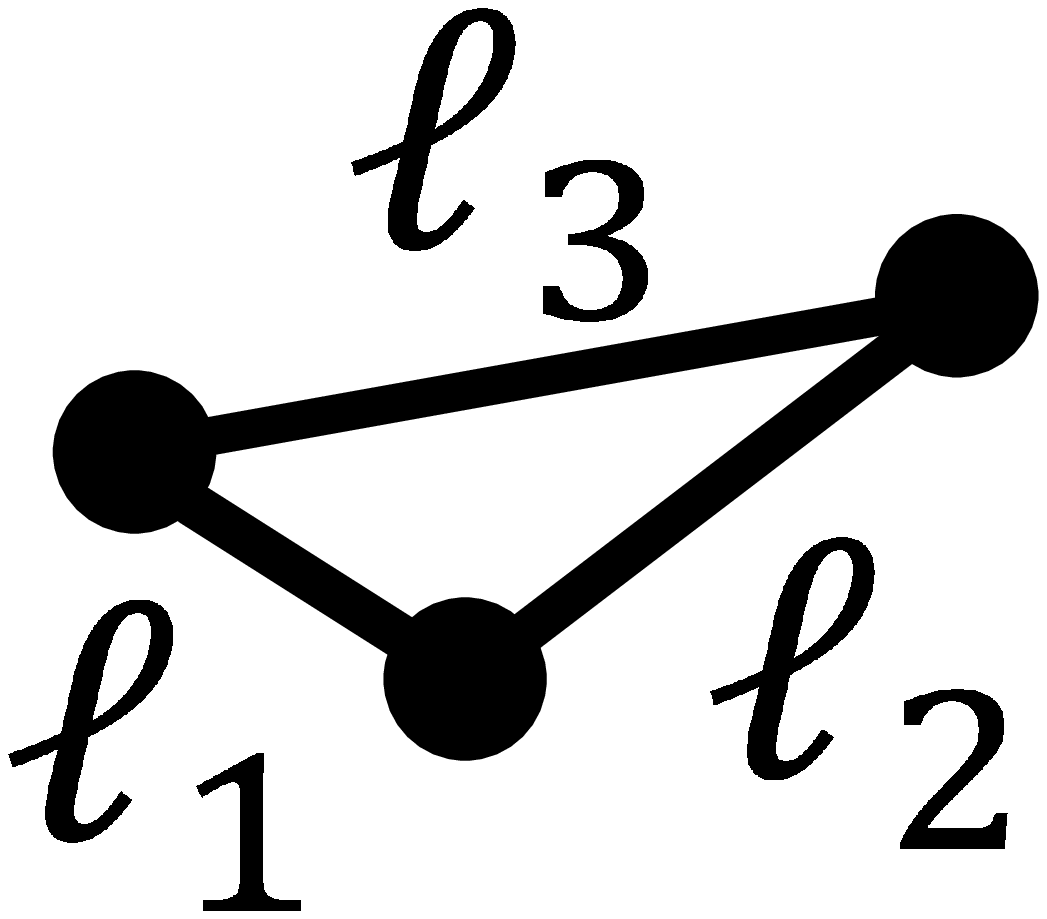}\quad}
  \subfloat[$\tilde{C}_\ell^{(2)}$]{\quad\includegraphics[width=0.06\textwidth]{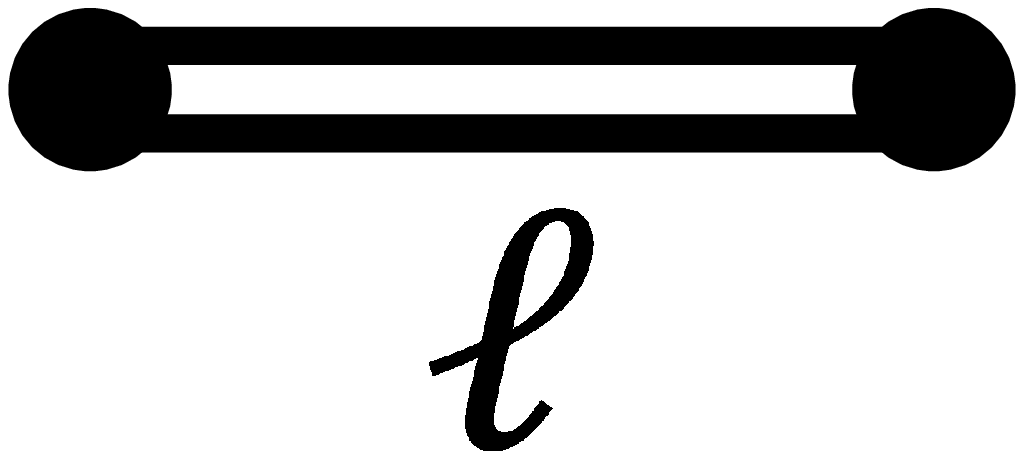}\quad}
  \subfloat[$\tilde{C}_\ell^{(3)}$]{\quad\includegraphics[width=0.06\textwidth]{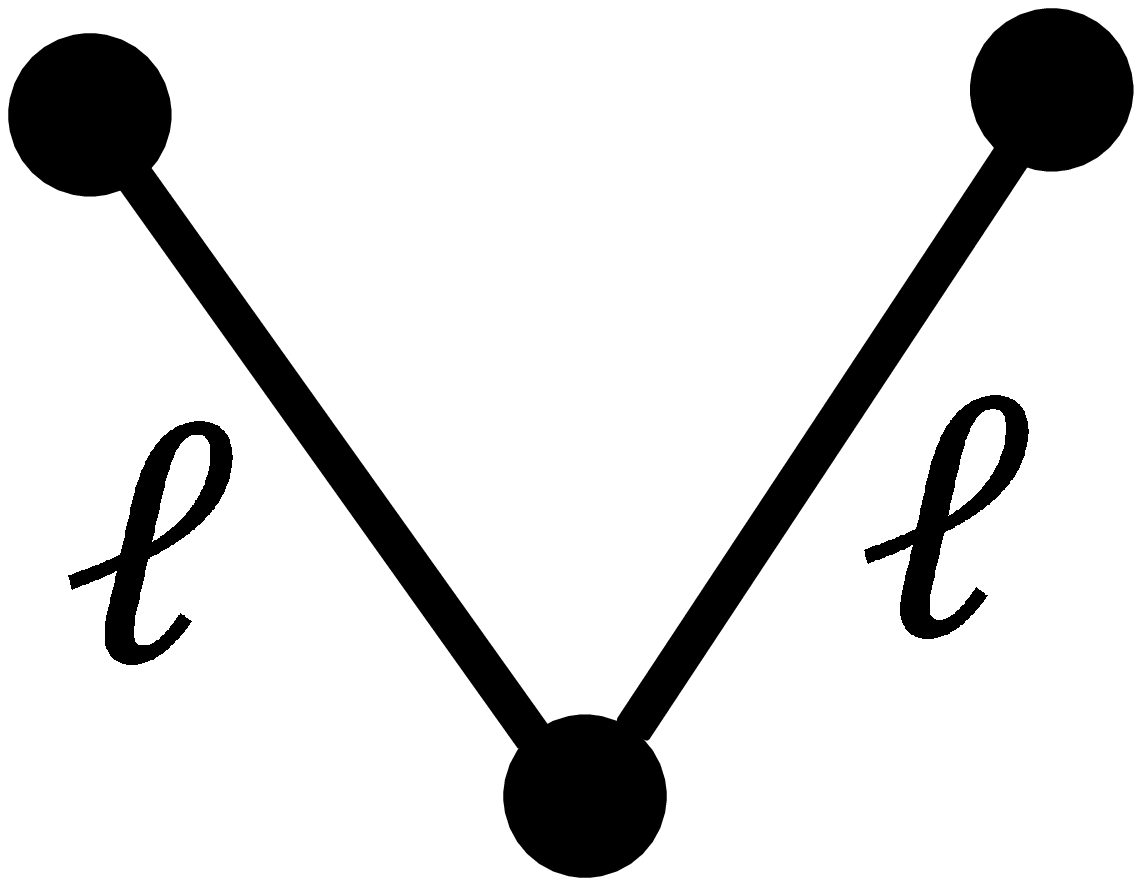}\quad}
  \subfloat[$\tilde{C}_\ell^{(4)}$]{\quad\includegraphics[width=0.06\textwidth]{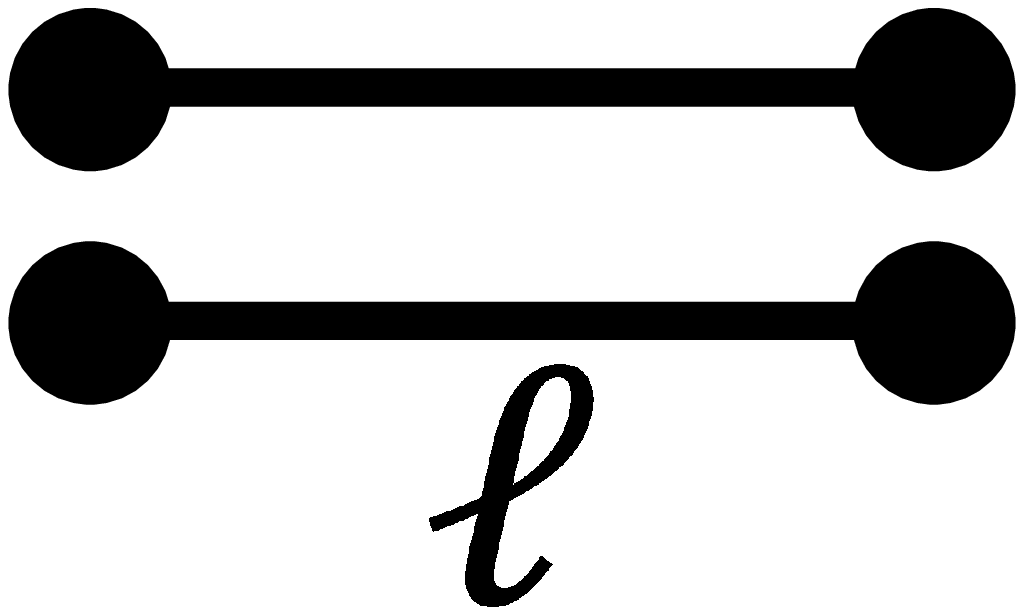}}
  \caption{\label{fig:diaspec}Diagrammatic representations of rotation-in\-var\-i\-ant spherical tensors.}
\end{figure}

Rank~0 tensors that are higher order than $\tilde{C}_\ell$ contribute to the statistical variance of $\tilde{C}_\ell$ estimates made from a finite random sampling of $\tilde{S}(\bn)$. We will see that these second order tensors are all possible ways to connect two field configurations with two correlators, as shown in Fig.~\ref{fig:diaspec}. They also represent all possible ways to form rank~0 tensors from two contractions of rank 1 and rank 2 spherical harmonic transforms.

\hspace{-\parindent}\textbf{Composite Angular Power Spectrum:} There are multiple ways to contract two rank~2 spherical transforms, all giving the same result, denoted here as
\begin{align}
  \tilde{C}_\ell^{(2)}&=\frac{1}{(2\ell+1)^2}\sum_{m_1,m_2}\tilde{S}_\ell^{\ m_1m_2}\tilde{S}_{\ell m_1m_2}\nonumber\\
  &=\frac{1}{(2\ell+1)^2}\sum_{m_1,m_2}\tilde{S}_{\ell\ \ \ \,m_2}^{\ m_1}\tilde{S}_{\ell m_1}^{\ \ \ \ m_2}\nonumber\\
  &=\frac{1}{(2\ell+1)^2}\sum_{m_1,m_2}\tilde{S}_{\ell\ \ \ \,m_2}^{\ m_1}\tilde{S}_{\ell\ \ \ \,m_1}^{\ m_2}\nonumber\\
  &=\frac{1}{(2\ell+1)^2}\sum_{m_1,m_2}\left|\tilde{S}_{\ell m_1m_2}\right|^2.
\end{align}
This tensor is unitless. Applying the harmonic addition theorem gives
\begin{align}
  \tilde{C}_\ell^{(2)}&=\int\frac{\der\bn_1}{4\pi}\frac{\der\bn_2}{4\pi}[\tilde{S}(\bn_1)-1]P_\ell^2(\bn_1\cdot\bn_2)[\tilde{S}(\bn_2)-1]\\
  &=\int\frac{\der\bn_1}{4\pi}\frac{\der\bn_2}{4\pi}\tilde{S}(\bn_1)P_\ell^2(\bn_1\cdot\bn_2)\tilde{S}(\bn_2)-\frac{1}{2\ell+1}.
\end{align}
This shows that the composite spectrum can be interpreted as two field configurations connected by two corellators, the latter accounting for its moniker of being composite. The second line shows that taking the central transform removes a term of $(2\ell+1)^{-1}$ from each component. This term is, in fact, what is responsible for the shot noise contribution to the variance of angular power spectrum measurements. Writing central transforms removes this from the composite power spectrum, so that the shot noise appears explicitly as its own term of the variance.

Using \eqn{eq:transform1totransform2} and the orthogonality of the Wigner 3-j symbols, 
\begin{align}
  \sum_{m_1=-\ell}^\ell&\sum_{m_2=-\ell}^\ell\wignerthreej{\ell}{\ell}{\ell'_1}{m_1}{m_2}{m'_1}\wignerthreej{\ell}{\ell}{\ell'_2}{m_1}{m_2}{m'_2}\nonumber\\
  &=\frac{1}{2\ell'_1+1}\delta_{\ell'_1\ell'_2}\delta_{m'_1m'_2}
\end{align}
we find
\begin{equation}
  \label{eq:cl2fromcl}
  \tilde{C}_\ell^{(2)}=\sum_{\ell'=0}^{2\ell}\frac{2\ell'+1}{4\pi}\wignerthreej{\ell}{\ell}{\ell'}{0}{0}{0}^{\!\!2}\tilde{C}_{\ell'},
\end{equation}
that the composite angular power spectrum is a weighted sum of neighboring angular power spectrum components.

For example, a white noise power spectrum that is multipole independent with amplitude $\tilde{C}_\ell=4\pi\mathscr{N}(1-\delta_{\ell,0})$ has composite power spectrum
\begin{equation}
  \tilde{C}_\ell^{(2)}=\frac{2\ell}{2\ell+1}\frac{\tilde{C}_\ell}{4\pi}\qquad\text{(white noise)}
\end{equation}
A Sachs-Wolfe spectrum from a scale-invariant (Harrison-Zel'dovich) primordial 3-D power spectrum, with $\tilde{C}_\ell\propto[\ell(\ell+1)]^{-1}$ (see, e.g., \citet{Durrer2008}), has
\begin{equation}
  \label{eq:swcl2}
  \tilde{C}_\ell^{(2)}=\frac{2\ell(\ell+1)}{2\ell+1}\kappa_\ell\frac{\tilde{C}_\ell}{4\pi}\qquad\text{(scale-invariant Sachs-Wolfe)}
\end{equation}
where
\begin{equation}
  \kappa_\ell\equiv\sum_{\ell'=1}^\ell\frac{1}{(2\ell'-1)(2\ell')(2\ell'+1)}
\end{equation}
which is written conveniently in terms of the harmonic numbers $H_n=\sum_{k=1}^nk^{-1}$ as
\begin{equation}
  \kappa_\ell=H_{2\ell}-H_{\ell}-\frac{\ell}{2\ell+1}.
\end{equation}

\hspace{-\parindent}\textbf{Open Angular Bispectrum:} Another tensor that is rotation-invariant is found by contracting two rank~1 transforms with one rank~2 transform,
\begin{equation}
  \label{eq:cl3}
  \tilde{C}_\ell^{(3)}=\frac{1}{(2\ell+1)^2}\sum_{m_1,m_2}\tilde{S}_\ell^{\ m_1}\tilde{S}_\ell^{\ m_2}\tilde{S}_{\ell m_1m_2},
\end{equation}
producing another tensor with units of solid angle. The addition theorem reveals
\begin{align}
  \tilde{C}_\ell^{(3)}=\ &4\pi\int\frac{\der\bn_1}{4\pi}\frac{\der\bn_2}{4\pi}\frac{\der\bn_3}{4\pi}[\tilde{S}(\bn_1)-1]P_\ell(\bn_1\cdot\bn_2)\nonumber\\
  &[\tilde{S}(\bn_2)-1]P_\ell(\bn_2\cdot\bn_3)[\tilde{S}(\bn_3)-1]\\
  =\ &4\pi\left[\int\frac{\der\bn_1}{4\pi}\frac{\der\bn_2}{4\pi}\frac{\der\bn_3}{4\pi}\tilde{S}(\bn_1)P_\ell(\bn_1\cdot\bn_2)\tilde{S}(\bn_2)\right.\nonumber\\
  &P_\ell(\bn_2\cdot\bn_3)\tilde{S}(\bn_3)-\delta_{\ell,0}\bigg]-\frac{\tilde{C}_\ell}{2\ell+1},\label{eq:openbisp}
\end{align}
where, importantly, the last term containing the angular power spectrum is generated from the term containing the contraction of two connected correlators
\begin{equation}
  \label{eq:sumleg}
  \int\frac{\der\bn_3}{4\pi}P_\ell(\bn_1\cdot\bn_3)P_{\ell'}(\bn_3\cdot\bn_2)=\frac{\delta_{\ell\ell'}}{2\ell+1}P_\ell(\bn_1\cdot\bn_2).
\end{equation}

Since $\tilde{C}_\ell^{(3)}$ is an irreducible correlation of three sources, it must therefore be related to the angular bispectrum, which is often defined from three rank~1 transforms as
\begin{equation}
  \tilde{B}_{\ell_1\ell_2\ell_3}=\sum_{m_1,m_2,m_3}\wignerthreej{\ell_1}{\ell_2}{\ell_3}{m_1}{m_2}{m_3}\tilde{S}_{\ell_1m_1}\tilde{S}_{\ell_2m_2}\tilde{S}_{\ell_3m_3}.
\end{equation}
The angular bispectrum is a rank~(0,0,0) spherical tensor. One condition for non-zero components is that the three multipoles must be able to form a closed triangle with sides having lengths equal to the values of the multipoles. 

Applying \eqn{eq:transform1totransform2} to \eqn{eq:cl3} shows that indeed
\begin{equation}
  \tilde{C}_\ell^{(3)}=\frac{1}{2\ell+1}\sum_{\ell'=0}^{2\ell}\sqrt{\frac{2\ell'+1}{4\pi}}\wignerthreej{\ell}{\ell}{\ell'}{0}{0}{0}\tilde{B}_{\ell\ell\ell'}.
\end{equation}
We can see that $\tilde{C}_\ell^{(3)}$ is a weighted sum of the isosceles configurations of the angular bispectrum with $\ell$ being the length of the equal sides. It coherently compares all allowed open angles of two edges of length $\ell$, and so it is called the open angular bispectrum. It may alternatively be calculated from three rank~1 transforms via
\begin{align}
  \tilde{C}_\ell^{(3)}=\ &\frac{1}{(2\ell+1)^2}\sum_{\ell'}\sum_{m'}\sum_{m_1,m_2}\tilde{S}_{\ell m_1}\tilde{S}_{\ell m_2}\tilde{S}_{\ell'm'}\\
  &\int\der\bn\, Y_\ell^{\,m_1}\!(\bn)\,Y_\ell^{\,m_2}\!(\bn)\,Y_{\ell'}^{\ m'}\!(\bn).\nonumber
\end{align}

\hspace{-\parindent}\textbf{Disjoint Angular Trispectrum:} The final rank~0 tensor at this order is found by contracting four rank~1 transforms,
\begin{equation}
  \tilde{C}_\ell^{(4)}=(\tilde{C}_\ell)^2.
\end{equation}
It connects four field configurations, two joined with one correlator and another two joined with a second correlator. Since this is a disjoint diagram and is a four-point function, it is called the disjoint angular trispectrum. It has units of solid angle squared. Note that this function has connected terms corresponding to the parts of the integrand where any of the four positions coincide.

\section{Maximum Likelihood Analysis of a Gaussian Distribution on a Sphere}
\label{ap:likelihood}
This appendix reviews the derivation of the minimum variance estimator of the angular power spectrum of an angular-Gaussian distribution $\mathcal{G}$ that is statistically stationary and isotropic \citep{Hinshaw2007}. Let $S(\bn)$ be such an observable distributed in this way.

In this discussion, it is convenient to partition the sky into pixels of uniform solid angle $\Omega_{\text{pix}}$ positioned at $\bn_i$ for $i=1,2,\ldots,N_{\text{pix}}$, where $N_{\text{pix}}$ is the number of pixels. Then 
\begin{equation}
  S(\bn_i)=\int_{\Omega_i}\frac{\der\bn'_i}{\Omega_{\text{pix}}}S(\bn_{i'})
\end{equation}
is the value of $S$ in pixel $i$ covering the solid area $\Omega_i$. The distribution $\mathcal{G}[\bar{S}_i,\mathcal{C}_{ij}]$ is uniquely specified by the values 
\begin{equation}
  \bar{S}_i\equiv\meang{S(\bn_i)},
\end{equation}
the ensemble average of $S(\bn_i)$ at each pixel, and the covariance matrix
\begin{align}
  \mathcal{C}_{ij}&=\mathcal{C}(\bn_i,\bn_j)\equiv\meang{\Big(S(\bn_i)-\bar{S}_i\Big)\Big(S(\bn_j)-\bar{S}_j\Big)}\\
  &=\sum_{\ell,m}\sum_{\ell',m'}\meang{S_{\ell}^{\ m}S_{\ell' m'}}Y^*_{\ell m}(\bn_i)Y_{\ell'}^{\ m'}(\bn_j).
\end{align}
Statistical isotropy is defined to mean that the covariance depends only on the angle between the positions, so that
\begin{equation}
  \meang{S_{\ell}^{\ m}S_{\ell' m'}}=\int\der\bn_i\der\bn_j\,Y_\ell^{\,m}(\bn_i)Y^*_{\ell' m'}(\bn_j)\mathcal{C}(\bn_i\cdot\bn_j).
\end{equation}
This is rotation-invariant only if it transforms coordinates as a scalar in multipole indices, requiring the covariance to be such that
\begin{align}
  &\!\!\!\!\!\meang{S_{\ell}^{\ m}S_{\ell' m'}}=\delta_{\ell \ell'}\,\delta^m_{\ \ m'}\frac{1}{2\ell+1}\sum_{m_1=-\ell}^{\ell}\meang{S_{\ell}^{\ m_1}S_{\ell m_1}}\\
  &\qquad=4\pi\delta_{\ell \ell'}\delta^m_{\ \ m'}\!\!\int\frac{\der\bn_i}{4\pi}\frac{\der\bn_j}{4\pi}P_\ell(\bn_i\cdot\bn_j)\mathcal{C}(\bn_i\cdot\bn_j),\nonumber
\end{align}
where the last step made use of the addition theorem in \eqn{eq:harmadd}. Thus, the spherical harmonic transforms are statistically orthogonal, with each mode contributing equally to the mean angular power spectrum for distribution $\mathcal{G}$,
\begin{equation}
  \meang{S_{\ell}^{\ m}S_{\ell' m'}}=\meang{C_\ell}\delta_{\ell \ell'}\,\delta^m_{\ \ m'}.
\end{equation}
These expressions reveal that under the specified conditions, the covariance is simply the Legendre polynomial transform of the mean angular power spectrum,

\begin{equation}
  \label{eq:covarfromcl}
  \mathcal{C}(\bn_i\cdot\bn_j)=\sum_\ell(2\ell+1)P_\ell(\bn_i\cdot\bn_j)\frac{\meang{C_\ell}}{4\pi}.
\end{equation}

Different realizations of a fixed distribution $\mathcal{G}$ on the sphere do not in general have the same angular power spectrum. The variance of $C_\ell$ for different realizations is referred in cosmological contexts to cosmic variance. Though a complete understanding of the distribution $\mathcal{G}$ predicts a well-defined mean spectrum $\meang{C_\ell}$, it cannot predict $C_\ell$ for a single realization. Cosmological experiments can only sample from one realization of $\mathcal{G}$ (we have access to only one Universe), and are therefore limited by cosmic variance in their ability to make inferences on $\mathcal{G}$. Determining the magnitude of cosmic variance for a given distribution is clearly of great importance for scrutinizing physical interpretations of $C_\ell$. This theoretical uncertainty, however, does not place a limit on the precision to which our sky can be sampled and measured, although there generally exist other systematics that do place limitations.

Using Eqns.~(\ref{eq:Plclosure}) and (\ref{eq:sumleg}), the inverse covariance is seen to be
\begin{equation}
  \mathcal{C}^{-1}(\bn_i\cdot\bn_j)=\sum_\ell(2\ell+1)P_\ell(\bn_i\cdot\bn_j)\frac{4\pi}{\meang{C_\ell}}.
\end{equation}

Then the Gaussian probability distribution function of $S(\bn)$ is
\begin{align}
  &P\Big(S(\bn_i)\Big|\bar{S}_i,\meang{C_\ell}\Big)=\left[|\mathcal{C}|(2\pi)^{N_{\text{pix}}}\right]^{-1/2}\\
  &\exp\left\{-\frac{1}{2}\sum_{i,j}\left[S(\bn_i)-\bar{S}_i\right]\mathcal{C}^{-1}(\bn_i\cdot\bn_j)\left[S(\bn_j)-\bar{S}_j\right]\right\},\nonumber
\end{align}
where $|\mathcal{C}|$ is the determinant of the covariance matrix.

Now that we understand how the PDF depends on $C_\ell$, the likelihood analysis of $C_\ell$ begins by writing down the log-likelihood function
\begin{align}
  &\mathcal{L}\Big(\meang{C_\ell}\Big|S(\bn_i)\Big)=-2\ln P\Big(S(\bn_i)\Big|\meang{C_\ell}\Big)\\
  &\!=\!\sum_{i,j}\left[S(\bn_i)-\bar{S}_i\right]\!\mathcal{C}^{-1}(\bn_i\cdot\bn_j)\!\left[S(\bn_j)-\bar{S}_j\right]+\ln|\mathcal{C}|+\cdots\nonumber\\
  &\!=\sum_{\ell, m}\left(S_{\ell}^{\ m}S_{\ell m}\frac{1}{\meang{C_\ell}}+\ln \meang{C_\ell}\right)+\cdots\nonumber
\end{align}
up to terms that do not depend on $C_\ell$.

The maximum likelihood estimator (MLE) for $C_\ell$ is the value of $\meang{C_\ell}$ that maximizes $\mathcal{L}$,
\begin{equation}
  C_{\ell}^{\text{MLE}}=\frac{1}{2\ell+1}\sum_m S_{\ell}^{\ m}S_{\ell m}.
\end{equation}

The variance of the MLE is determined from the convexity of the likelihood function at its maximum
\begin{align}
  \var{C_{\ell}^{\text{MLE}}}&=\left[-\frac{1}{2}\frac{\der^2\mathcal{L}(C_\ell^{\text{MLE}})}{\der(\meang{C_\ell})^2}\right]^{-1/2}\nonumber\\
  &=\frac{2}{2\ell+1}(C_\ell^{\text{MLE}})^2.\label{eq:cosmicvar}
\end{align}
This quantifies the cosmic variance of the power spectrum for realizations of a Gaussian distribution.

It turns out that this estimator has optimal variance since the Fisher matrix (giving the optimum inverse covariance of the power spectrum; see, e.g., \citet{Vallisneri2008}) is 
\begin{align}
  \mathcal{C}_{\text{opt}}^{-1}(C_\ell^{\text{MLE}},C_{\ell'}^{\text{MLE}})&=\frac{1}{2}\frac{\der^2\mathcal{L}(C_\ell^{\text{MLE}})}{\der\meang{C_\ell}\der\meang{C_{\ell'}}}\nonumber\\
  &=\frac{2\ell+1}{2(C_\ell^{\text{MLE}})^2}\delta_{\ell\ell'},
\end{align}
or
\begin{equation}
  \mathcal{C}_{\text{opt}}(C_\ell^{\text{MLE}},C_{\ell'}^{\text{MLE}})=\var{C_\ell^{\text{MLE}}}\,\delta_{\ell\ell'}.
\end{equation}

The presence of noise $\mathscr{N}_\ell$ that is also diagonal in multipole space such that
\begin{equation}
  \meang{S_\ell^{\ m}S_{\ell' m'}}=(\meang{C_\ell}+\mathscr{N}_\ell)\,\delta_{\ell\ell'}\delta^m_{\ \ m'}
\end{equation}
has the simple effect of replacing $C_\ell^{\text{MLE}}$ in the cosmic variance with $C_\ell^{\text{MLE}}+\mathscr{N}_\ell$.

\section{Gaussian Central Moments}
\label{ap:moments}

For convenient reference, here is reviewed the determination of Gaussian central moments. Let $\mathcal{G}$ be a Gaussian distribution of some observables $S_i$ with known means $\meang{S_i}$ and covariance $\mathcal{C}_{i_1i_2}=\meang{S_{i_1}S_{i_2}}-\meang{S_{i_1}}\meang{S_{i_2}}$. Let $T_i\equiv S_i-\mean{S_i}$ so that $\mathcal{C}_{i_1i_2}=\meang{T_{i_1}T_{i_2}}$. Thus, the covariance is the central 2-moment of $S$.

Calculation of the central $n$-moment is done conveniently with generating function techniques (see, for example, \citet{Ramond2001}). Recognize that the central $n$-moment
\begin{align}
  \meang{T_{i_1}T_{i_2}\dotsm T_{i_n}}=\mathcal{A}\int&\left[\prod_k \der T_k\right]T_{i_1}T_{i_2}\dotsm T_{i_n}\nonumber\\
  &\exp\left[-\frac{1}{2}\sum_{i,j}T_i\mathcal{C}^{-1}_{ij}T_j\right]
\end{align}
is equivalent to
\begin{align}
  &\meang{T_{i_1}T_{i_2}\dotsm T_{i_n}}=\lim_{Z\rightarrow 0}\frac{\partial^n}{\partial Z_{i_1}\dotsm\partial Z_{i_n}}\mathcal{A}\int\left[\prod_k \der T_k\right]\nonumber\\
  &\qquad\qquad\exp\left[-\frac{1}{2}\sum_{i,j}T_i\mathcal{C}^{-1}_{ij}T_j+\sum_iT_iZ_i\right]\\
  &=\lim_{Z\rightarrow 0}\frac{\partial^n}{\partial Z_{i_1}\dotsm\partial Z_{i_n}}\exp\left[\frac{1}{2}\sum_{i,j}Z_i\,\mathcal{C}_{ij}\,Z_j\right]\nonumber\\
  &\qquad\qquad\int\left[\prod_k \der \mathcal{T}_k\right]\mathcal{A}\exp\left[-\frac{1}{2}\sum_{i,j}\mathcal{T}_i\,\mathcal{C}^{-1}_{ij}\,\mathcal{T}_j\right],
\end{align}
for arbitrary `generators' $Z_i$ which are all taken to 0 at the end. In the last equality, the substitution
\begin{equation}
  \mathcal{T}_i=T_i-\sum_j\mathcal{C}_{ij}Z_j
\end{equation}
was made. Since the integrand is just the PDF, the integral is 1 and the central moments are generated as
\begin{equation}
  \meang{T_{i_1}T_{i_2}\dotsm T_{i_n}\!}=\!\lim_{Z_i\rightarrow 0}\frac{\partial^n}{\partial Z_{i_1}\!\dotsm\partial Z_{i_n}}\exp\!\left[\!\frac{1}{2}\!\sum_{i,j}Z_i\,\mathcal{C}_{ij}\,Z_j\!\right]\!,
\end{equation}
from which it is straightforward to determine that
\begin{equation}
  \meang{T_{i_1}T_{i_2}\dotsm T_{i_n}\!}=\begin{cases} \mathcal{C}_{i_1i_2}\dotsm\mathcal{C}_{i_{n-1}i_n}+\text{perms.}, &n\text{ even,}\\0, &n\text{ odd},\end{cases}
\end{equation}
where all unique permutations of products of covariances contribute. In particular, $\meang{T_1T_2T_3}=0$ and
\begin{equation}
  \label{eq:4moment}
  \meang{T_1T_2T_3T_4}=\mathcal{C}_{12}\,\mathcal{C}_{34}+\mathcal{C}_{13}\,\mathcal{C}_{24}+\mathcal{C}_{14}\,\mathcal{C}_{23}.
\end{equation}

\end{document}